\RequirePackage[2020-02-02]{latexrelease}
\documentclass[aip,
amsmath,amssymb,preprint,%
author-numerical,%
]{revtex4-2}
\usepackage{graphicx}% Include figure files
\usepackage{dcolumn}% Align table columns on decimal point
\usepackage{bm}% bold math
\usepackage[T1]{fontenc} % optional
\usepackage[utf8]{inputenc}
\usepackage{textalpha}

\usepackage{graphicx}
\usepackage{epstopdf, epsfig}
\usepackage{amssymb}
\usepackage{amsmath}
\usepackage{subfloat}
\usepackage{subcaption}
\def\vec#1{\mbox{\boldmath $#1$}}

\def\Otnm1f{\Omega^\mathrm{f}(t^{\mathrm{n}-1})}
\def\Otnm12f{\Omega^\mathrm{f}(t^{\mathrm{n}-\frac{1}{2}})}

\def\Otnm1s{\Omega^\mathrm{s}_\mathrm{i}(t^{\mathrm{n}-1})}
\def\Otnm12s{\Omega^\mathrm{s}_\mathrm{i}(t^{\mathrm{n}-\frac{1}{2}})}

\def\usnp1{{\vec u}^\mathrm{s,n+1}}

\def\vphinp1{\vec{\varphi}^\mathrm{s,n+1}}

\def\b1{\mbox{\boldmath $1$}}

\def\Otnm1f{\Omega^\mathrm{f}(t^{\mathrm{n}-1})}
\def\Otnm12f{\Omega^\mathrm{f}(t^{\mathrm{n}-\frac{1}{2}})}

\def\Otnm1s{\Omega^\mathrm{s}(t^{\mathrm{n}-1})}
\def\Otnm12s{\Omega^\mathrm{s}(t^{\mathrm{n}-\frac{1}{2}})}

\newcommand{\nwc}{\newcommand}

\nwc{\qref}[1]{(\ref{#1})} %\eqref seems to have a spacing problem

\nwc{\ip}[1]{\langle #1 \rangle}

\nwc{\ta}{\tilde{a}}

\def\Otnm1f{\Omega^\mathrm{f}(t^{\mathrm{n}-1})}
\def\Otnm12f{\Omega^\mathrm{f}(t^{\mathrm{n}-\frac{1}{2}})}

\def\Otnm1s{\Omega^\mathrm{s}(t^{\mathrm{n}-1})}
\def\Otnm12s{\Omega^\mathrm{s}(t^{\mathrm{n}-\frac{1}{2}})}

\begin{document}
%============================================================================ 

\preprint{AIP/123-QED}
\title[Vortex-Induced Vibrations of Circular Cylinder Placed Close to a Wall]{Two-Dimensional Numerical Analysis on Vortex-Induced Vibrations of an Elastically Mounted Circular Cylinder Placed Very Close to a Wall}
% Force line breaks with \\

\author{V Brahmini Priya P}
\author{Aditya Karthik}
\author{Vedanth N K}
\author{Supradeepan K}
\author{P S Gurugubelli}
\email{pardhasg@hyderabad.bits-pilani.ac.in}
\affiliation{Department of Mechanical Engineering, Birla Institute of Technology \& Science - Pilani, Hyderabad Campus, India 500078}%
%\address{\affilnum{1}Department of Mechanical Engineering, National University of Singapore, Singapore 119077.}
%\corauth{Corresponding author. Tel.: + 65 6601 2547; fax: +65 6779 1459.
%\\\textsl{E-mail address:} mperkj@nus.edu.sg (Rajeev Kumar Jaiman).}

%============================================================================ 
\linespread{1.6}
\begin{abstract}
Two-dimensional numerical simulations have been performed to understand the effect of wall proximity on the vortex-induced vibrations (VIV) of an elastically mounted circular cylinder having two degrees of freedom (2-DOF) for gap ratios $g/D = \{0.1,0.2,0.3,0.4,0.5,0.6\}$, where $D$ is the cylinder diameter, and $g$ is the gap between the cylinder's bottom surface and the wall. Parametric simulations have been performed using a quasi-monolithic coupled fluid-structure interaction solver with exact interface tracking for $Re = 100$ and $m^*=10$ over a range of $U^*$ to present the effect of the wall's proximity on the vibration response, vortex structures, force dynamics, pressure distribution, the phase between the hydrodynamic forces and displacements, and finally different VIV branching response regimes. The numerical simulations reveal that as the gap ratio reduces, the maximum transverse vibration amplitude reduces, and the lock-in region widens. Periodic vortex shedding with a single ``S" vortex street has been observed for all gap ratios $g/D = 0.1$ to $0.6$. It has also been observed that the wall proximity affects the mean lift coefficient not only in the lock-in region but also in the pre and post lock-in regions. In the lock-in region, the response dynamics of the vibrating cylinder can be characterized into two branches, with the transition between the branches marked by a sudden jump in the phase angle $\Phi$ between the lift and transverse displacement from $0$ to a value slightly greater than $\pi$. As the gap ratio reduces, the widening of the lock-in region is accompanied by the widening of the $\mathrm{II^{nd}}$ response branch and a narrowing down of the $\mathrm{I^{st}}$ response branch. For the gap ratio of $g/D = 0.1$, the $\mathrm{I^{st}}$ branch vanishes entirely. This work is directly relevant to the design, maintenance, and fatigue life estimation of underwater pipelines/cables laid on the seabed.

% This analysis might help us to get an improved understanding of dynamic responses of subsea pipelines with varying gap ratios, which can lead to a better design and enhanced fatigue life.

%The first challenge in designing sub-sea pipelines is the seabed profile’s uneven nature, leading to spanning along the cylinder length. Vortex-induced vibrations (VIV) can induce dynamic stresses in the channels, which can be detrimental to fatigue life and lead to catastrophic failure. The presence of a wall results in an asymmetry in the development of vortices on the two sides of the cylinder, which impacts flow dynamics, and periodic vortex-induced forces and thereby affects the vibration response of the structures. Proximity distance, boundary layer thickness, and Reynolds number are three parameters affecting the flow past a cylinder near a fixed wall. Vortex-induced vibrations for the near-wall cases have primarily been researched in moderate to high Reynolds number regimes. The purpose of this work is to investigate the effects of vortex-induced vibrations at a low Reynolds number on a $2$-DOF oscillating cylinder at low gap ratios by characterizing structural responses, lift and drag forces, distribution of pressure, phase differences between hydrodynamic forces and displacements, and finally, portraying the different modes of vortex shedding.

\end{abstract}

%\begin{keyword} 
%Near-wall vortex-induced vibration, Circular cylinder, Transverse vibrations, Gap ratio, Lock-in range, Amplitude branches
%\end{keyword}

\maketitle

%\end{frontmatter}
% main text
\section{Introduction}
%\doublespacing
\linespread{1.6}
\label{sec_intro}
% motivation
In the 21st century, the ever-increasing global energy demands and diminishing on-shore oil \& gas resources have resulted in the oil \& gas industries shifting their focus towards deeper offshore resources. One of the main challenges with offshore oil \& gas production is to economically transport the oil \& gas back to the on-shore refineries. Subsea pipelines are typically used for this purpose. However, the complex, uneven seabed surface profiles and the scouring action around the pipes may result in free spanning along the length of the pipeline. The flow action around the pipe may lead to the pipe being suspended above the seabed with a small gap $g$. While span lengths are typically found to be around $O(10^2)$ times the diameter of the cylinder ($D$), the gaps are usually in the range of $g = (0.1 - 1)D$ \citep{Sumer2006}.

Long elastic cylindrical structures when subjected to external fluid flow will experience flow-induced forces due to the vortex shedding phenomenon. Such structures can also undergo large amplitude vibrations due to the resonance synchronization of the structure's natural frequency with the vortex-shedding frequency \citep{Sumer2006,Li2016,gurugubelliCAF2018}. This resonance phenomenon is popularly referred to as vortex-induced vibrations (VIV). VIV is strongly effected by non-dimensional parameters: reduced velocity $(U^*)$, Reynolds number $(Re)$, mass-ratio $m^*$ and the combined mass-damping coefficient $m^*-\zeta$. Due to the numerous practical application of VIV in civil, thermal, and naval engineering, it has been investigated extensively \citep{Williamson2004}. Traditionally, VIV of long elastic pipelines is modelled numerically and experimentally as a spring-mounted cylinder for both 1-DOF (transverse vibrations only) and 2-DOF (cylinder free to vibrate in both the stream-wise and transverse directions with respect to the fluid flow) systems. Most of the previous studies related to VIV have focused on isolated cylinders with no near-wall effects. [\onlinecite{Bearman1978,govardhan2000,Williamson2004,Sarpkaya2004}] have summarized the problem of vortex-induced vibrations for an isolated cylinder and studied the effect of high and loss mass-damping for high Reynolds $(Re)$ numbers. They also provided a thorough analysis of the different modes of vortex shedding for an isolated cylinder, namely the 2S (2 single vortices), 2P (2 pairs of vortices) and P+S (single vortex + pair of vortices) mode, under different conditions. [\onlinecite{govardhan2000}] identified that the response characteristics for the high and low mass-damping cases had distinctly different synchronization regimes. Two response branches, initial and lower were identified for the case of high mass–damping parameter while for the case of low mass-damping parameter, three response branches: initial, high, and low were identified. It was also noted that mode transitions are marked by a jump in the amplitude and phase angle between the transverse displacement and lift force. For the low mass-damping parameter, the upper branch is characterized by high transverse amplitudes and a phase angle close to zero while the lower branch is characterized by lower transverse amplitudes and a phase angle around $\pi$.
Experimental and numerical investigation at low $Re$ for an isolated circular cylinder were carried out by [\onlinecite{Anagnostopoulos1992,Anagnostopoulos1994,Leontini2006,prasanth_mittal_2008}]. Through numerical simulations [\onlinecite{Leontini2006}] identified for the case of low $Re = 200$, two synchronization regimes similar to the upper and lower branch response observed for high $Re$ in [\onlinecite{govardhan2000}]. Unlike [\onlinecite{govardhan2000}], the two regimes of synchronous response identified by [\onlinecite{Leontini2006}] for low $Re$ numbers were not apparent from the maximum transverse displacement response, however, the distinction between the branches was characterized by a shift in the phase angle between the transverse displacement and lift force. 

The presence of a wall strongly impacts the flow around the structure thereby influencing the forces acting on it. The near-wall effects of flow past a stationary circular cylinder were investigated by~[\onlinecite{Taneda1965}] at a low $Re = 170$. Regular vortex shedding was observed at a gap ratio of $g/D = 0.6$, however, only a single line of vortices were shed for the gap ratio $g/D = 0.1$. Similar observations about the suppression of vortex shedding from the bottom surface of the cylinder were reported by~[\onlinecite{Bearman1978}].~[\onlinecite{Zdravkovich1985}] performed wind tunnel experiments for high $Re$ and gap ratios of $g/D \le 1.5$. It was observed that the lift coefficient depended strongly on gap ratio, increasing with decreasing gap ratios, especially for $g/D < 0.5$. On the other hand, the drag coefficient was affected by the ratio of gap to the thickness of boundary layer, decreasing with decreasing gap to thickness ratio. Similar observations were made by~[\onlinecite{Buresti1992}] who performed experiments and noted that vortex shedding from the inner shear layer was suppressed completely at gap ratios less than the critical gap ratio of 0.4, irrespective of boundary layer thickness while~[\onlinecite{Lei1999}] observed that the critical gap ratio decreases with an increase in $\delta/D$ where $\delta$ is the boundary layer thickness at the wall.~[\onlinecite{Price2002}] divided the vortex shedding patterns into four regions, for $g/D \le 0.125$, no vortex shedding was observed but a periodicity was noted in the outer shear-layer. For $0.25 \le g/D \le 0.375$, the vortex characteristics are similar to the earlier case however, a pairing now exists between the shear layer from the bottom of the cylinder and the wall shear layer. On set of vortex shedding is observed for $0.5 \le g/D \le 0.75$ and regular vortex shedding is restored above $g/D \ge 1$. It was also noted that Strouhal number is strongly dependent on $Re$. While the Strouhal number is strongly effected by the gap ratio for low $Re < 2600$, it seems to be insensitive for high $Re \ge 4000$. 
[\onlinecite{Wang2008}] noted that consistent with the earlier observations, vortex shedding is suppressed for $g/D \le 0.3$. It was also presented that the Strouhal number is insensitive to the gap ratio when vortex shedding occurs and remains constant for changing $g/D$. Several numerical simulations too were performed for the problem of a stationary cylinder close to the wall. [\onlinecite{Ong2010, Ong2012}] investigated numerically the flow around a stationary cylinder close to the wall by using a 2D-URANS equation with high $Re$ $k-\epsilon$ model for $0.1 \le g/D \le 1$. It was observed that the mean friction velocity at the seabed was higher for small gap ratios than the higher gap ratios meaning that the sediment transport is higher for the smaller gap ratios. [\onlinecite{Prisc2016}] performed LES simulations for high $Re = 1.3 \times 10^4$ for gap ratios of $g/D = 0.2,0.6$ and $1$ and different boundary layer thicknesses. It was noted that the effects of a thicker boundary layer are similar to the effects of reducing $g/D$.

//
The complexity of the fluid-structure interaction problem increases many fold for an elastically mounted circular cylinder in close proximity of the wall as compared to the stationary cylinder. A number of experimental works focusing on the VIV of a cylinder near a plane wall were found. [\onlinecite{Tsahalis1981}] conducted 2-DOF experiments of an elastically mounted cylinder at high $Re$ for a near-wall case and noted that in the case of the sub-sea pipelines laid over the seabed, the proximity of the seabed surface strongly affected the vortex shedding from the bottom surface thereby modifying the VIV phenomenon. [\onlinecite{Tsahalis1981}] further noted that the proximity of the wall resulted in reduced transverse amplitudes and the the lock-in region started for higher reduced velocities. [\onlinecite{Fredsoe1987}] conducted experiments for 1-DOF VIV and observed that for small gap ratios $0 < g/D < 1$, the vibration frequency is larger than the frequency of vortex shedding in the lock-in region. Further the lift force and transverse displacement were found to increase drastically for very small gap ratios $g/D < 0.1$. [\onlinecite{Yang2009}] experimentally investigated 1-DOF VIV for different gap ratios between $0 < g/D < 5$ and observed that the behaviour for gap ratios $g/D > 0.66$ was distinctly different from the behaviour for $g/D < 0.3$. Further, the width of the lock-in range and the frequency ratio become larger as the mass ratio reduces. [\onlinecite{Hsieh2016,Daneswar2020}] also experimentally investigated 1-DOF VIV of a circular cylinder close to the plane wall. [\onlinecite{Barbosa2017}] performed 2-DOF experiments for the VIV of a near wall circular cylinder. Based on the observations, an existing wake-oscillator model was enhanced to account for the wall-boundary using a damper/spring set up. [\onlinecite{Feher2021}] proposed a wake-oscillator empirical model based the observations from [\onlinecite{Barbosa2017}] to model 2-DOF VIV of a cylinder close to a wall. The model is applicable only for cases with same mass and damping in the transverse and stream-wise directions. Other empirical works were performed by [\onlinecite{Jin2016}] for 1-DOF VIV. 

In addition to the experimental and empirical studies, there have been numerical studies that have investigated the vortex induced vibrations of a circular cylinder placed close to the wall \citep{Zhao2011,Tham2015,Li2016,Chen2020,Chen2021}. [\onlinecite{Zhao2011}] performed 2-DOF two-dimensional numerical simulations using Reynolds Averaged Navier Stokes (RANS) equation and an ALE (Arbitrary Lagrangian Eulerian) scheme with the turbulence model of $k-\omega$ for gap ratios $g/D = 0.002$ and $0.3$ for a range of $Re$ between $1000$ and $15000$. In order to avoid excessive mesh distortion a bounce back coefficient was used and the cylinder would bounce away from the cylinder if the gap reduces below $0.002$D. %Three different modes of vortex shedding were observed: single vortex mode when the reduced velocity is very small; vortex shedding after bounce back mode where vortices are shed from both the top and bottom surface of the cylinder after bounce back; and the vortex shedding before bounce back mode for reduced velocity values in the lock-in region where vortices are shed from the bottom surface of the cylinder before bounce back. 
It was also observed that the resonance region widened with increasing initial condition current velocity. [\onlinecite{Tham2015}] performed a 2-DOF two-dimensional numerical study at low $Re = 100$ for gap ratios of $10 \le g/D \le 0.5$ using a fully implicit combined field scheme based on Petrov-Galerkin formulation. New correlations to characterize the forces and peak amplitudes as a function of the gap ratio were introduced for an elastically mounted circular cylinder vibrating close to a wall. [\onlinecite{Li2016}] used a similar numerical scheme as [\onlinecite{Tham2015}] to perform 2D and 3D simulations for 2-DOF VIV of a cylinder close to the wall and compared the near-wall case with a gap ratio of $g/D = 0.9$ with an isolated cylinder for $Re = 200$ and $1000$. It was observed that stream-wise vibration frequency reduces to half during vortex suppression due to the wall proximity. It was noted that the wall proximity enhanced the mean lift coefficient but did not have a considerable effect on the mean drag coefficient. It further discusses the wall proximity enhanced stream-wise vibrations for the near wall cylinder. [\onlinecite{Chen2019}] performed two-dimensional 2-DOF numerical simulations using Immerse boundary method at $Re = 100$ for gap ratios $g/D = 0.6$ to $3$ and boundary layer thickness $\delta/D = 0$ to $3$. A thorough analysis on the effect of $\delta/D$ is performed and three regimes of flow were identified based on the combined effect of $g/D$ and $\delta/D$. [\onlinecite{Chen2020}] performed two-dimensional 1-DOF numerical simulations for $Re = 200$ and gap ratios $g/D \le 0.5$. Periodic displacement and lift force and a single ``S" vortex street was observed for all gap ratios $0.1 \le g/D \le 0.5$. [\onlinecite{Chen2021}] performed two-dimensional 2-DOF numerical studies for the gap ratio $g/D = 0.1$ for $55 \le Re \le 200$. It was observed that the vortex force component dominates the total fluid force. Further, the oscillation frequency was identified to be larger than the natural frequency for high transverse amplitudes for all $Re$. An equivalent stiffness is proposed to rectify the ratio of the vibration frequency to the natural frequency in resonance. [\onlinecite{Chen2022}] performed 3D Direct Numerical Simulations (DNS) for gap $g/D = 0.8$ at $Re = 500$. It was noted that as the amplitude increased, the three-dimensionality became more prominent resulting in considerable variations in the vortex dynamics. Further, four regions of synchronization were identified by [\onlinecite{Chen2022}] which exhibited different flow and three-dimensionality characteristics. %\vspace{10mm}\newline 

It is evident from the experimental works presented above that the gap ratio plays an important role in defining the VIV response of the free-span sub-sea pipes for $g/D = (0.1 - 1)D$. It is to be noted that there exist numerical simulations that have investigated 2-DOF VIV for $g/D \ge 0.5$ and 1-DOF for $g/D \ge 0.1$. Similar systematic parametric numerical simulations are required to understand the effect of wall proximity on the 2-DOF VIV of the circular cylinder for $g/D<0.5$. The present study performs two-dimensional numerical simulations for the 2-DOF VIV of an elastically mounted circular cylinder in close proximity to the wall for gap ratios $g/D = ~\{0.1,0.2,0.3,0.4,0.5,0.6\}$ at $Re = 100$ and $m^*=10$ over a range of $U^*$ to characterize the structural vibration response, vortex structures, lift and drag forces acting on the structure, phase differences between hydrodynamic forces and displacements, distribution of pressure and finally VIV branching response regimes. The simulations have been performed using a finite-element based quasi-monolithic coupled fluid-structure interaction solver that is $2^\mathrm{nd}$ order accurate in time and $3^\mathrm{rd}$ order accurate in space with exact interface tracking at $Re = 100$ and $m^*=10$ over a range of $U^*$.
%no work presents the 2-DOF VIV of a circular cylinder placed close to the wall for $g/D \le 0.5$. %it was identified that there is a need for the understanding of the vibration and flow dynamics of a cylinder as the gap ratio is varied systematically below $g/D \le 0.6$. 
%While [\onlinecite{Chen2020} performed 1-DOF simulations for $g/D \le 0.5$, it is known that the wall proximity considerably effects the vibration response in the stream-wise direction. Further, it is known that for low mass ratios, that the transverse vibration response is significantly influenced by the stream-wise direction especially in close proximity to the wall. This is the main motivation behind 

The present work is organised as follows: Section \ref{sec:problem_defination} defines the problem statement and the computational domain for the numerical investigation. Section \ref{NumMeth} describes the governing equations, coupled fluid-structure scheme employed and the validation of the scheme used for the investigation. Section \ref{sec:results} presents the the transverse and stream-wise vibration response, force dynamics, vortex structures, frequency characteristics, and phase relations. 

\section{Problem Definition}\label{sec:problem_defination}
\begin{figure}
   \centering
    \includegraphics[scale=0.8]{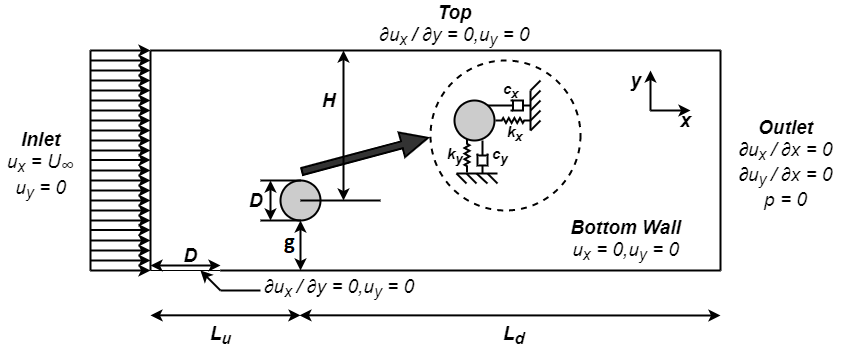}
    \caption{Schematic of computational domain and boundary conditions for freely vibrating cylinder near stationery wall}
    \label{fig:computational domain}
\end{figure}
To understand the vortex-induced vibrations of a circular cylinder that is placed very close to a wall and having two degrees of freedom (2-DOF), two-dimensional incompressible flow over a rigid circular cylinder having a diameter $D$ separated by a small gap $g$ from the wall and mounted on a pair of linear spring-damper setups as shown in figure~\ref{fig:computational domain} is simulated. Since VIV is the synchronization of the structure's natural frequency with the vortex shedding frequency, primary non-dimensional parameters that would influence VIV are: the reduced velocity $U^*=U/f_nD$ that characterizes the natural frequency of the elastically mounted cylinder $(f_n)$, Reynolds number $Re= \rho^\mathrm{f}U_\infty D/\mu^\mathrm{f}$ which characterizes the vortex shedding, and the mass ratio  $m^* = 4M / \rho^{f}\pi D^{2}$ that defines the inertial effects of the fluid-structure interactions. In addition to the three primary non-dimensional parameters that define the VIV, the wall gap-ratio $g/D$ and the non-dimensional boundary layer thickness $\delta/D$ account for the wall proximity induced effects. Here, $U_\infty$ is the free-stream fluid velocity, $f_n=1/2\pi\sqrt{k/M}$ is the natural frequency of an elastically mounted cylinder in the transverse and stream-wise directions where $M$ is the mass of the cylinder and $k$ would represent the spring stiffness, $\rho^\mathrm{f}$ and $\mu^\mathrm{f}$ represent the incompressible fluid density and dynamic viscosity respectively. In this work, damping effects are neglected and both the stream-wise and transverse springs are considered to be linear and have identical stiffness. In this work, the non-dimensional boundary layer thickness near the cylinder is approximately $1.936D$, which is $g/\delta < 0.3 $ and the cylinder would be completely immersed inside the boundary layer.

Figure~\ref{fig:computational domain} presents a rectangular computational domain of size $[41D \times (12D+e)]$ along with the boundary conditions considered for the simulations. A uniform freestream flow having the velocity $(U_\infty,0)$ enters the computational domain from the left boundary and leaves the domain from the right boundary. The centre of cylinder is positioned at a distance $L_u=16D$ from the inlet and $L_d=25D$ from the outlet thereby ensuring sufficiently long downstream wake region to minimize the effect of outflow boundary on the vortex shedding phenomenon. A Dirichlet boundary condition representing the uniform free flow is employed along the inlet boundary and a traction-free outflow boundary condition exposed to atmosphere is considered at the outlet. No-slip wall boundary condition is employed on the bottom wall after a small length $D$ from the inlet boundary. Over this small length slip wall boundary condition is implemented and this boundary condition is enforced to prevent the influence of inlet boundary condition on the evolution of boundary layer. The top boundary of the domain is positioned at a distance $H=12D$ from the cylinder center and slip-wall boundary condition has been considered over this boundary.  
%at the downstream outlet, a traction-free Neumann boundary condition is used with $\partial u_{x}/ \partial x = 0$ \& $\partial u_{y}/ \partial x =0$, at the top boundary of the domain slip wall boundary with $\partial u_{ x}/ \partial y=0$  and $u_y=0$, and the lower boundary consists of two parts, where the first part is slip-wall condition and no-slip condition for the latter part. Boundary layer thickness $\delta = 5(L_u - D)/\sqrt{Re_L}$, where $Re_L = U(L_u - D)/\nu$ and $(L_u - D)$ is the upstream distance for which the no-slip condition condition is applied. 

 The instantaneous force and pressure coefficients are defined as:
\begin{equation}
 C_d= \frac{1}{\frac{1}{2}\rho^{f}U_\infty^{2}D}\int (\vec{\sigma}^{f}\cdot n)\cdot n_{x}d\tau
 \label{eq:1}
\end{equation}
\begin{equation}
 C_L= \frac{1}{\frac{1}{2}\rho^{f}U_\infty^{2}D}\int (\vec{\sigma}^{f}\cdot n)\cdot n_{y}d\tau
\label{eq:2}
\end{equation}
where $n_x$ and $n_y$ are unit normal in $x$ and $y$, respectively. The drag coefficient and the lift coefficient are evaluated from the Cauchy stress function $\vec{\sigma}$. The pressure coefficient is given by:
\begin{equation}
 C_p= \frac{P-P_\infty}{\frac{1}{2}\rho^{f}U_\infty^{2}}
 \label{eq:3}
\end{equation}
Where $P$ and $P_\infty$ represents the pressure at the point and free stream pressure, respectively.
Additionally, typical results such as dimensionless mean transverse amplitude $A_y$, root-mean-square of stream-wise displacement $X_{rms}$, drag $C_D^{rms}$ and lift $C_L^{rms}$ force coefficients are respectively defined as:
\begin{equation}
    A_Y = \frac{1}{2N_Y}\sum_{i=1}^{N_Y} \left(Y_{max}^{i}- Y_{min}^{i}\right), \hspace{2mm} X_{rms} = \sqrt{\frac{1}{N_X}\sum_{i=1}^{N_X} \left(X_{i}- \bar{X_i}\right)^2},
    \label{Amplitude}
\end{equation}
\begin{equation}
     C_D^{rms} = \sqrt{\frac{1}{N_X}\sum_{i=1}^{N_X} \big(C_{D}^{j}- \overline{C_D})^2},
     \hspace{2mm}
     C_L^{rms} = \sqrt{\frac{1}{N_X}\sum_{i=1}^{N_X} \big(C_{L}^{j}- \overline{C_L})^2}
    \label{Drag Coefficients}
\end{equation}
\section{Numerical Methodology}\label{NumMeth}
The VIV of a rigid circular cylinder that placed very close to wall in both stream-wise and transverse direction due to a two-dimensional in-compressible flow field is simulated using an in-house finite element based solver that is $2^\mathrm{nd}$ order accurate in time and $3^{rd}$ order accurate in space. The solver considered for the work modifies the quasi-monolithic approach presented in [\onlinecite{jieJCP}] that is numerically stable for very low mass-ratios for simulating fluid-flexible structures interactions to simulate the fluid-rigid structure interactions. In this section, we first present a brief overview of the quasi-monolithic formulation that has been developed for the simulating fluid-rigid body interactions and then validate the ability of the solver to simulate the VIV of circular cylinder placed close to a wall. 

\subsection{Governing Equations and Formulation}\label{formulation}
The quasi-monolithic formulation developed for the current work considers an arbitrary Lagrangian Eulerian (ALE) description for modeling the fluid flow domain that is deforming due to the VIV of the elastically mounted cylinder. The Navier-Stokes equations describing the incompressible fluid flow in an ALE reference frame over the fluid domain $\Omega^{\mathrm{f(\mathrm{t})}}$ are given as 
 \begin{align}
 \rho^\mathrm{f}\frac{\partial\vec{u}^\mathrm{f}}{\partial t} + \rho^\mathrm{f}(\vec{u}^\mathrm{f}-\vec{w})\cdot \nabla\vec{u}^\mathrm{f} = \nabla \cdot \vec{\sigma}^\mathrm{f} + {b}^\mathrm{f}\hspace{5mm}on\hspace{1mm} \Omega^\mathrm{f}(\mathrm{t})
  \label{navier} \\
  \nabla\cdot\vec{u}^\mathrm{f} = 0 \hspace{5mm}on\hspace{1mm}\Omega^\mathrm{f}(\mathrm{t})
\label{continuity}
\end{align}
where $\rho^\mathrm{f}$ is the fluid density, $\vec{u}^\mathrm{f} =\vec{u}^\mathrm{f}({x},t)$ is the fluid velocity defined at an ALE point ${x}$ in $\Omega^\mathrm{f} (0)$ that would be moving arbitrarily inside the domain to accommodate the vibration of the rigid cylinder. Here $ w = w ({x},t)$ denotes velocity with which the point ${x}$ is moving, $b^\mathrm{f}$ is the body force acting per unit volume of fluid and $\sigma^\mathrm{f}$ is the Cauchy stress tensor for a Newtonian fluid which is defined as 
\begin{equation}
    \sigma^\mathrm{f} = -p\vec{I}+ \mu^\mathrm{f}\big(\Delta \vec{u}^\mathrm{f} + (\Delta \vec{u}^\mathrm{f})^T\big)
    \label{cauchy stress}
\end{equation}
where $p$ is the fluid pressure, $\vec{I}$ is the identity tensor and $\mu$ is fluid dynamic viscosity.

The dynamics of a rigid cylinder mounted on a spring-damper system and free to move in the two-dimensional space due to the fluid dynamic forces acting on it can be represented using:
\begin{equation}
    {m}\frac{\partial\vec{u}^\mathrm{s}}{\partial t} + \vec{c}\cdot\vec{u}^\mathrm{s} + \vec{k}\cdot\big(\varphi^\mathrm{s}(\vec{z}_0,t) - {z}_0\big) = \vec{F}^\mathrm{s} + \vec{B}^\mathrm{s} , \hspace{5mm} on \hspace{1mm} \Omega^\mathrm{s}
    \label{rigid body motion}
\end{equation}
where $\vec{u}^\mathrm{s}$ is the velocity with which the elastically mounted rigid body is vibrating. In the above equation, $m$ denotes the mass of the structural system, $\vec{c}$ and $\vec{k}$ represent damping and stiffness vectors for each translation degree of freedom respectively. $\vec{F}^\mathrm{s}$ and $\vec{B}^\mathrm{s}$ denote the fluid dynamic forces and body forces acting on the rigid body. $\varphi^\mathrm{s}$ and $\vec{z}_0$ are the position vector and the initial position of rigid body. 

The fluid forces $\vec{F}^\mathrm{s}$ acting on the rigid body can be determined using the traction continuity condition along the fluid-structure interface $\Gamma$
\begin{equation}
  \int_{\Gamma(t)}\sigma^\mathrm{f}({x},t)\cdot n d\Gamma + \vec{F}^\mathrm{s} = 0  \label{Traction_bc}.
\end{equation}
In addition to the traction continuity, the interface between the fluid-structure $\Gamma$ should also satisfy the no-slip condition given by
\begin{equation}
     \vec{u}^{f}({x},t)=\vec{u}^{s}({z}_{0},t) \quad \forall {x}\in\Gamma.
     \label{velocity_bc}
\end{equation}

The arbitrary movement of an ALE point ${x}$ inside the fluid domain due to the VIV of elastically mounted cylinder can be assumed as a pseudo-linearly elastic material given by [\onlinecite{Hughes1981}]: 
\begin{equation}
    {\nabla}\cdot (1 + k_m)\Bigg[\big({\nabla}{\eta}^\mathrm{f} +({\nabla}{\eta}^\mathrm{f})^T\big)+ ({\nabla}\cdot {\eta}^\mathrm{f}) I\Bigg] =  0
    \label{meshequation}
\end{equation}
where ${\eta}^\mathrm{f}$ denotes the displacement of ${x}$ and $k_m$ is the local variable stiffness which can be chosen as a function of area of the finite element mesh to minimize the distortion of the mesh. For more details about choosing the $k_m$ refer [\onlinecite{donea2004}].
The weak variational form of the combined fluid-structure system can be constructed by multiplying the governing equation~\ref{navier} with the fluid velocity weight function $\phi^\mathrm{f}$, equation \ref{continuity} with $q$ the weight function for fluid pressure, adding them together along with equations \ref{rigid body motion} and  \ref{Traction_bc}, we get
\begin{align}
    \int_{\Omega^\mathrm{f}(\mathrm{t})}\rho^\mathrm{f}\big(\partial_t  u^\mathrm{f} + ( u^\mathrm{f} -  w)\cdot (\nabla) \vec{u}^\mathrm{f}\big)\cdot (\phi)^\mathrm{f}d\Omega + 
    \int_{\Omega^\mathrm{f}(\mathrm{t})} {\sigma}^\mathrm{f}:\nabla \phi^\mathrm{f} d\Omega \nonumber\\ - 
    \int_{\Omega^\mathrm{f}(\mathrm{t})} \nabla\cdot  u^\mathrm{f} q d\Omega \nonumber\\ 
+ \big[{m}\frac{\partial\vec{u}^\mathrm{s}}{\partial t} + \vec{c} \cdot\vec{u}^\mathrm{s} + \vec{k}\cdot\big(\varphi^\mathrm{s}(\vec{z}_0,t) - {z}_0\big)\big] = \nonumber\\
    \int_{\Omega^\mathrm{f}(\mathrm{t})}  b^\mathrm{f}\cdot(\phi)^\mathrm{f}d\Omega + \int_{\Gamma_h^\mathrm{f}(\mathrm{t})}  h^\mathrm{f} \cdot (\phi)^\mathrm{f}d\Gamma + \vec{B}^\mathrm{s}, \label{CFEI}
\end{align}
where $\Gamma_h^\mathrm{f}(\mathrm{t})$ denotes the Neumann boundary along which ${\sigma}^\mathrm{f}( x,t)\cdot n^\mathrm{f} =  h^\mathrm{f}$, and $\partial_t$ is partial time derivative that is approximated using $2^\mathrm{nd}$ order backward Euler. The above weak form implicitly satisfies the traction, displacement and the velocity continuity conditions. Hence, the coupled equation offers unconditional numerical stability even for very low structure-to-fluid mass ratios. Additionally, it should be noted that the mesh deformation algorithm is not part of the combined weak variational form in equation~\ref{CFEI} and instead it is decoupled from the coupled equation by explicitly predicting the structural positions at the start of each time step. Decoupling of the mesh deformation algorithms reduces the complexity of the coupled fluid-structure system of linear equations that are required to be solved at the end of each time step substantially. For detailed step-by-step derivation of the above weak form, numerical stability and computational accuracy refer [\onlinecite{jieJCP}] and [\onlinecite{jaiman2015fully}].
%\subsection{ALE reference frame}\vspace{5mm}An arbitrary 
%\vspace{1mm}

\subsection{Validation}\label{subsec: validation}
\begin{figure} 
\centering
{\includegraphics[width = 1\linewidth]{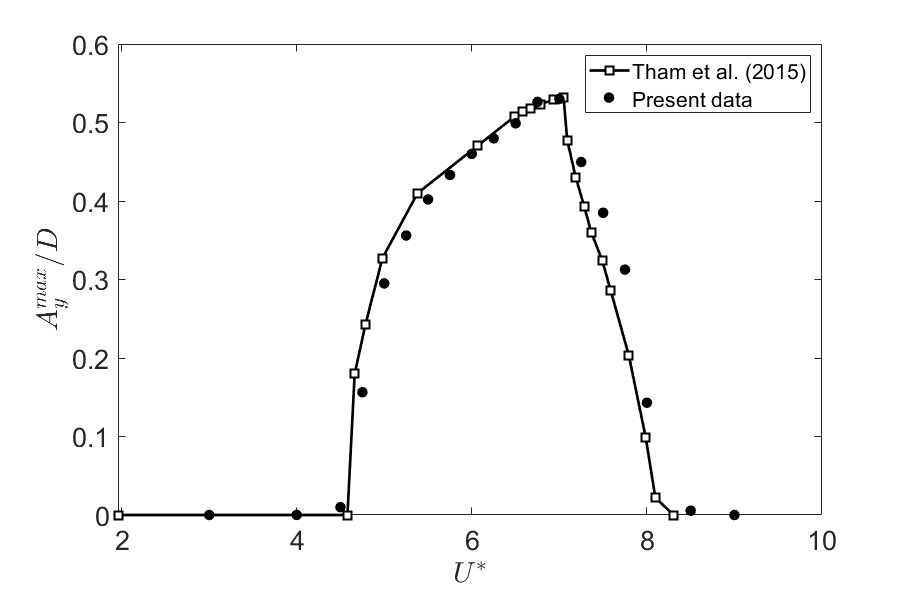}}
\caption{Comparison of maximum transverse amplitude with [\onlinecite{Tham2015}] for a cylinder near the wall at $m^* = 10$, $\zeta = 0$, $g/D = 0.75$ \& $Re = 100$.}
\label{validation_ymax}
\end{figure}

Before numerically investigating the VIV of a 2-DOF elastically mounted cylinder that is in close proximity to a wall, it is necessary to validate the numerical methodology presented in Section~\ref{formulation} and show its capability to accurately simulate the selected problem described in Section~\ref{sec:problem_defination}. To validate the present numerical methodology, flow past a circular cylinder with $m^* = 10$ at $g/D = 0.75$ and $Re = 100$  for a range of $U^*$ from 2 to 9. The computational domain and boundary conditions considered for the validation are the same as the one present in [\onlinecite{Tham2015}]. Figure~\ref{validation_ymax} compares the non-dimensional maximum transverse amplitude, i.e. $A_y^{max}/D$, from the parametric simulations against the data reported. The figure shows that the results produced using the formulation presented in Section~\ref{formulation} are in good agreement with those in the literature. 

% \begin{figure}
% 		\centering
% 		\begin{subfigure}
% 		\centering
% 			\includegraphics[trim={5mm 0mm 30mm 0mm},clip,width=0.49\textwidth]{Pictures/validation_Ay.eps}
% 		\end{subfigure}
% 		\begin{subfigure}
% 			\centering
% 			\includegraphics[trim={5mm 0mm 30mm 0mm},clip,width=0.49\textwidth,]{Pictures/validation_Xrms.eps}
% 		\end{subfigure}\\
% 		\caption{Comparison of maximum transverse amplitude \& root-mean-squared stream-wise amplitude with [\onlinecite{CHEN2021103247} \& [\onlinecite{THAM2015103} for a cylinder near the wall at $m^* = 10$, $\zeta = 0$, $g/D = 0.5$ \& $Re = 100$.}
% \end{figure}
\section{Results and Discussion}\label{sec:results}
\begin{figure}
    \includegraphics[width=0.63\textwidth,trim={35mm 7mm 3cm 0cm}, clip]{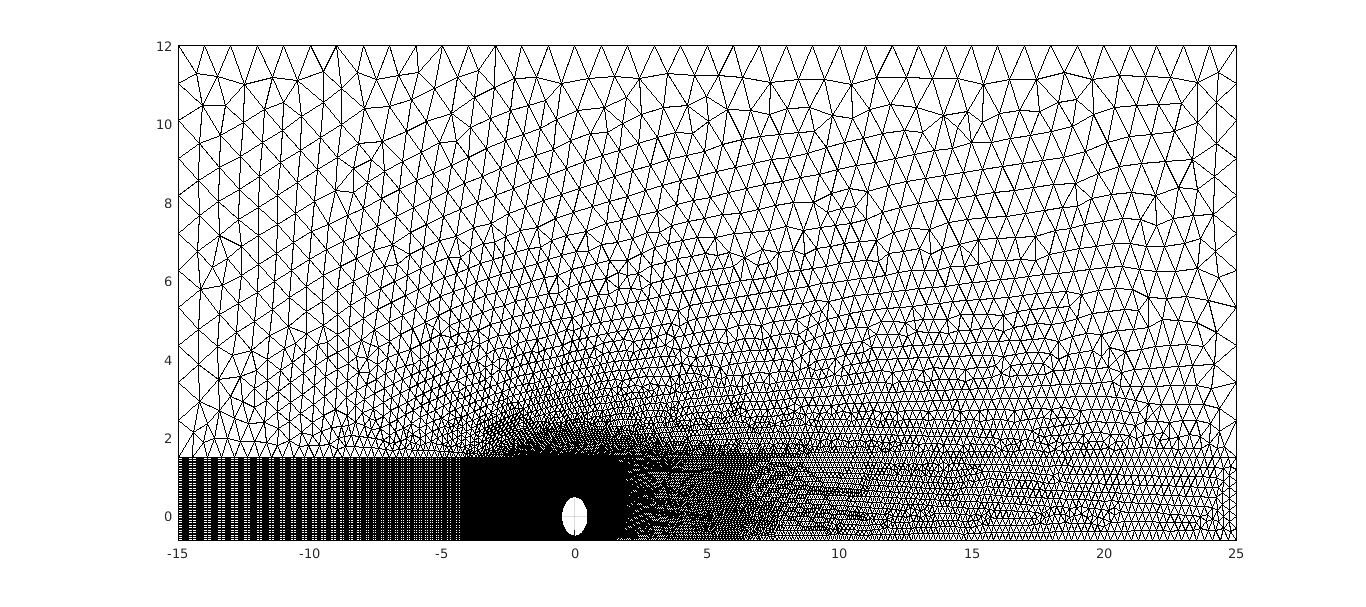}
    \includegraphics[width=0.36\textwidth,trim={10cm 6mm 9cm 2mm}, clip]{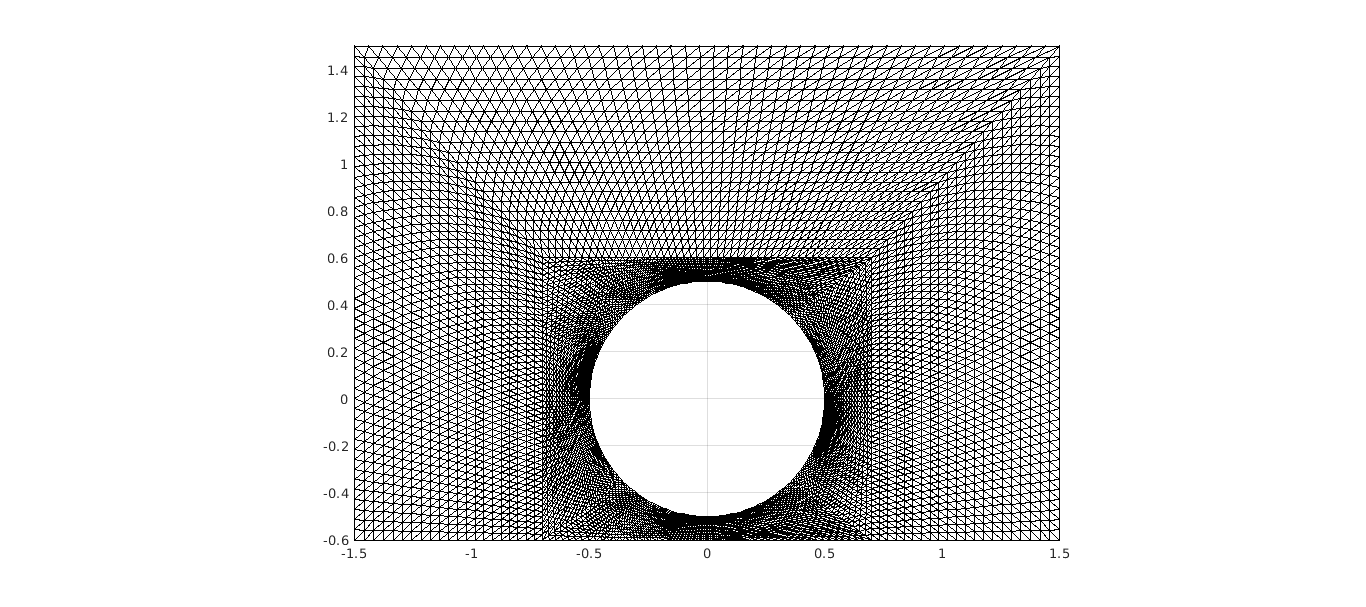}
    \caption{The typical quadratic six-noded triangular fluid mesh considered for the gap-ratio of $g/D=0.1$ (left) and closeup view of the mesh around the circular cylinder (right).}
    \label{fig:mesh}
\end{figure}

Systematic parametric two-dimensional numerical simulations are performed for the elastically mounted circular cylinder described in section~\ref{sec:problem_defination} at $g/D\in\{0.1,0.2,0.3,0.4,0.5,0.6\}$.  Simulations were performed until the non-dimensionless time of $tU_0/D = 400$ with a time step size ($\Delta t U_0/D$) of 0.01, and the results are discussed over a time range of $tU_0/D = 150-400$ where the response dynamics have stabilized. The fluid mesh is constructed by considering the resolution around the cylinder and in the wake similar to that of the validation presented in section~\ref{subsec: validation}. The figure~\ref{fig:mesh} presents the typical fluid mesh and the closeup view of the mesh around the cylinder for the case of gap-ratio $g/D=0.1$. The fluid mesh in figure~\ref{fig:mesh} consists of 65,902 mesh nodes and 32,642 $\mathbb{P}_2|\mathbb{P}_1$ triangular mesh elements describing fluid velocity and pressure respectively. To further confirm if the mesh considered is resolved the cylinder maximum response and force parameters at $g/D=0.1,~m^*=10,~Re=100$ and $U^*=20$ are compared against a mesh consisting of roughly double the number of elements (i.e. 65,231). The maximum difference in the parameters was observed to be less than $1.2\%$.

\subsection{Response Kinematics}\label{subsec: res_kin}
\begin{figure}
\centering
{\includegraphics[width = 0.85\linewidth]{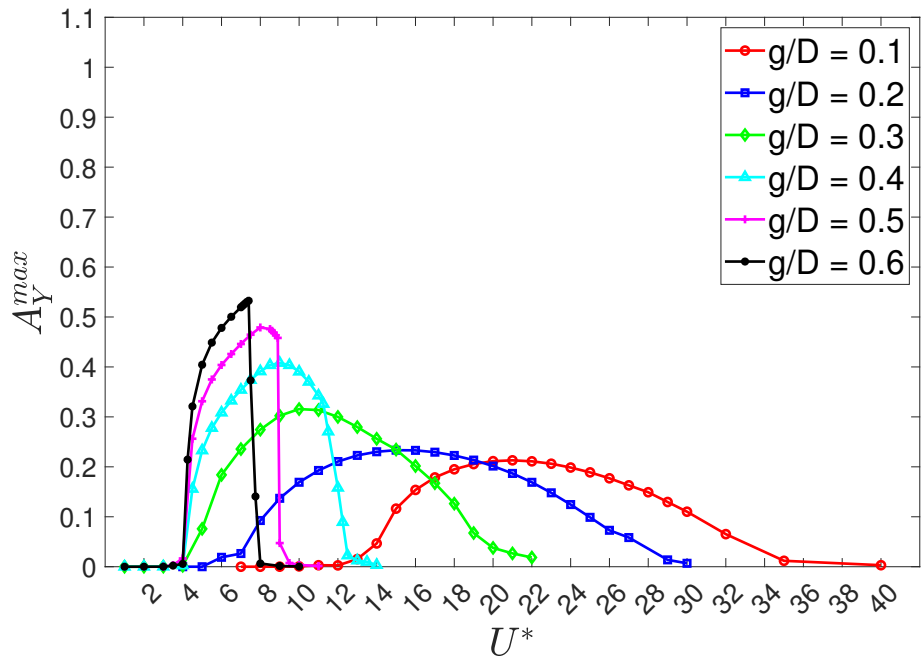}}
\caption{Variation of maximum transverse displacements$(A_{y}^{max}/D)$ with $U^*$ for $Re=100$, $m^*=10$ and $\zeta=0$.}
\label{ay_max}
\end{figure}

\begin{figure}
\centering
{\includegraphics[width = 0.85\linewidth]{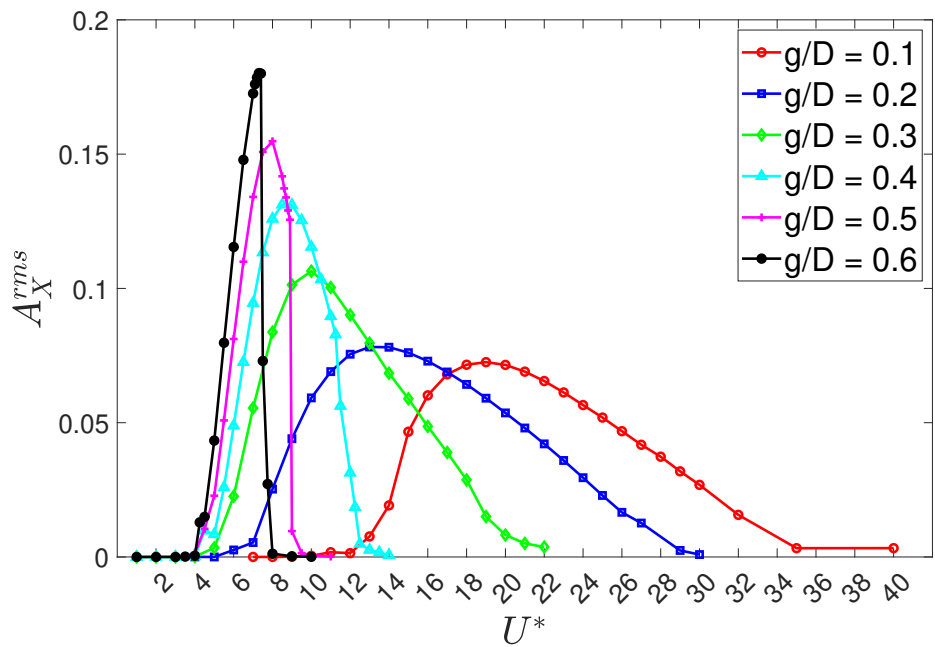}}
\caption{Variation of root-mean-squared stream-wise displacement$(A_{X}^{rms}/D)$ with $U^*$ for $Re=100$, $m^*=10$ and $\zeta=0$.}
\label{axrms}
\end{figure}

Figures~\ref{ay_max} and \ref{axrms} present the dimensionless maximum transverse displacements $(A_{y}^{max}/D)$ and the dimensionless root-mean-squared stream-wise displacements $(A_{x}^{rms}/D)$ as a function of different $U^*$ values for different gap ratios. Both these figures are distinctly different from their counterparts for gaps $g/D>0.6$, wherein the cylinder exhibits large transverse amplitudes for $4 \le U^* \le 8$ [\onlinecite{Tham2015}]. In figure~\ref{ay_max}, for smaller gap ratios of $g/D < 0.5$ the high amplitude response sets in for much higher $U^*$ values and the response smoothens i.e. the vibration range widens and the amplitude reduces. This effect becomes more pronounced for $g/D \le 0.3$. For example at $g/D = 0.1$, the large amplitude response begins only from $U^*=12$ and the lock-in region ranges over $12 < U^* <35$ with no sharp peaks. Similarly for $g/D=0.2$, the transverse vibration response starts for $U^*\ge 6$ and lasts up to $U^* < 30$. This widening of the lock-in region might be a consequence of the increased added mass component due to the wall proximity-induced effects, which in turn increase the effective mass of the system and reduce the natural frequency of the vibrating cylinder system. A similar hypothesis was proposed in [\onlinecite{Chen2021}] to account for the effect of wall proximity on the natural frequency. It can also be observed from figure~\ref{ay_max} that with decreasing $g/D$ the $U^*$ values for which peak $(A_{y}^{max}/D)$ occurs shift rightward. For $g/D=0.6$ and $0.5$, $(A_{y}^{max}/D)$ increases gradually over a range of $U^*$ before decreasing steeply to zero which is unlike the vortex-induced vibrations of an isolated cylinder where $(A_{y}^{max}/D)$ showed a steep increase over a small range of reduced velocity before gradually decreasing over a range of $U^*$ values. This observation also conforms with the observations reported in [\onlinecite{Tham2015}] for $1 \le g/D \le 0.5$. 


Similar effects of the wall proximity can be observed for the non-dimensional root-mean-squared stream-wise amplitude $A_X^{rms}/D$ presented in figure~\ref{axrms} as a function of $U^*$. As the gap ratio reduces, the lock-in region widens and the vibration amplitude reduces. A rightward shift of the lock-in region accompanied by the smoothening of the curve can also be observed as $g/D$ reduces. These observations are in accordance with the results observed in [\onlinecite{Sumer2006}] and [\onlinecite{Tsahalis1981}]. 

\begin{figure} 
\centering
{\includegraphics[width = 0.85\linewidth]{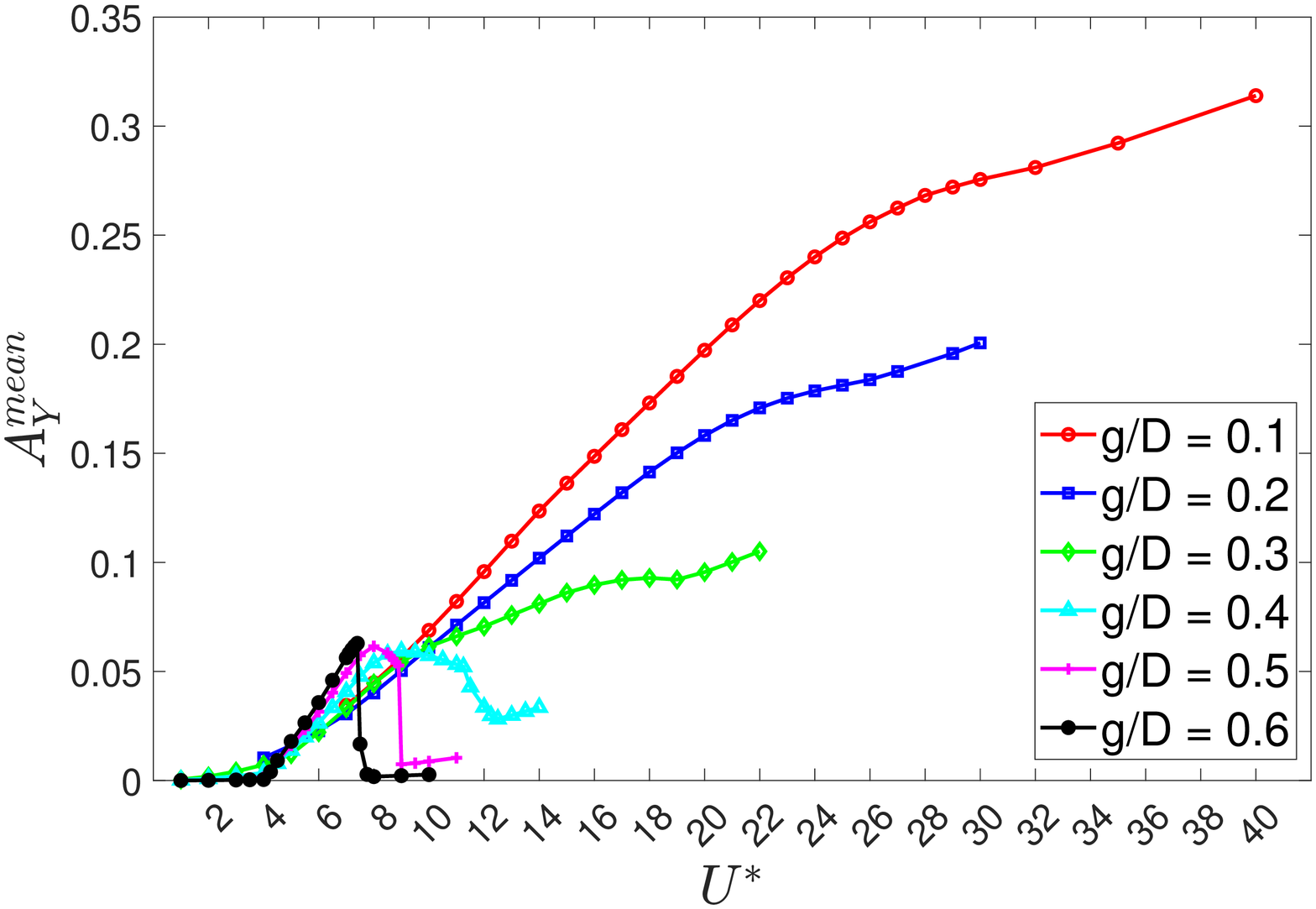}}
\caption{Variation of dimensionless mean transverse displacement$(A_{y}^{mean}/D)$ with $U^*$ for $Re=100$, $m^*=10$ and $\zeta=0$.}
\label{mean_ay}
\end{figure}

%\begin{figure} 
%\centering
%{\includegraphics[width = 0.85\linewidth]{Pictures/mean_ax.eps}}
%\caption{Variation of dimensionless mean stream-wise displacement$(A_{x}^{mean}/D)$ with $U^*$ for $Re=100$, $m^*=10$ and $\zeta=0$.}
%\label{mean_ax}
%\end{figure}
Figure~\ref{mean_ay} presents the variation of non-dimensional mean transverse amplitude ${A}_Y^{mean}/D$ as a function of $U^*$ for different gap ratios. For relatively large gap ratios of $g/D = 0.5$ and $0.6$, the effect of the wall proximity can predominantly be observed in the lock-in region. However, as the gap ratio reduces below $g/D < 0.5$, large $A_Y^{mean}$ values can be observed in both lock-in and post lock-in regions. As the gap ratio reduces, there is a substantial increase in the maximum $A_Y^{mean}$ values observed in the lock-in region. Additionally for $g/D \le 0.3$, the transition from the lock-in to the post lock-in region is not marked by a sudden drop in the $A_Y^{mean}$.

\subsection{Vortex Dynamics}\label{subsec: res_vorDyn}

Vorticity contours have been plotted to better understand the variation of transverse amplitudes with decreasing gap ratios. It is known from previous works that the wall proximity results in the hindered development of the shear layer on the bottom surface of the cylinder, which in turn results in the suppression of regular vortex shedding. For $g/D = 0.9$, [\onlinecite{Li2016}] observed a single ``S" vortex street for low transverse amplitudes and a ``2S" vortex street for higher transverse amplitudes. However, for very small gap ratios of $g/D \le 0.5$, [\onlinecite{Chen2020}] observed a single ``S" vortex street over the entire lock-in region. Similar results have been observed in the present study where a single ``S" vortex street is observed for the gap ratios of $g/D = 0.1 - 0.6$ over the complete lock-in region.

\begin{figure} 
    \begin{subfigure}{0.49\textwidth}
    \includegraphics[width=0.99\columnwidth]{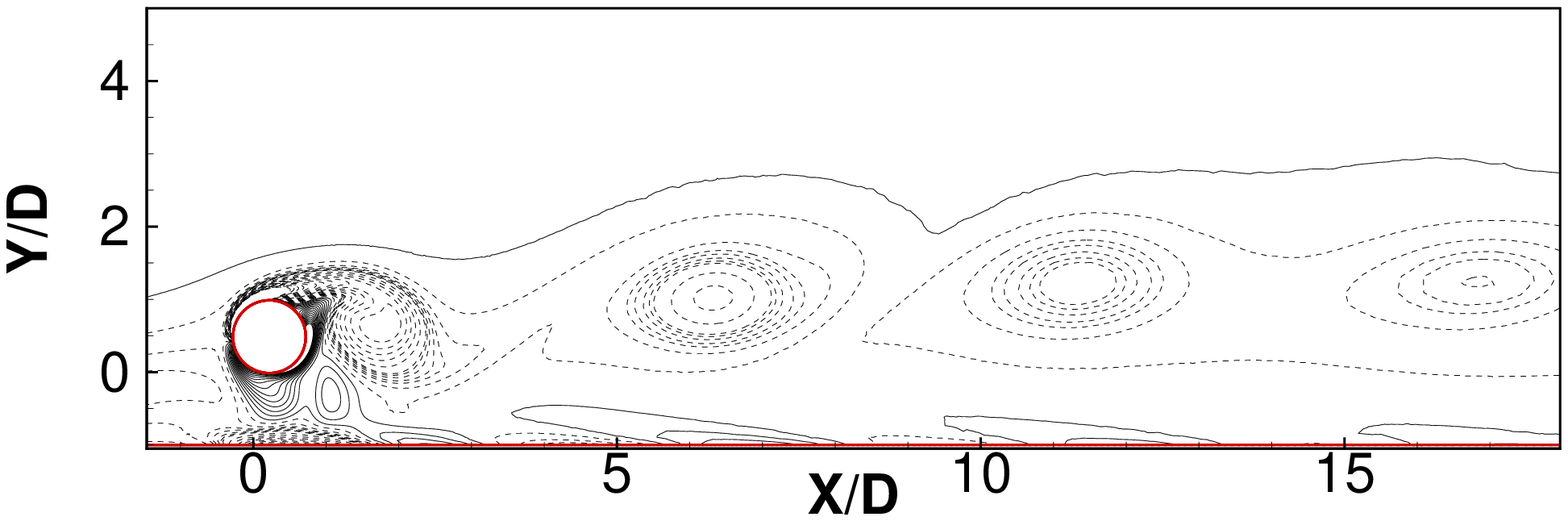}
    \caption{$t=0$}\label{vor_5_1}
    \end{subfigure}
    \begin{subfigure}{0.49\textwidth}
    \includegraphics[width=0.99\columnwidth]{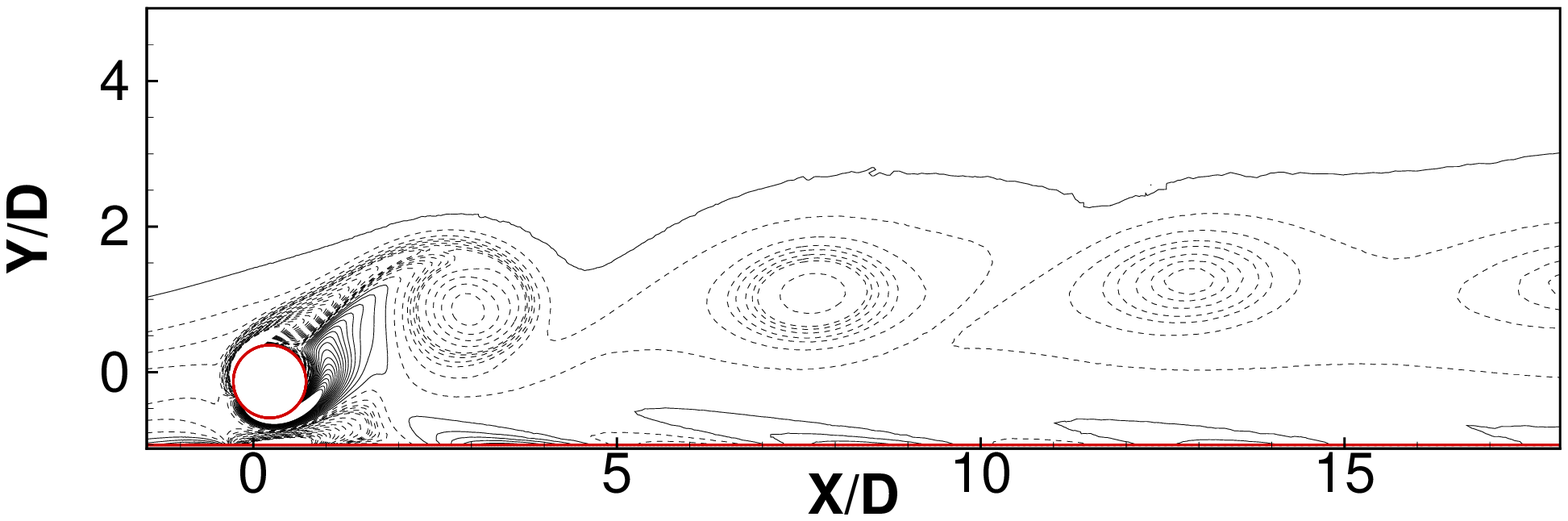}
    \caption{$t=T/4$}\label{vor_5_2}
    \end{subfigure}
    \begin{subfigure}{0.49\textwidth}
    \includegraphics[width=0.99\columnwidth]{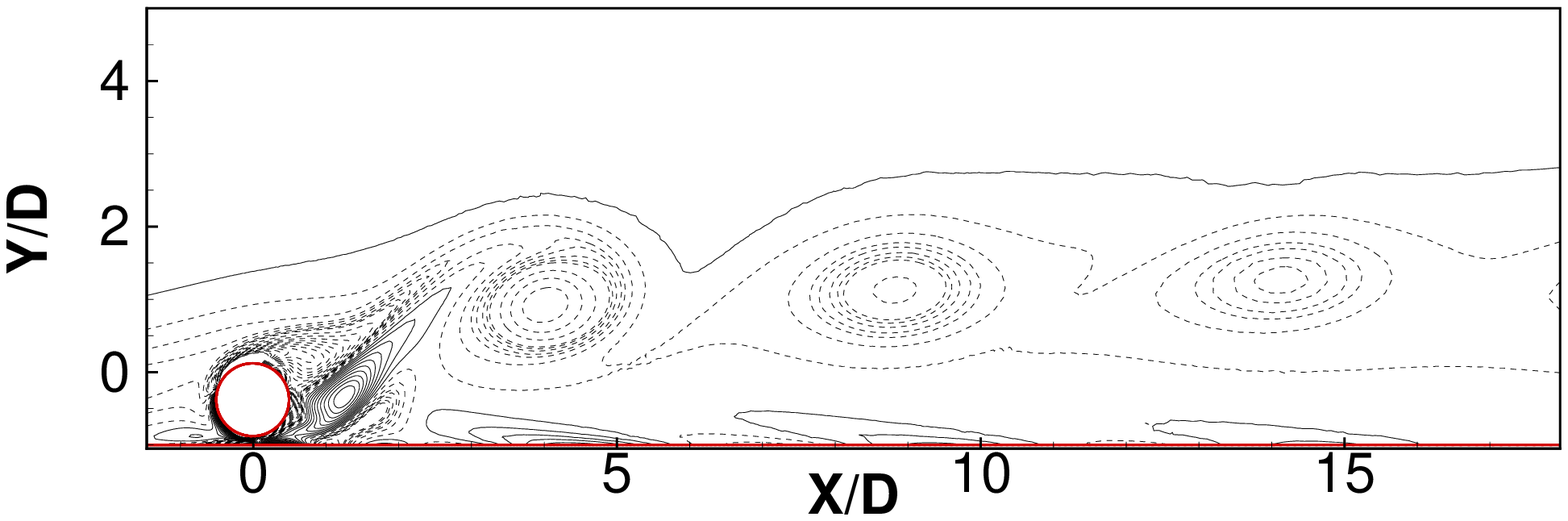}
    \caption{$t=T/2$}\label{vor_5_3}
    \end{subfigure}
    \begin{subfigure}{0.49\textwidth}
    \includegraphics[width=0.99\columnwidth]{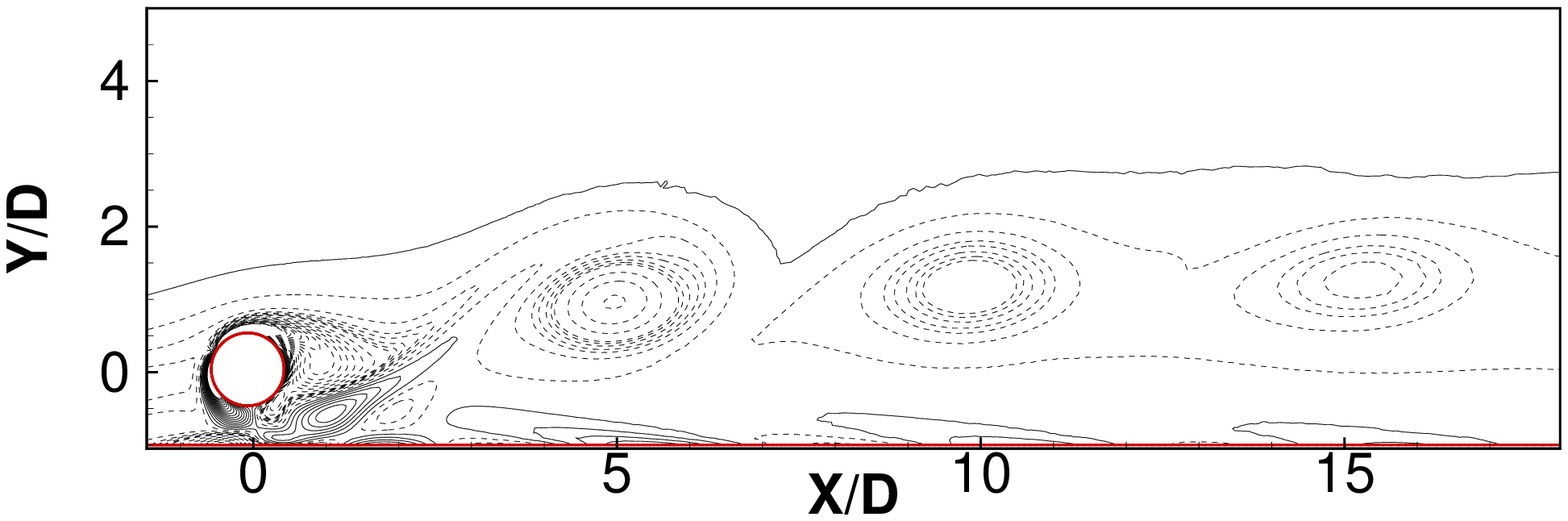}
    \caption{$t=3T/4$}\label{vor_5_4}
    \end{subfigure}
    \caption{Vortex shedding pattern at four equally spaced time snapshots a) $t = 0$, b) $t = T/4$, c) $t = T/2$ and d) $t = 3T/4$  for $U^* = 7$ and $g/D = 0.5$ at $Re = 100$, where $T$ represents time period of one transverse oscillation of the cylinder}\label{vortex_G5}
\end{figure}
\begin{figure} 
    \begin{subfigure}{0.24\textwidth}
    \includegraphics[width=0.99\columnwidth]{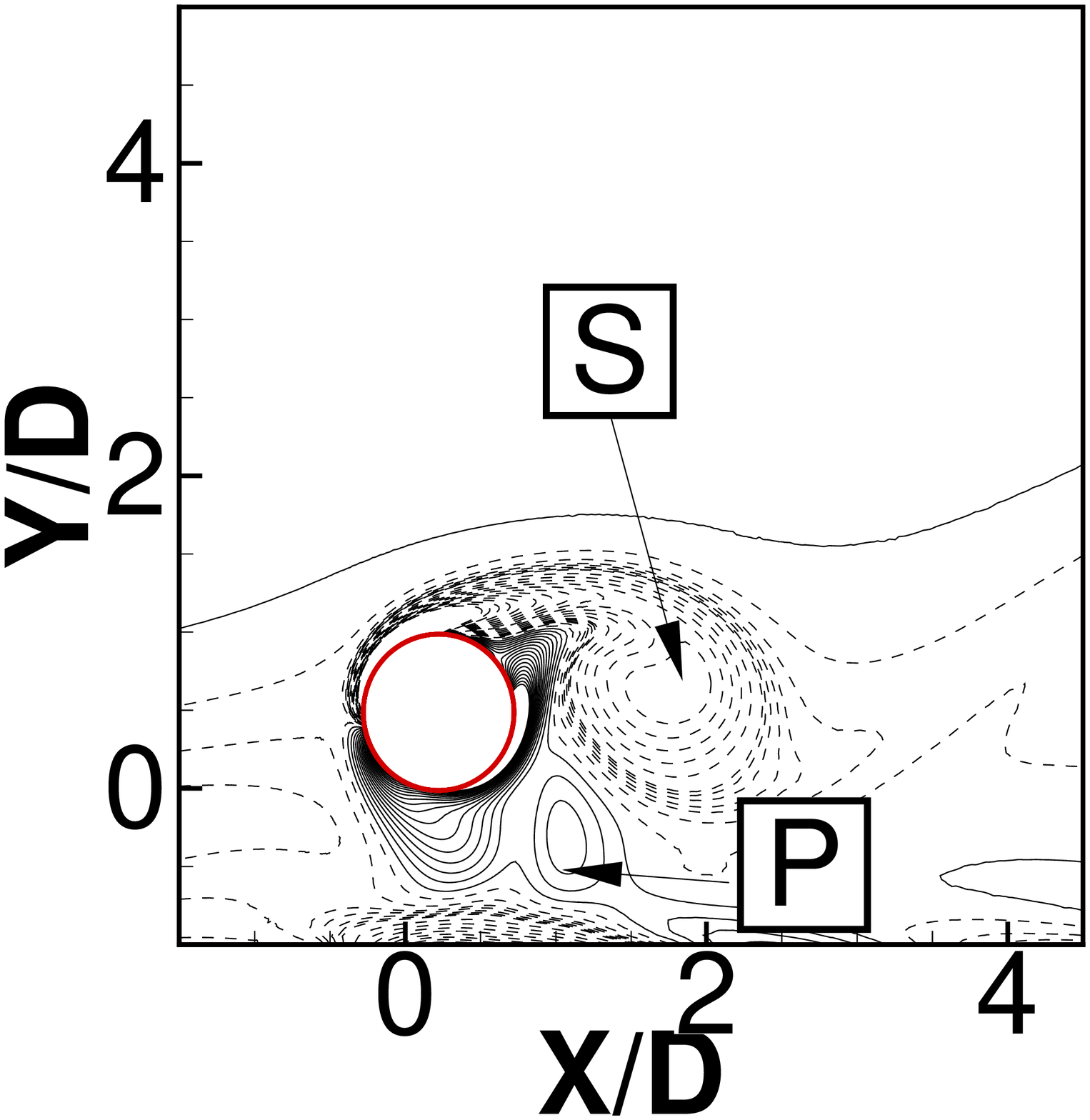}
    \caption{$t=0$}\label{vorc_5_1}
    \end{subfigure}
    \begin{subfigure}{0.24\textwidth}
    \includegraphics[width=0.99\columnwidth]{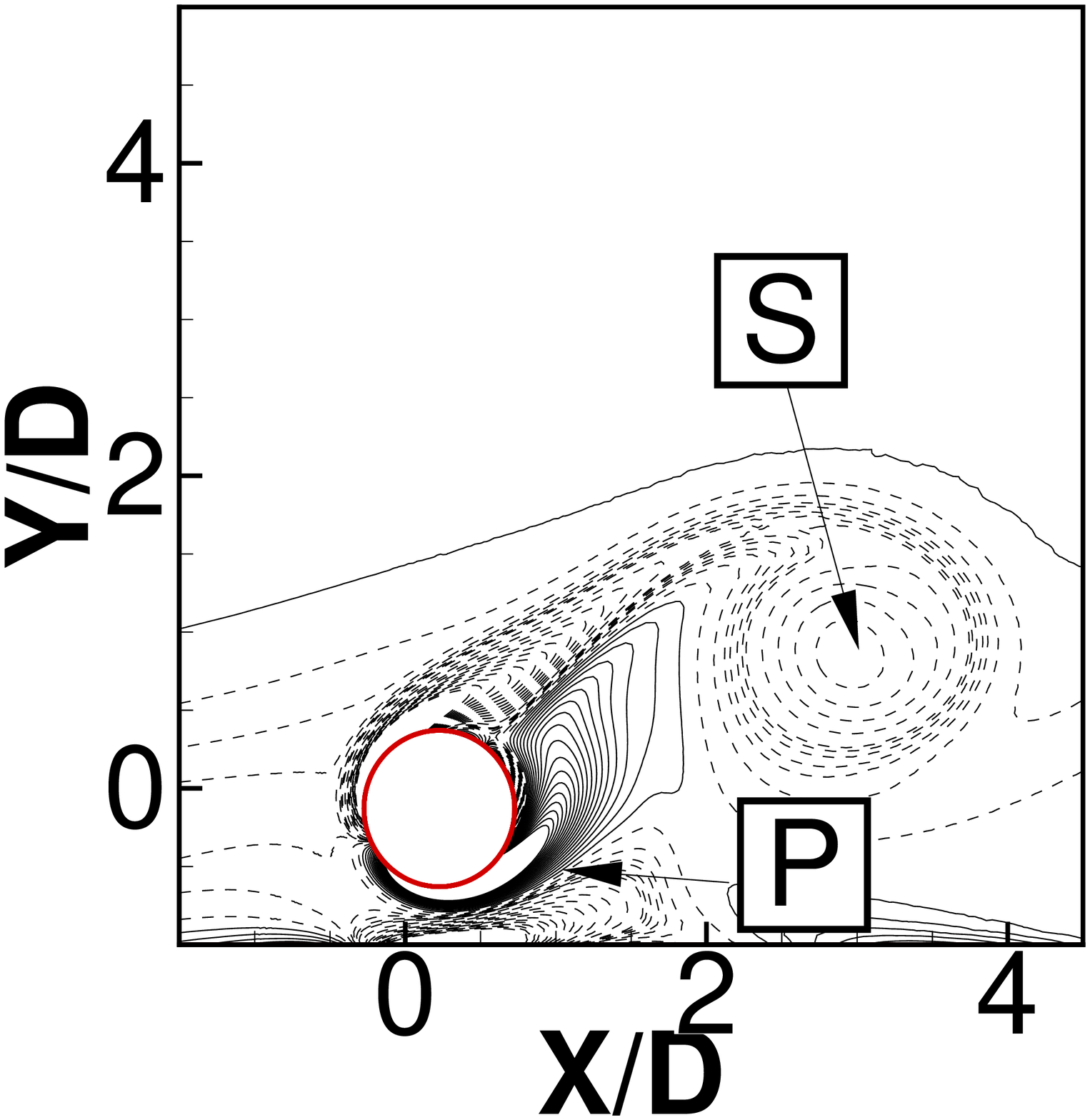}
    \caption{$t=0$}\label{vorc_5_2}
    \end{subfigure}
    \begin{subfigure}{0.24\textwidth}
    \includegraphics[width=0.99\columnwidth]{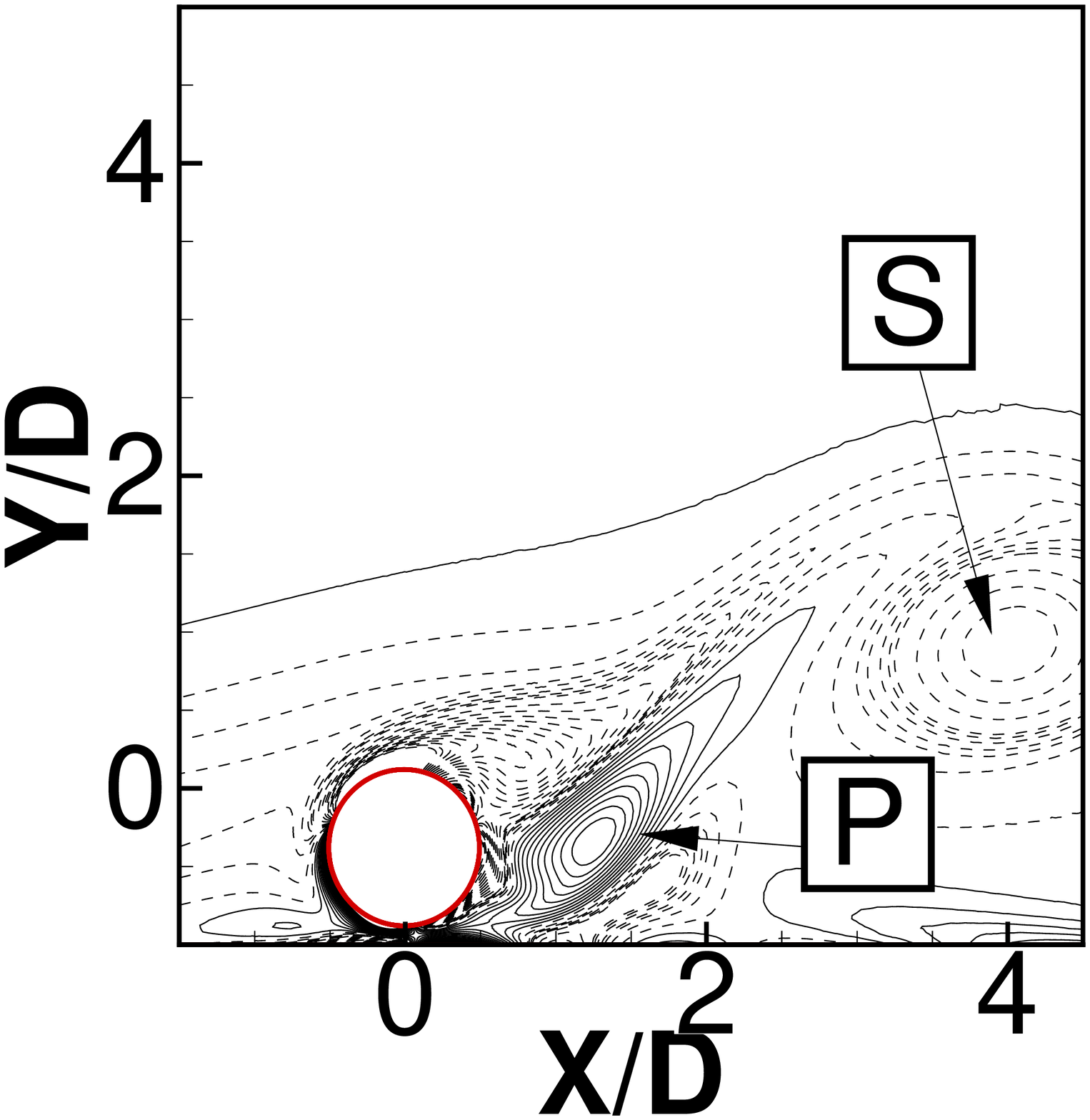}
    \caption{$t=0$}\label{vorc_5_3}
    \end{subfigure}
    \begin{subfigure}{0.24\textwidth}
    \includegraphics[width=0.99\columnwidth]{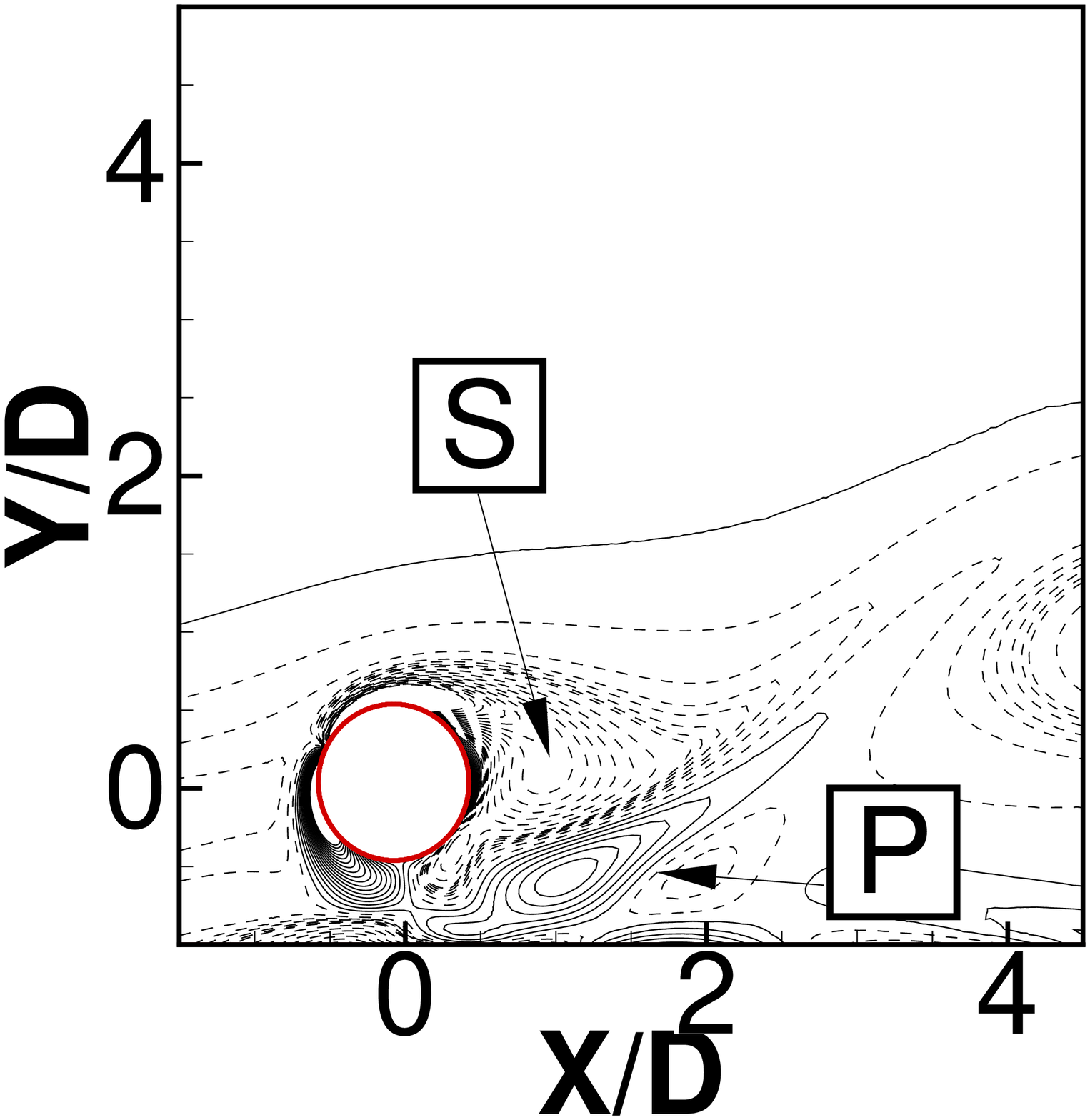}
    \caption{$t=0$}\label{vorc_5_4}
    \end{subfigure}
    \caption{Vortex shedding pattern at four equally spaced time snapshots a) $t = 0$, b) $t = T/4$, c) $t = T/2$ and d) $t = 3T/4$  for $U^* = 7$ and $g/D = 0.5$ at $Re = 100$, where $T$ represents time period of one transverse oscillation of the cylinder}\label{vortexc_G5}
\end{figure}

Figures~\ref{vortex_G5} and \ref{vortexc_G5} present the vortex shedding pattern at four equally spaced snapshots over an oscillation cycle for $U^* = 7$ and  $g/D = 0.5$. Figure \ref{vortexc_G5} takes a closer look at the gap between the cylinder and the wall for the same snaps presented in figure \ref{vortex_G5}. It can be observed that the shear layer from the top surface of the cylinder rolls into a clockwise rotating vortex and sheds downstream to form a single ``S" vortex street. On taking a closer look into the shear layer roll-up phenomenon on the bottom surface, it can be observed that a counter-clockwise rotating vortex from the bottom surface rolls up a clockwise vortex from the wall shear layer to form a pair of counter-rotating vortex ``P". However, this vortex pair ``P" instead of shedding downstream dissipates. A detailed experimental analysis on the dissipation of the vortex pair can be found in [\onlinecite{Jiankang2021}] for the case of a stationary circular cylinder placed near a wall. This formation and dissipation of the ``P" vortex pair at the bottom of the cylinder could be the reason behind the multiple frequency lift force response that will be discussed in Section \ref{subsec: phase}.

\begin{figure} 
    \raggedright
    \begin{subfigure}{0.49\textwidth}
    \includegraphics[width=0.99\columnwidth]{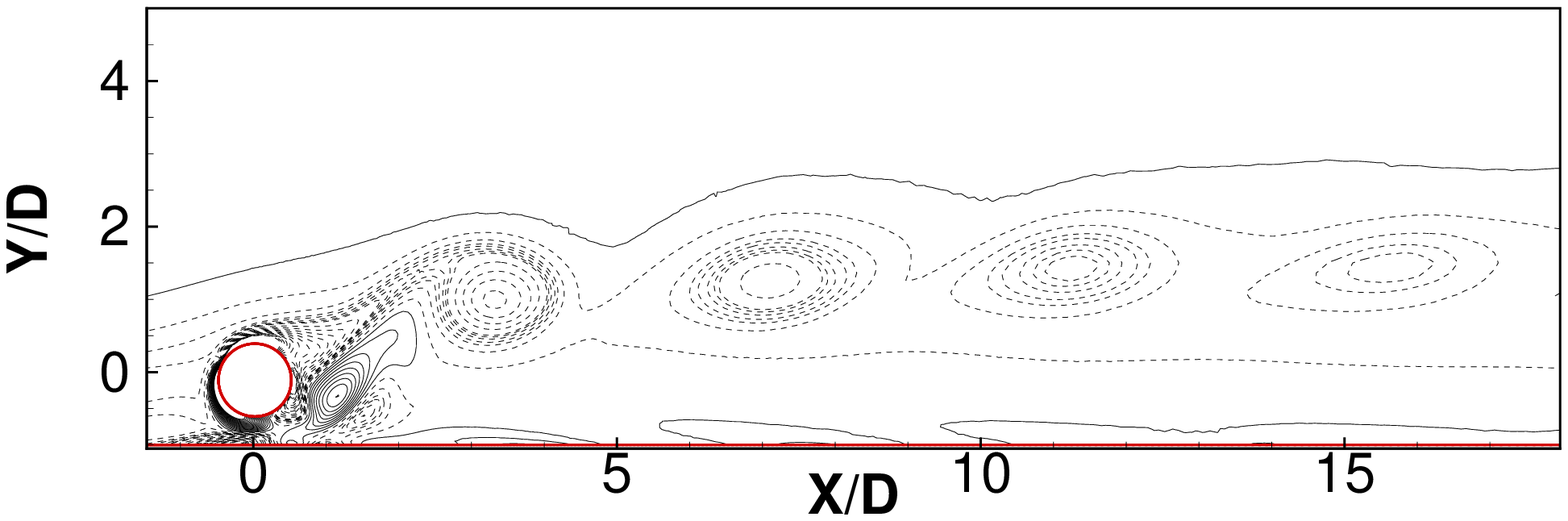}
    \caption{$U^* = 5$}\label{vor_51}
    \end{subfigure}\\
    \begin{subfigure}{0.49\textwidth}
    \includegraphics[width=0.99\columnwidth]{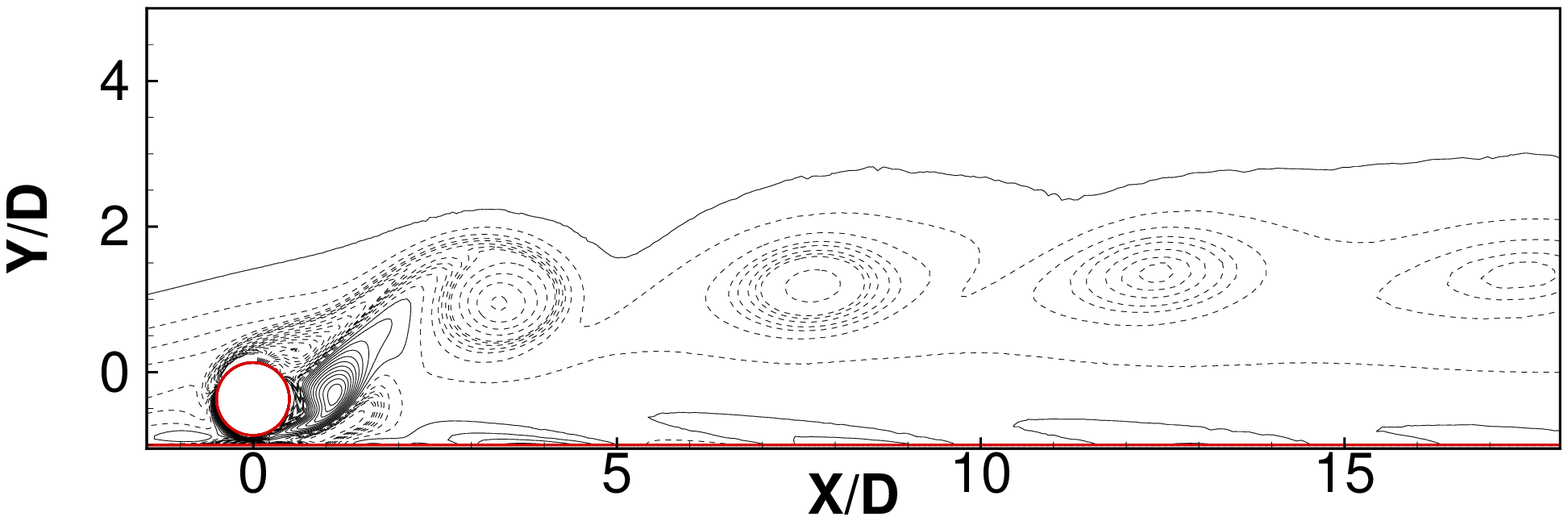}
    \caption{$U^* = 6$}\label{vor_52}
    \end{subfigure}\\
    \begin{subfigure}{0.49\textwidth}
    \includegraphics[width=0.99\columnwidth]{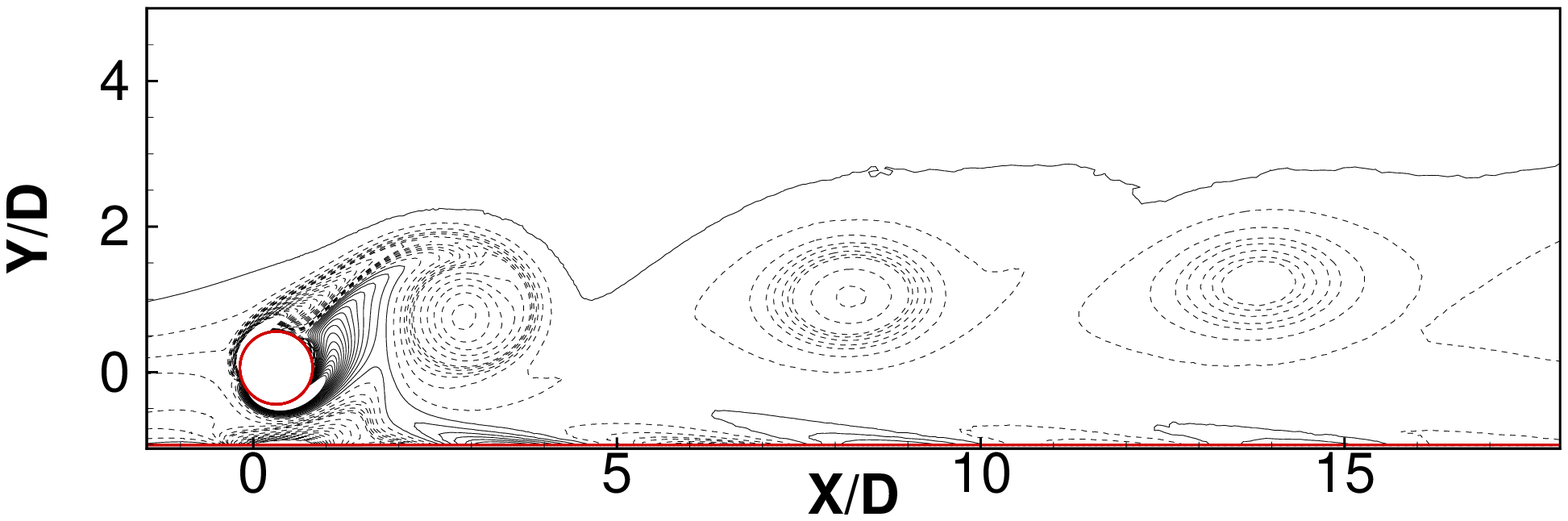}
    \caption{$U^* = 8$}\label{vor_53}
    \end{subfigure}\\
    \begin{subfigure}{0.49\textwidth}
    \includegraphics[width=0.99\columnwidth]{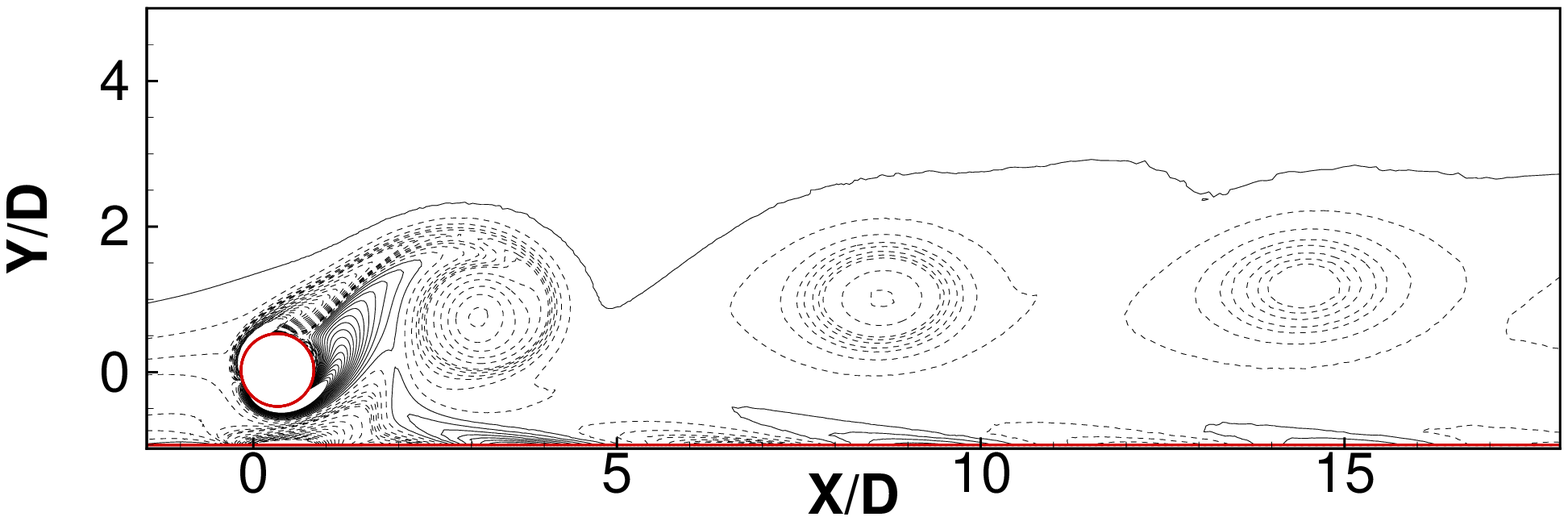}
    \caption{$U^* = 8.5$}\label{vor_54}
    \end{subfigure}\\
    \begin{subfigure}{0.49\textwidth}
    \includegraphics[width=0.99\columnwidth]{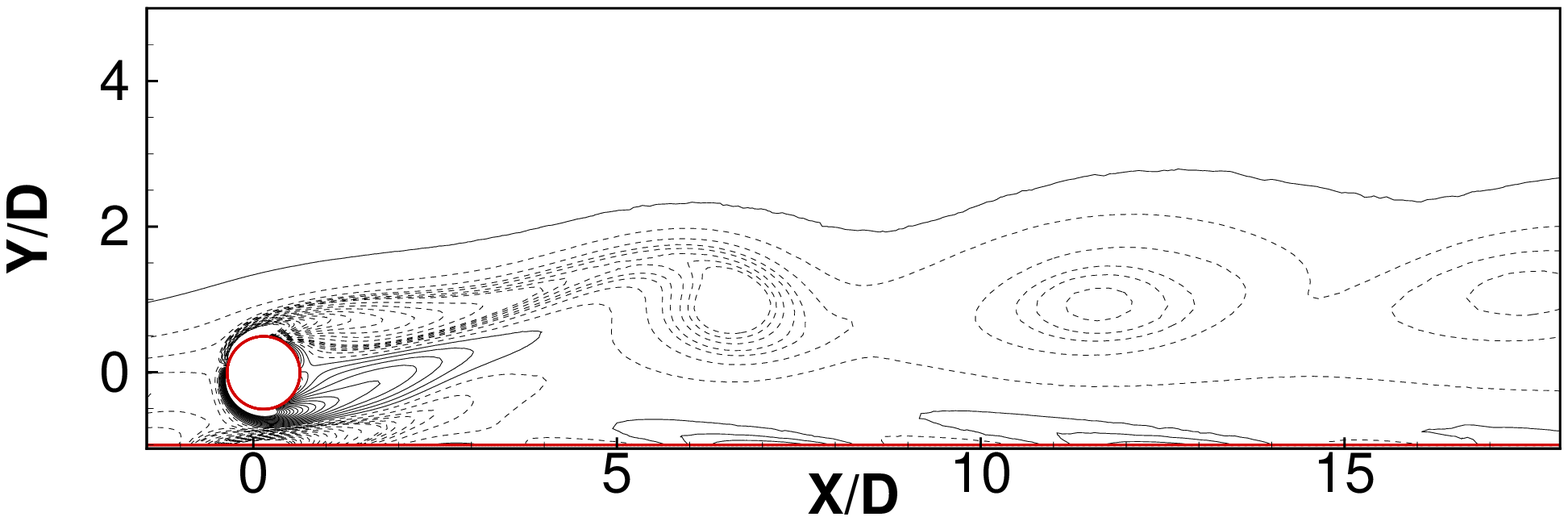}
    \caption{$U^* = 9$}\label{vor_55}
    \end{subfigure}\\
    \raggedleft
    \vspace{-0.81\textheight}
    \begin{subfigure}{0.49\textwidth}
    \includegraphics[width=0.99\columnwidth]{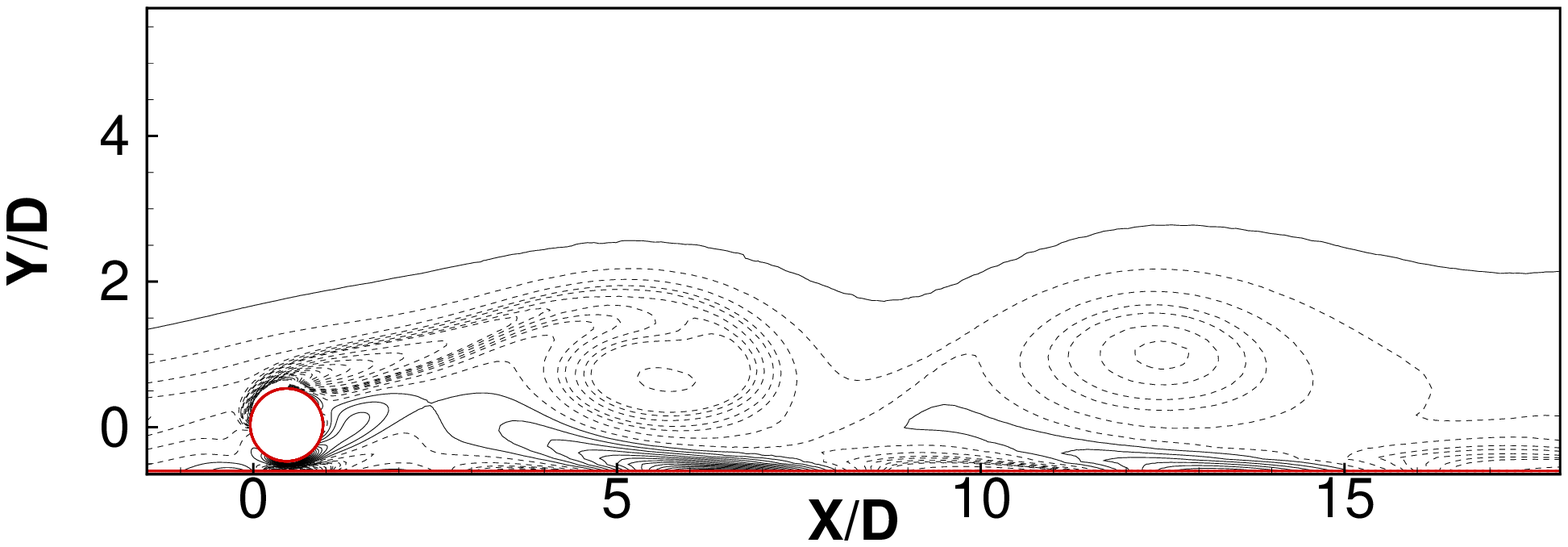}
    \caption{$U^* = 16$}\label{vor_11}
    \end{subfigure}\\
    \begin{subfigure}{0.49\textwidth}
    \includegraphics[width=0.99\columnwidth]{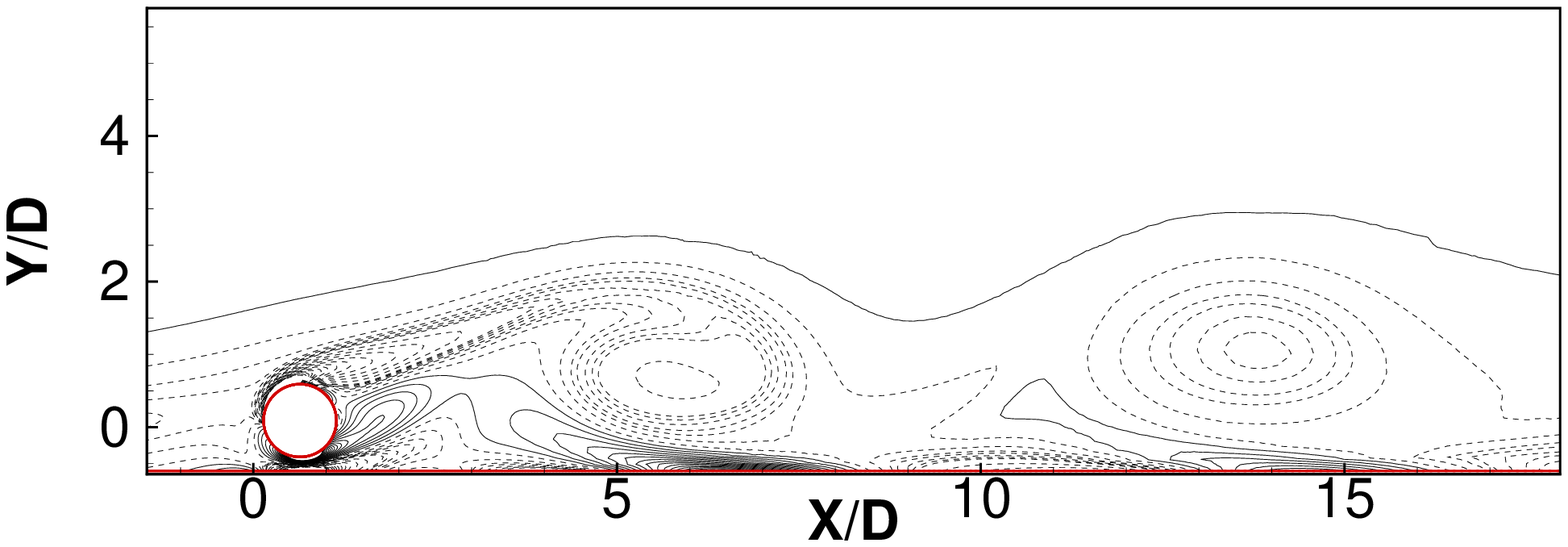}
    \caption{$U^* = 19$}\label{vor_12}
    \end{subfigure}\\
    \begin{subfigure}{0.49\textwidth}
    \includegraphics[width=0.99\columnwidth]{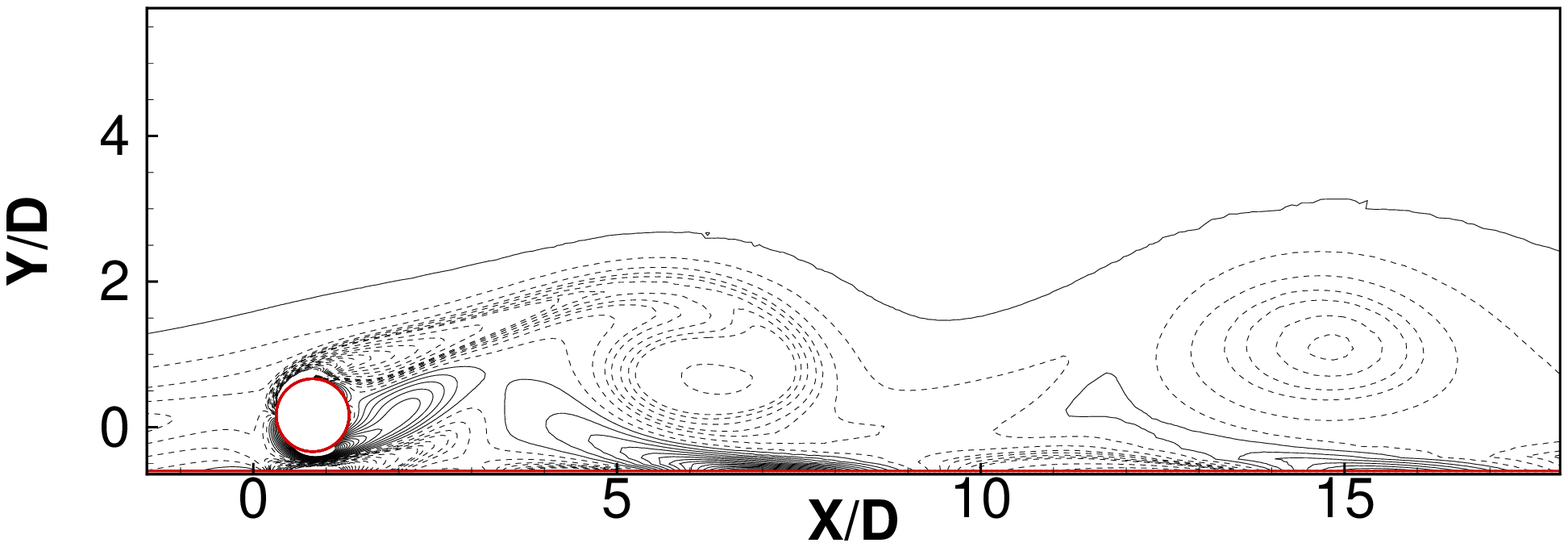}
    \caption{$U^* = 22$}\label{vor_13}
    \end{subfigure}\\
    \begin{subfigure}{0.49\textwidth}
    \includegraphics[width=0.99\columnwidth]{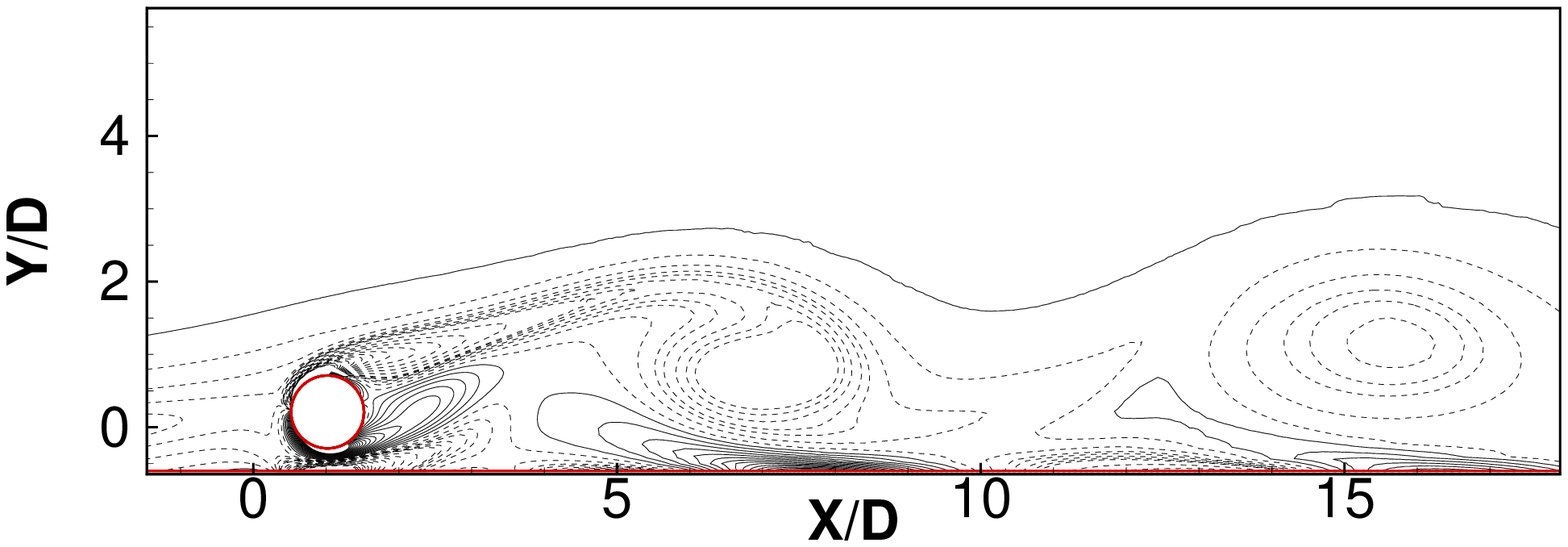}
    \caption{$U^* = 25$}\label{vor_14}
    \end{subfigure}\\
    \begin{subfigure}{0.49\textwidth}
    \includegraphics[width=0.99\columnwidth]{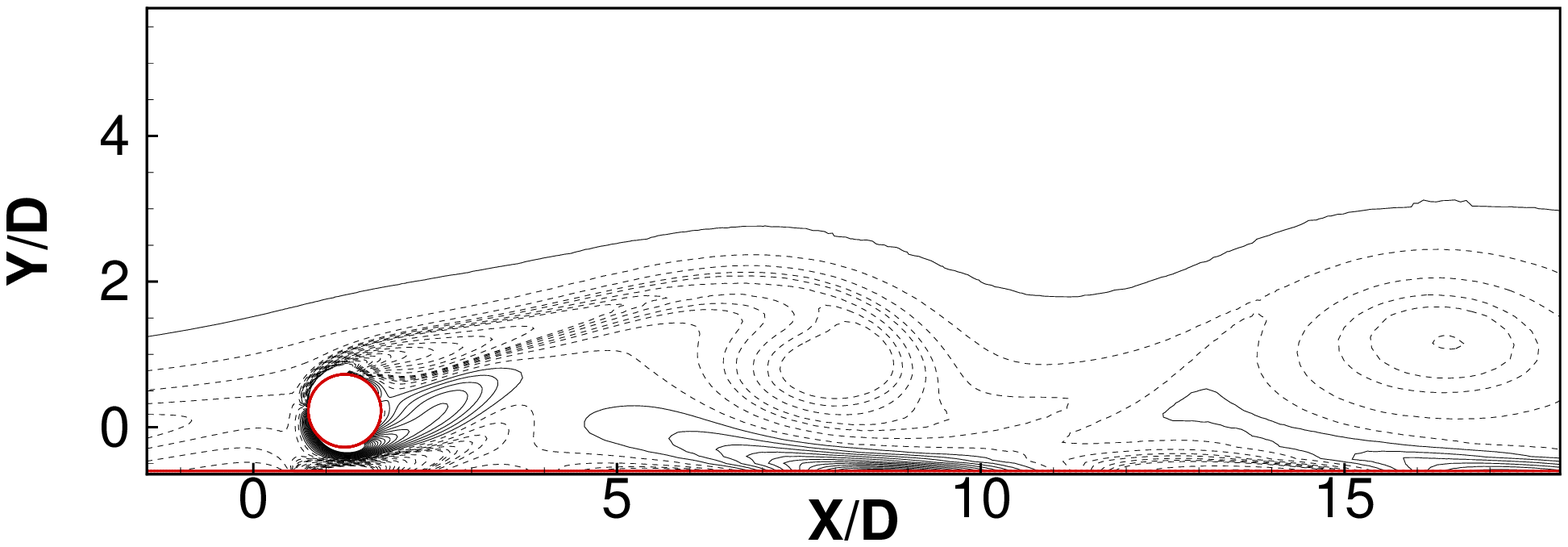}
    \caption{$U^* = 28$}\label{vor_15}
    \end{subfigure}
    \caption{The vorticity contours presented on the left side of the figure correspond to $U^* = \{5,6,8,8.5,9\}$ for the gap ratio $g/D = 0.5$ while the contours presented on the right side correspond to the $U^* = \{16,19,22,25,28\}$ values at $g/D = 0.1$. }\label{vortex_U}
\end{figure}

To understand how the vortex dynamics change with increasing $U^*$ and decreasing $g/D$, figure \ref{vortex_U} presents the vorticity contours snapped at the instant of vortex shedding for different $U^*$ values. The sub-figures on the right side, sub-figures \ref{vor_51}-\ref{vor_55} correspond to increasing $U^*$ values in the lock-in region for gap $g/D = 0.5$. As $U^*$ increases from $5$ to $8.5$ for $g/D = 0.5$, it can be observed that the vortex is shed closer to the surface of the cylinder. It can also be observed that the vortex is biased slightly away from the center-line of the cylinder in these cases. The proximity of the vortex to the cylinder and the slight upward bias results in a low pressure region being formed over the top surface and behind the cylinder. As $U^*$ further increases to $9$, from sub-figure \ref{vor_55}, it can be noted that the vortex is now shed far away from the surface of the cylinder. This shift in the vortex results in a smaller low pressure region being formed over the top of the cylinder. It is to be noted from figure \ref{ay_max} that the vortices are shed closer to the cylinder for the cases corresponding to higher transverse amplitudes and as the vortex sheds away from the cylinder, lower transverse amplitudes are observed. Similar vortex structures have been observed for $g/D = 0.3,0.4$ and $0.6$.

The contours on the left side of figure \ref{vortex_U} present the vorticity contours at the instant of vortex shedding for $U^* = 16,19,22,25,28$ and $g/D = 0.1$. Two main observations from these contours is that the vortices are shed farther away from the cylinder as the gap ratio reduces. For $g/D = 0.1$, it can be observed that for increasing $U^*$, vortices are subsequently shed further away from the cylinder. This results in a smaller low pressure region being observed over the top surface of the cylinder. Further, the shedding of the vortex farther away from the cylinder is accompanied by lower transverse amplitudes observed in figure \ref{ay_max} for the case of $g/D = 0.1$ than $g/D = 0.5$. Similar vortex structures are observed for the case of $g/D = 0.2$. This variation of the vortex structures and the resulting variation in the pressure with $U^*$ and $g/D$ will be used to explain the force dynamics presented in the next section.

\subsection{Force Dynamics}\label{subsec: res_dyn}

Before a detailed analysis of the force dynamics is presented, it should be noted that the previous works [\onlinecite{Li2016}] have shown that the wall proximity effects on $C_L^{mean}$ can only be observed in the lock-in region and the $C_L^{mean}$ values in the pre and post lock-in regions are close to zero. To understand how this behaviour changes as the cylinder moves further close to the wall, the mean lift coefficient variation for different gap ratios is presented in this section.

\begin{figure}
\centering
{\includegraphics[width = 1\linewidth]{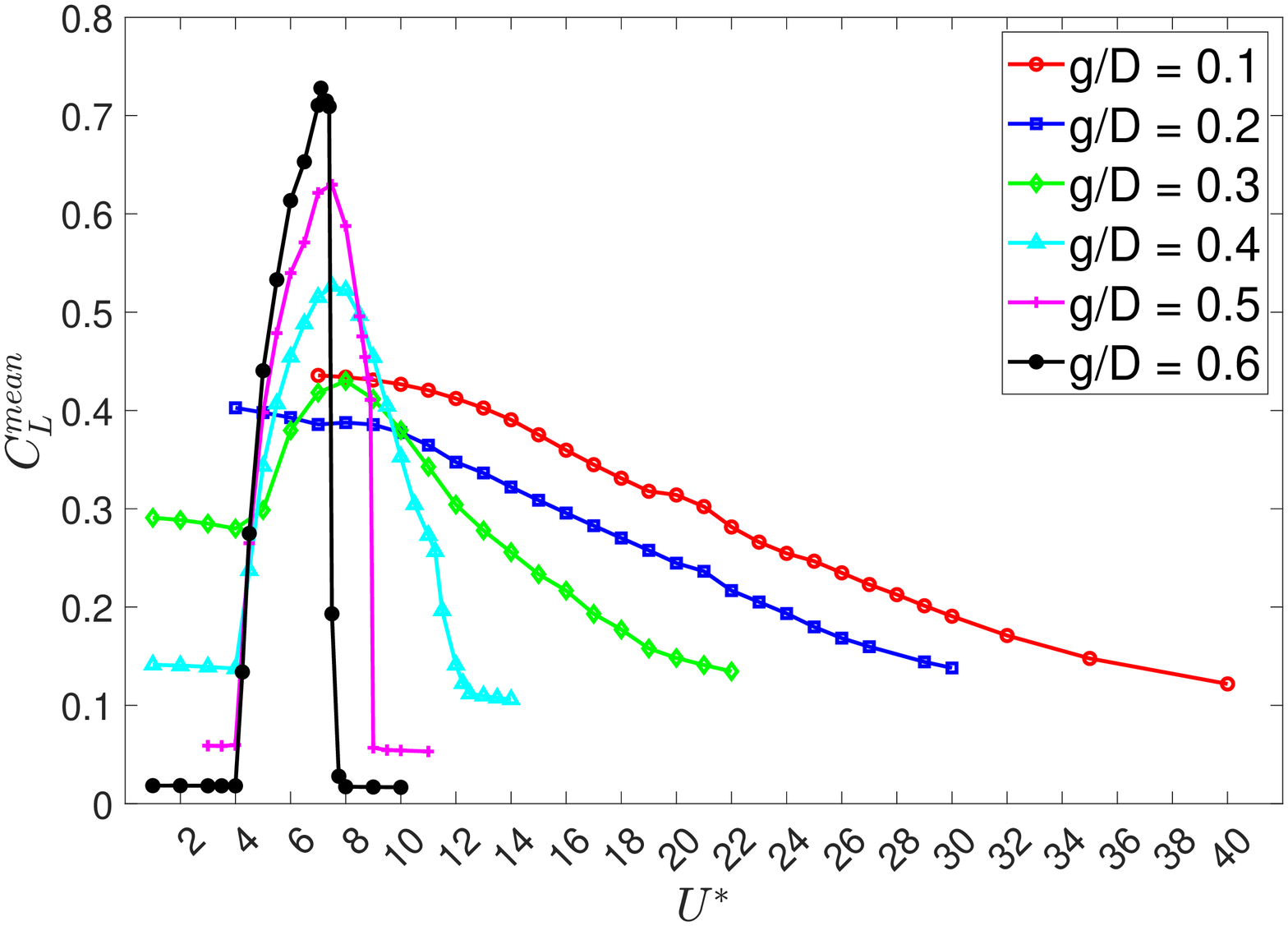}}
\caption{Variation of mean lift coefficient$(C^{mean}_L)$ with $U^*$ for $Re=100$, $m^*=10$ and $\zeta=0$.}
\label{mean_cl}
\end{figure}

Figure~\ref{mean_cl} presents the mean lift coefficient ($C^{mean}_L$) versus $U^*$ for different gap ratios. The key observations from the figure~\ref{mean_cl} are:
\begin{itemize}
    \item{The wall proximity results in a non-zero $C_L^{mean}$ even in the pre and post lock-in regions for $g/D <0.6$. Further, $C_L^{mean}$ values in the pre and post lock-in regions increase with decreasing gap ratios.}
    \item{For $0.3 \le g/D \le 0.6$, the lock-in region is marked by a sudden increase in the $C_L^{mean}$ values followed by a sharp decrease. The highest $C_L^{mean}$ values are observed for $g/D = 0.6$, and the peak $C_L^{mean}$ reduces with decreasing $g/D$.}
    \item{For gap ratios $g/D = 0.2$ and $0.1$, a single contiguous decrease in the $C_L^{mean}$ values is observed over the pre lock-in, lock-in, and post lock-in regions.}
    \item{For $g/D \le 0.3$, the $C_L^{mean}$ in the post lock-in region is much lower than the $C_L^{mean}$ in the pre lock-in region. Further, a steeper decrease in the $C_L^{mean}$ values is observed with increasing $U^*$.}
\end{itemize} 

\begin{figure}
\centering
{\includegraphics[width = 1\linewidth]{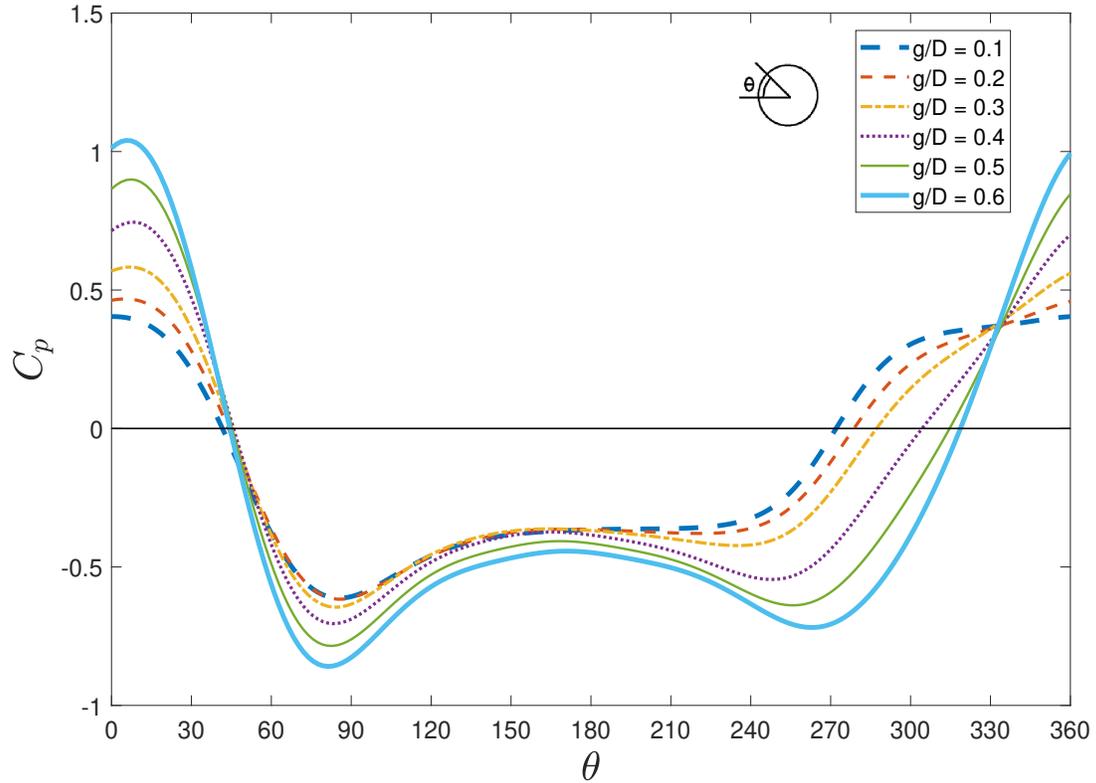}}
\caption{Variation of Cp along the surface of the circular cylinder in the pre lock-in region for $Re=100$, $m^*=10$ and $\zeta=0$. Here $\theta = 0$ starts from the negative x axis and varies in the clockwise direction.}
\label{prelock}
\end{figure}

In figure~\ref{mean_cl}, for $g/D < 0.6$, a non-zero $C_L^{mean}$ is observed not just in the lock-in region but also in the pre and post lock-in regions. This observation is distinctly different from the $C_L^{mean}$ for $g/D \ge 0.6$ where the $C_L^{mean}$ in the pre and post lock-in region is close to zero [\onlinecite{Tham2015,Li2016}]. For the pre lock-in region, figure~\ref{mean_cl} shows that the $C_L^{mean}$ value increases as the gap ratio decreases. To further understand this behaviour, the variation of mean coefficient of pressure $(C_P)$ along the surface of the cylinder is presented in figure~\ref{prelock} for the pre lock-in region. Traditionally, the distribution of mean $C_P$ is symmetric about the center line for an isolated cylinder. From figure~\ref{prelock}, it is apparent that the wall proximity results in an asymmetry in the mean $C_P$ distribution and this induced asymmetry becomes more prominent as the gap ratio reduces. 
From figure \ref{prelock}, a low-pressure zone is observed over the upper part of the cylinder for all gap ratios and the intensity of the low-pressure region increases slightly with the gap ratio. As the gap ratio reduces, due to the proximity of the wall, the high pressure region which would typically exists in front of the cylinder expands further down such that for the gap ratios of $g/D = 0.1$ and $0.2$, one can observe the high pressure region even at the bottom of the cylinder. The compounded effect of the extended high pressure region near the bottom surface of the cylinder and the low-pressure region at the top would result in a greater pressure difference being observed for smaller gap ratios. Hence, an increasing trend of $C_L^{mean}$ for decreasing gap ratios can be observed in the pre lock-in region.

%It can be observed from figure~\ref{prelock} that as the gap ratio reduces, the transition from the low-pressure zone to the high-pressure zone occurs at a lower angle. Consequently, the high-pressure zone begins at the bottom of the cylinder for smaller gap ratios of $g/D = 0.2$ and $0.1$. The combined effect of the extended high pressure region below and the low-pressure region at the top results in a higher $C_L^{mean}$ that is being observed for lower gap ratios. While figure~\ref{prelock} presents that an increased low-pressure zone at the top of the cylinder is observed for gap ratios of $g/D = 0.6$ and $0.5$, the combined effect of the low-pressure region at the top and the high pressure region at the bottom is dominant.

The trend of $C_L^{mean}$ observed in the lock-in region is distinctly different from that observed in the pre and post lock-in regions. The variation of both $C_L^{mean}$ and $C_L^{rms}$ can be explained based on the vorticity contours plotted in Section \ref{subsec: res_vorDyn}. It was observed for that higher gap ratios of $g/D = 0.6$ and $0.5$, vortices are formed and shed close to the surface of the cylinder as compared to the lower gap ratios where the vortices were formed and shed farther away from the cylinder. This behavior results in an extended low-pressure region on the top surface of the cylinder for higher gap ratios and hence results in the high $C_L^{mean}$ values for $g/D = 0.6$ $\&$ $0.5$. Further, it was observed in Section \ref{subsec: res_vorDyn} that as the gap ratio reduces, vortices are shed farther away from the cylinder. This behaviour can explain the reducing trend of the $C_L^{mean}$ that is observed with decreasing $g/D$ in the lock-in region from figure \ref{mean_cl}.
For $g/D = 0.5$ and $0.6$, vortices are shed close to the surface of the cylinder for $U^*$ values corresponding to high transverse amplitudes. However, for smaller transverse displacements, as $U^*$ increases, it was observed in \ref{subsec: res_vorDyn} that the vortex sheds farther away from the cylinder. This results in a smaller low-pressure region at the top of the cylinder and in turn, lower $C_L^{mean}$ at higher $U^*$. This behavior explains why the $C_L^{mean}$ exhibits a reducing trend for high $U^*$ values.  

As compared to the higher gap ratios, the proximity of the wall for the gap ratios $g/D \le 0.4$ results in the vortices being shed farther away from the wall and hence in a smaller low-pressure region above the cylinder. As the $U^*$ value increases, this effect is only more pronounced with the vortices being shed even farther away, which can explain the decreasing trend in the $C_L^{mean}$ observed for gap ratios of $g/D = 0.4$ to $0.1$. It further explains why the values observed at the end of the lock-in region reduce below the $C_L^{mean}$ values observed in the pre lock-in region for these gap ratios.

\begin{figure}
\centering
{\includegraphics[width = 1\linewidth]{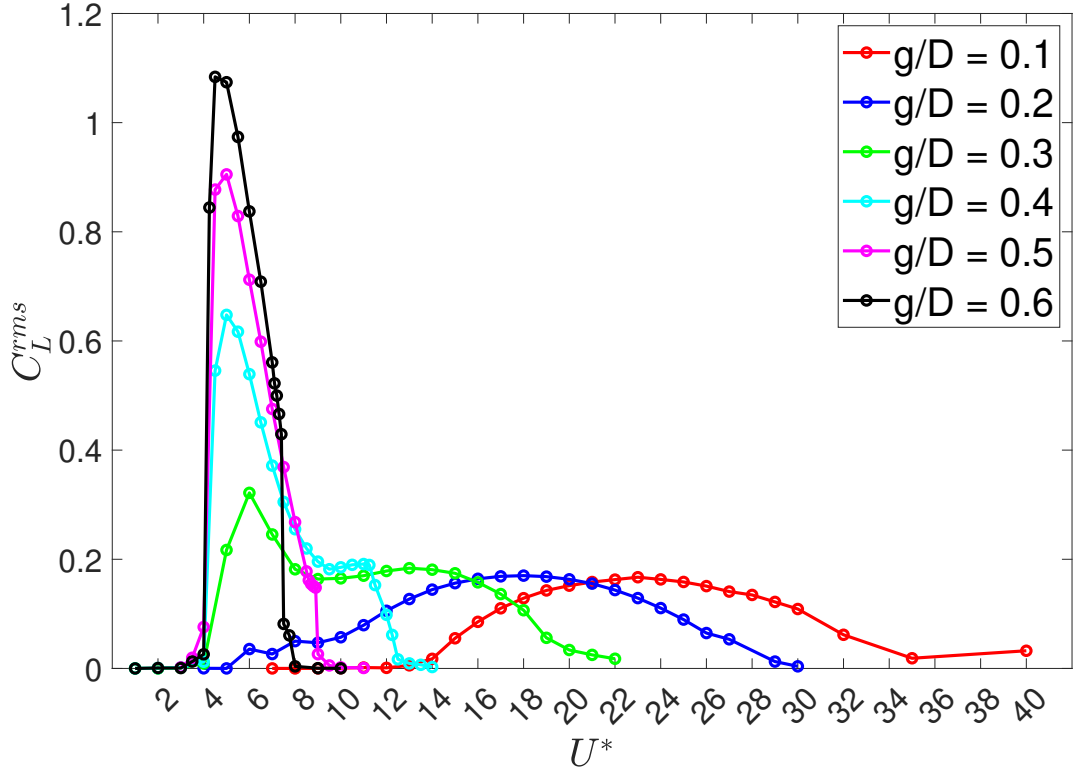}}
\caption{Variation of root-mean-squared lift coefficient$(C^{rms}_L)$ with $U^*$ for $Re=100$, $m^*=10$ and $\zeta=0$.}
\label{rms_cl}
\end{figure}

Figure~\ref{rms_cl} presents the root-mean-squared lift coefficient $C_L^{rms}$ of the cylinder versus $U^*$ values at different gap ratios. From the figure, it can be observed that large $C_L^{rms}$ values can only be observed in the lock-in region before they die down again for the de-synchronization regions. This trend is distinctly different from the observations in the case of an isolated cylinder where non-zero $C_L^{rms}$ values are also observed in the pre lock-in region. This phenomenon of the reduction in $C_L^{rms}$ values in the de-synchronization region as the cylinder moves closer to the wall has previously been observed in [\onlinecite{Tham2015, Li2016}] for gap ratios $g/D \le 0.9$. For gap ratios of $g/D = 0.6$ and $0.5$, $C_L^{rms}$ values peak at the beginning of the lock-in region before reducing gradually to zero. However, for $g/D = 0.3$ and $0.4$, $C_L^{rms}$ values do not die down after the initial sharp peak, instead $C_L^{rms}$ remains more or less a constant over a range of $U^*$ values before reducing to zero. This double-lobed response will be discussed in detail in Section \ref{subsec: phase}. For $g/D = 0.2$, only a small perturbation can be observed at the beginning of the lock-in region, followed by a smooth response over the remaining lock-in region. As the gap ratio reduces, it can be observed that the first peak diminishes and the second lobe widens. For $g/D = 0.1$, the first peak vanishes and a smooth variation can be observed in the $C_L^{rms}$. The $C_L^{rms}$ in the lock-in region reduces with decreasing gap ratios. As discussed earlier, vortices are shed farther away from the cylinder as the gap ratio reduces. As the vortex shedding occurs away from the cylinder, it results in lesser fluctuations in the pressure distribution around the cylinder and, in turn, reduces the oscillating components of the lift force.

\begin{figure} 

    \begin{subfigure}{0.85\textwidth}
    %\centering
    {\includegraphics[width = \columnwidth]{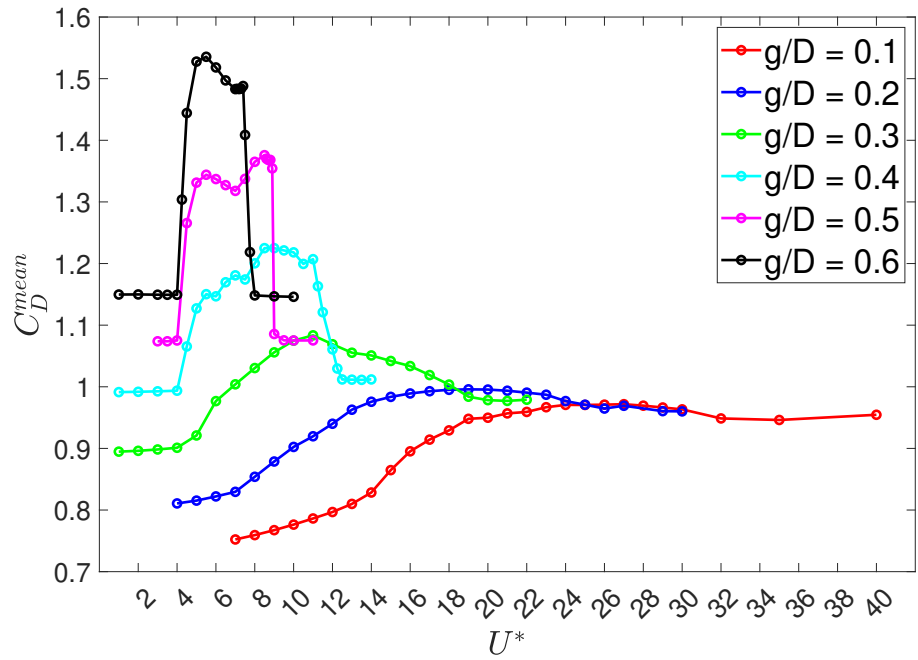}}
    \caption{}\label{mean_cd}
    \end{subfigure}
    \hfill
    \begin{subfigure}{0.85\textwidth}
    %\centering
    {\includegraphics[width = \columnwidth]{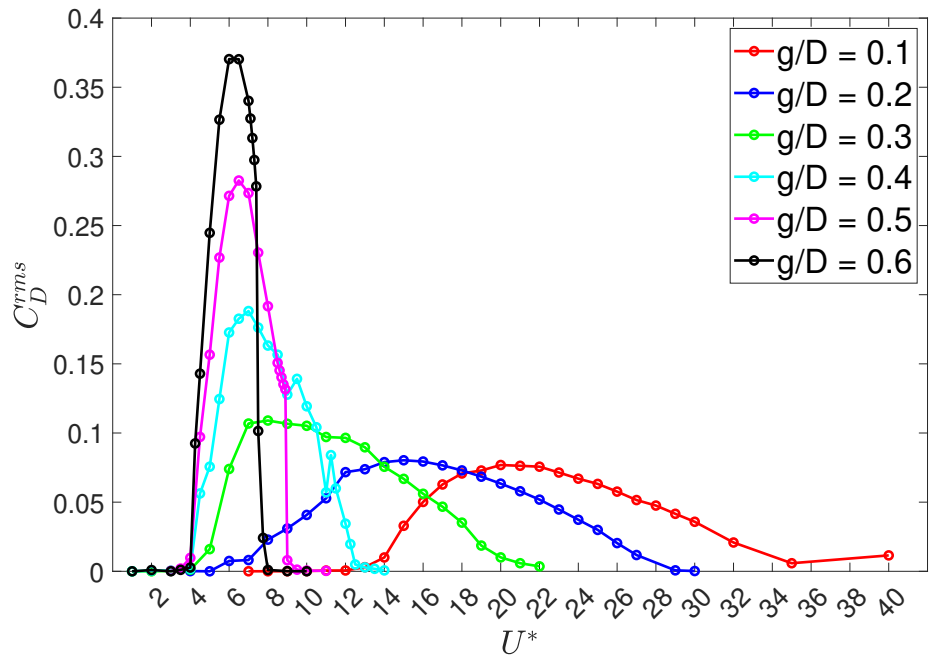}}
    \caption{}\label{rms_cd}
    \end{subfigure}
    \caption{Variation of mean drag coefficient$(C^{mean}_D)$ and root-mean-squared value of drag coefficient$(C^{rms}_D)$ with $U^*$ for $Re=100$, $m^*=10$ and $\zeta=0$.}
\end{figure}

Figure~\ref{mean_cd} presents the mean drag coefficient ({$C^{mean}_D$}) as a function of $U^*$ values at different gap ratios. As the gap ratio reduces, an overall decrease in the $C_D^{mean}$ values can be observed. This decrease is almost negligible in the post lock-in region for gap ratios of $g/D \le 0.3$ where the $C_D^{mean}$ value reaches a more or less constant value. In the lock-in region, a distinct peak can be observed for $g/D = 0.3$ to $0.6$. As the gap ratio reduces, the peak widens and smoothens, i.e., the increase and decrease in $C_D^{mean}$ as a function of $U^*$ is more gradual. For gap ratios $g/D = 0.2$ and $0.1$, the $C_D^{mean}$ values in the lock-in region increase slowly before reaching an almost constant value. This value remains constant even in the post lock-in region. %This behavior can explain the increase in $A_X^{mean}$ for increasing $U^*$ observed in figure~\ref{mean_ax}. As the stiffness increases for increased $U^*$ values, the force remains constant, increasing $A_X^{mean}$ values.   

The variation of the root mean squared drag coefficient $C_D^{rms}$ as a function of increasing $U^*$ values for different gap ratios is presented in figure~\ref{rms_cd}. Similar to the trend followed by $A_X^{rms}$ presented in figure~\ref{axrms}, a rightward shift and a smoothening of the curve is observed in the $C_D^{rms}$ curve with decreasing gap ratios. Further, the $C_D^{rms}$ values in the pre and post lock-in regions are zero for all gap ratios $g/D = 0.6-0.1$ and the $C_D^{rms}$ values in the lock-in region decrease with reducing gap ratios. 

\begin{figure} 
\centering
{\includegraphics[width = 0.85\textwidth]{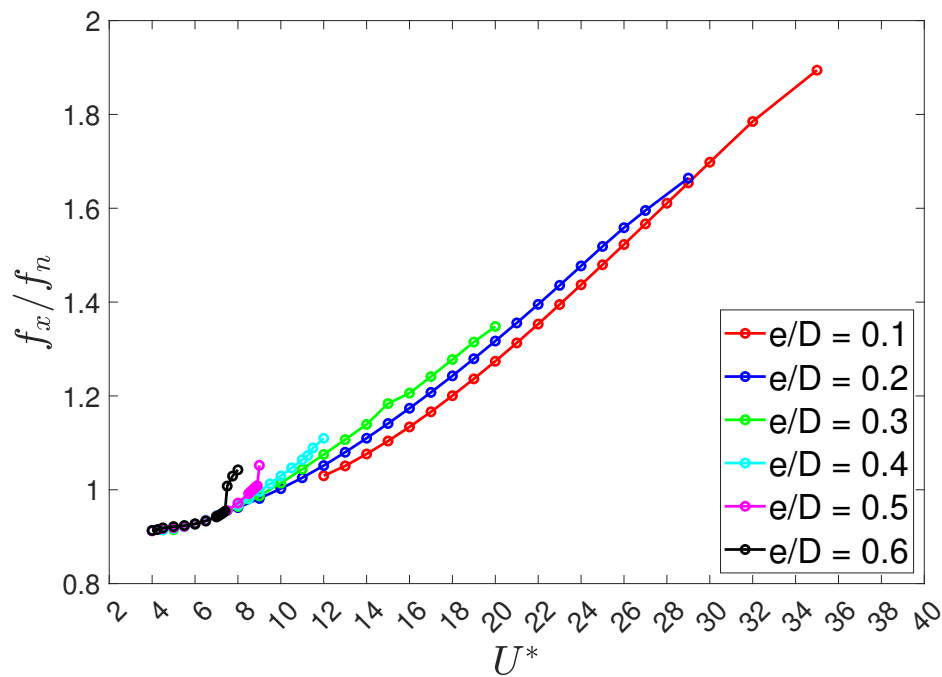}}
\caption{Variation of transverse frequency ratio $(f_{y}/f_{n})$ with $U^*$}
\label{freq_y}
\end{figure}

%In order to understand the widening of the lock-in region with decreasing gap ratios, the transverse frequency ratio is plotted in figure~\ref{freq_y}. 
Figure \ref{freq_y} presents the ratio of the transverse oscillation frequency to the natural frequency for different $U^*$. It can be observed that the oscillation frequency ``locks-in" with a frequency slightly less than the natural frequency for gap ratios of $g/D = 0.4$ to $0.6$. For gap ratios $g/D \le 0.3$, the $f_{y}/f_{n}$ values can be observed to be much higher than one in the lock-in region and increase with increasing $U^*$. These observations are in accordance with observations from [\onlinecite{Chen2020}]. Further, this increase in the transverse frequency ratio can be attributed to the fact that the calculation of $f_{n}$ does not take into account the effective stiffness and added mass effects of the proximity of the wall. 

\begin{figure} 
\centering
{\includegraphics[width = 0.85\textwidth]{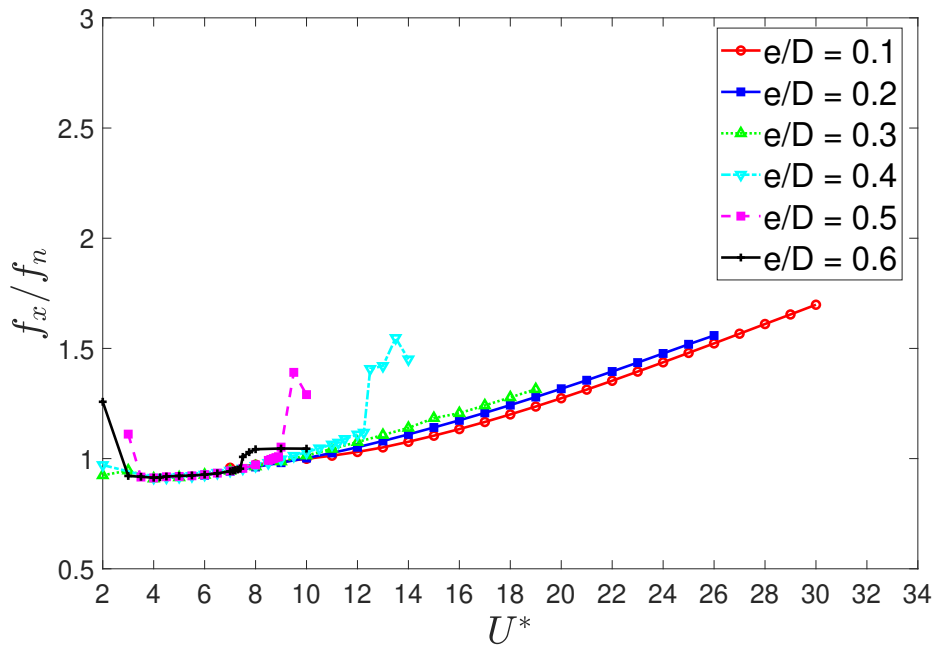}}
\caption{Variation of stream-wise frequency ratio $(f_{x}/f_{n})$ with $U^*$}
\label{freq_x}
\end{figure}

\begin{figure} 
\centering
{\includegraphics[width = 0.85\textwidth]{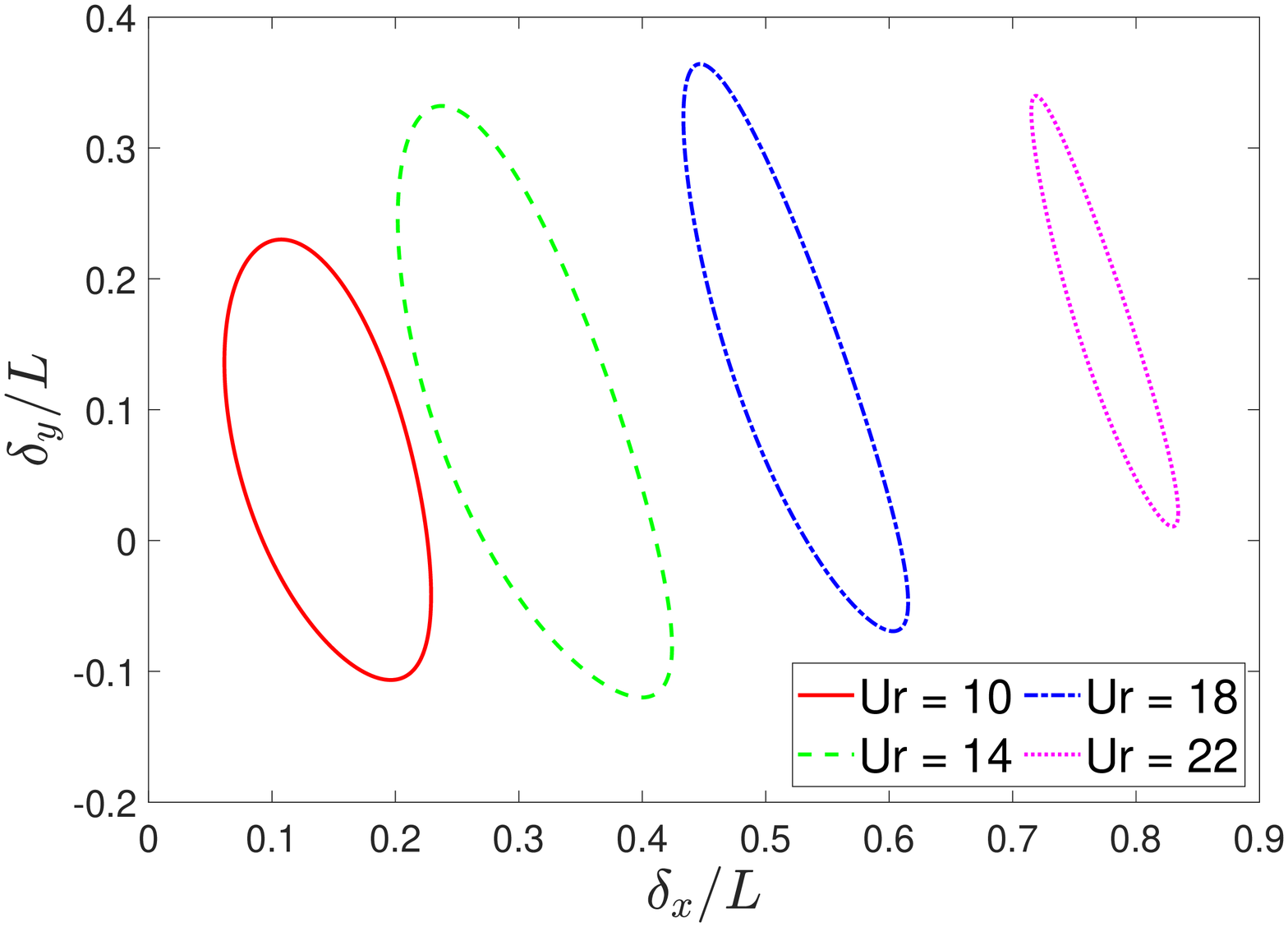}}
\caption{The x-y trajectory of the cylinder at different $U^*$ values at a gap ratio $g/D = 0.2$}
\label{x-y_trajectory_G2}
\end{figure}

Figure~\ref{freq_x} presents the ratio of stream-wise oscillation frequency to the natural frequency of the cylinder versus increasing $U^*$. It can be observed that the $f_{x}/f_{n}$ plots are very similar to the $f_{y}/f_{n}$ plots. In fact, the ratio of $f_{x}/f_{y}$ is observed to be one over the lock-in range unlike the $f_{x}/f_{y}$ ratio of an isolated circular cylinder where the $f_{x}$ frequency is usually twice the natural frequency in the stream-wise direction. This reduction in the $f_{x}$ value is a consequence of the suppression of vortices from the bottom surface of the cylinder. These observations are consistent with [\onlinecite{Tham2015,Li2016}] where the ratio of $f_{x}/f_{n}$ was observed to be one for $g/D \le 0.9$. This change in the $f_{x}/f_{n}$ value also impacts the trajectory of the cylinder. While the x-y trajectory of an isolated circular cylinder takes the shape of an eight, for a circular cylinder in close proximity to the wall, the trajectory takes the shape of an oblique elliptical configuration as can be observed in figure~\ref{x-y_trajectory_G2}.

The vibration response of a cylinder has been observed to be effected strongly by the close proximity to the wall as observed in the summary plots discussed above. It could also be observed that different regimes of flow existed for the same gap ratios at different $U^*$ values. To understand the effect of wall proximity on the different branching regimes figures~\ref{G6_branch}-\ref{G1_branch} summarize the dimensionless transverse amplitude $(A_{y}^{max}/D)$, phase angle $\Phi$ between the lift force and transverse displacement, and maximum lift coefficient $C_L^{max}$ as a function of $U^*$ for gaps $g/D=0.6$-$0.1$ respectively.

\subsection{Phase Relationship and Vortex Dynamics}\label{subsec: phase}
%Different regimes of synchronous response.

\subsubsection{Gap ratios: $g/D = 0.6$ and $0.5$}

\begin{figure} 
\centering
{\includegraphics[width = 1\linewidth]{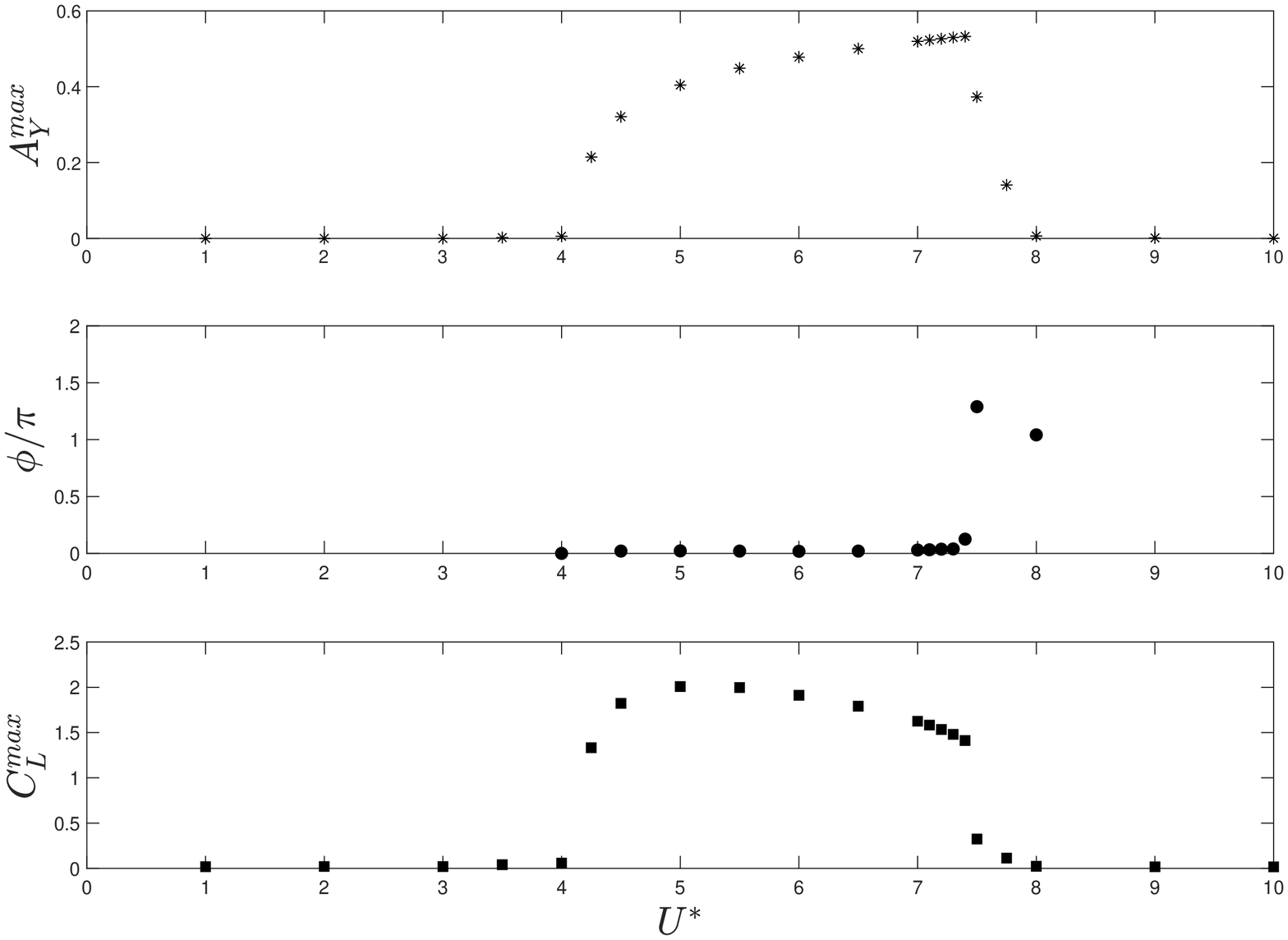}}
\caption{Variation of maximum transverse displacements$(A_{y}^{max}/D)$,  $\Phi$ and $C_L^{max}$ at a gap ratio of $g/D = 0.6$ with $U^*$ for $m^* = 10$ and $Re = 100$}
\label{G6_branch}
\end{figure}

\begin{figure} 
\centering
{\includegraphics[width = 1\linewidth]{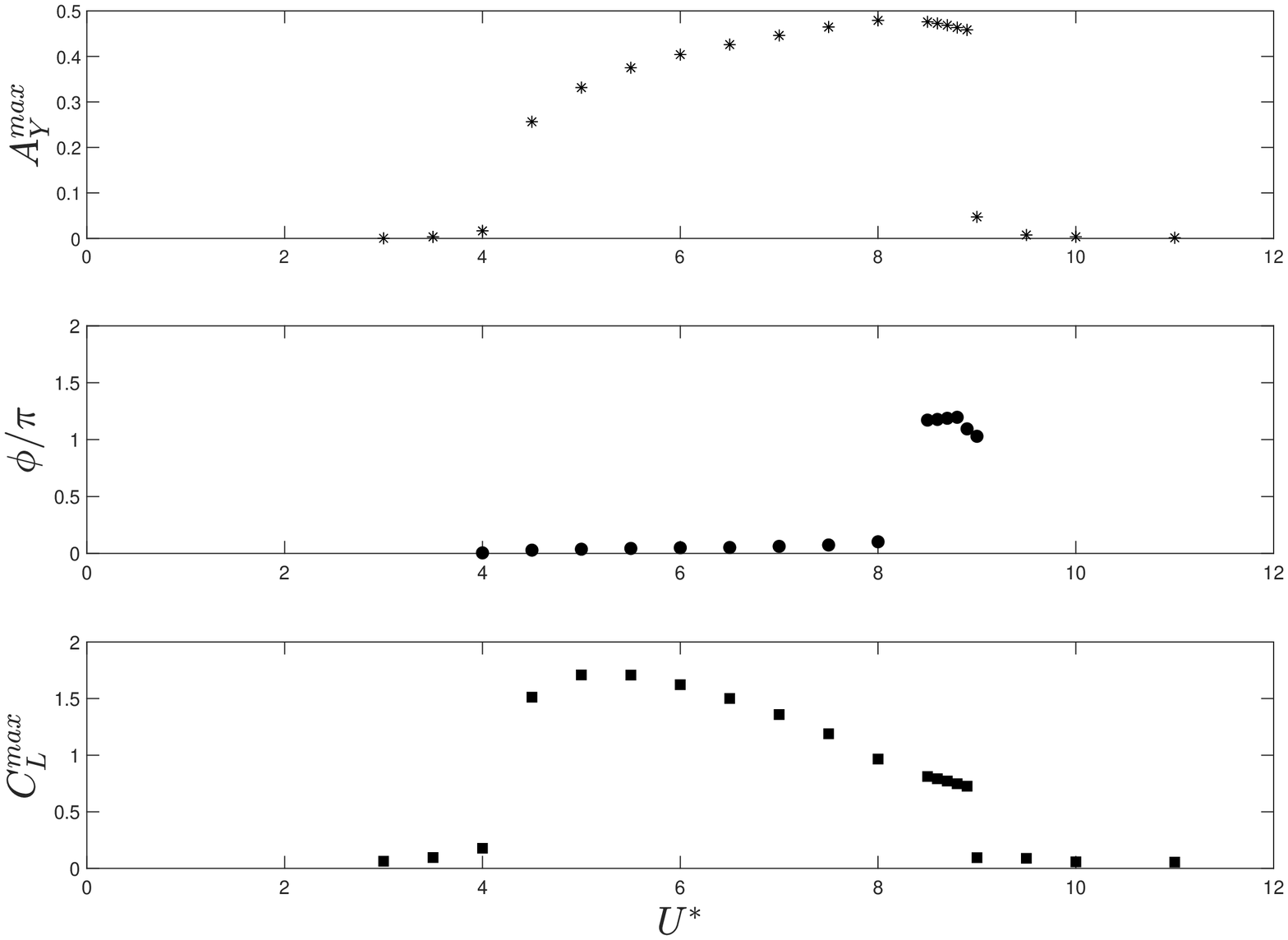}}
\caption{Variation of maximum transverse displacements$(A_{y}^{max}/D)$, $\Phi$ and $C_L^{max}$ at a gap ratio of $g/D = 0.5$ with $U^*$ for $m^* = 10$ and $Re = 100$}
\label{G5_branch}
\end{figure}
 %The synchronous vibration response of an elastically mounted circular cylinder free to oscillate in  2-DOF can be characterized into multiple response branches. %These regimes have been characterized on the basis of the amplitude response, the phase difference between the lift force and displacement and frequency of oscillation. The vibration and flow characteristcs have been observed to behave differently in the

The maximum dimensionless transverse amplitude $(A_{y}^{max}/D)$, phase angle between the lift force and transverse displacement ($\Phi$), and maximum lift coefficient ($C_L^{max}$) are presented as a function of $U^*$ in figure~\ref{G6_branch} for the gap ratio of $g/D = 0.6$. In the lock-in region, the $A_Y^{max}$ values presented in the first subplot increase gradually over a range of $U^*$ values between $4 < U^* \le 7.4$ before suddenly reducing to zero for $U^*\ge 8$. The transition of $A_{y}^{max}/D$ values from the peak at $U^*=7.4$ to close to zero at $U^*=8$ is accompanied by a jump in the $\Phi$ from zero to phase close to $\pi$.
%as can be observed by comparing the first and second subplot. 
Hence, the response dynamics can be characterized into two branches in the lock-in regime.
%This indicates towards a branched response in the lock-in region where
The $\mathrm{I^{st}}$ branch is characterized by increasing transverse vibration amplitude response with $\Phi = 0$. On the other hand the $\mathrm{II^{nd}}$ branch involves a sharp reduction of transverse amplitudes with $\Phi$ values about $\pi$. 
The third subplot in figure~\ref{G6_branch} further supports the distinction in the branched response. The $\mathrm{I^{st}}$ branch exhibits high $C_L^{max}$ values while the $\mathrm{II^{nd}}$ branch exhibits lower values. However, unlike the $A_Y^{max}$ plots, the $C_L^{max}$ value first peaks and then reduces over the range of the $\mathrm{I^{st}}$ branch. 

\begin{figure} 
    \begin{subfigure}{0.325\textwidth}
    {\includegraphics[width = 0.99\columnwidth,trim={28mm 0mm 4cm 0mm}, clip]{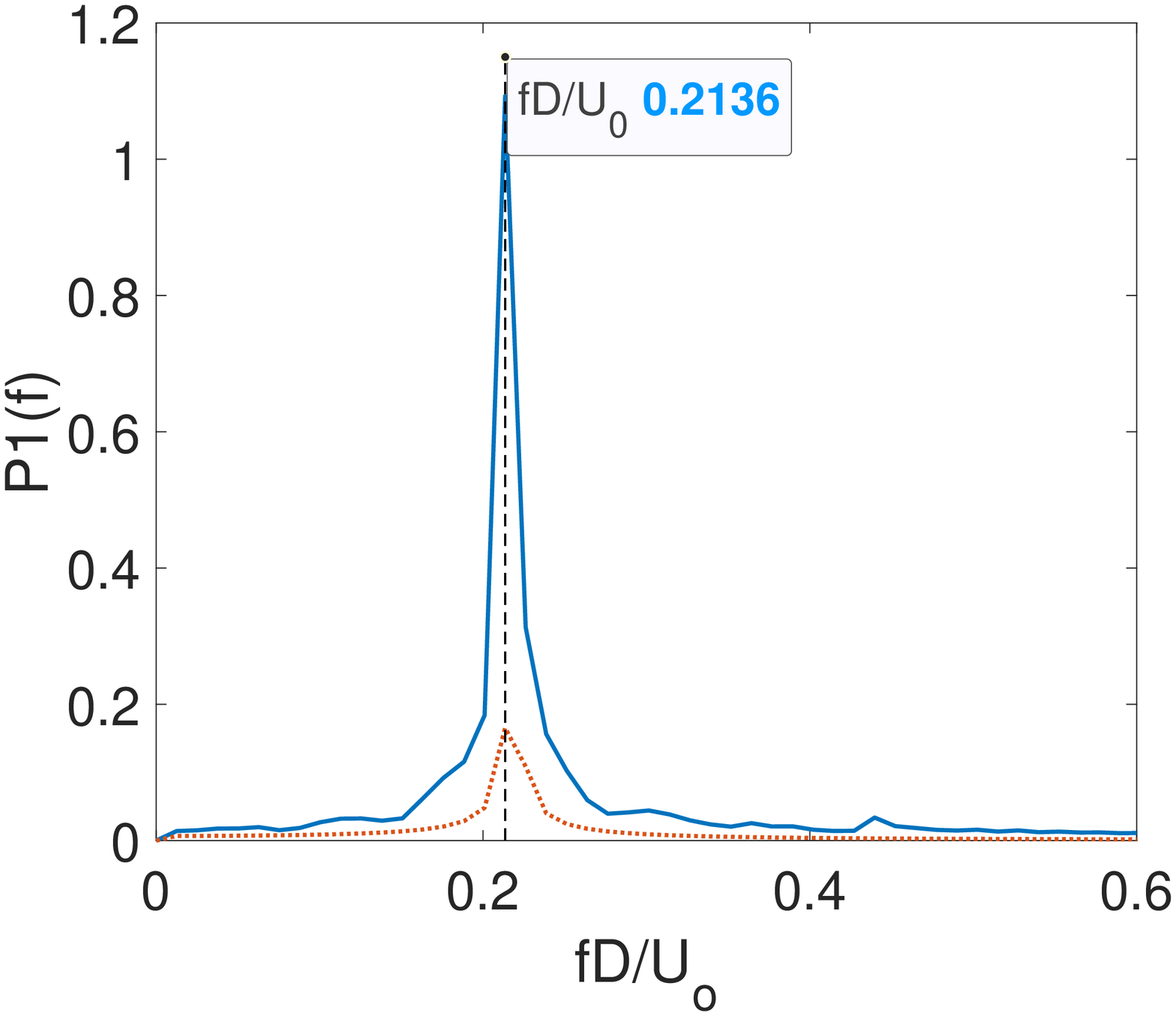}}
    \vspace{-12mm}
    \caption{}\label{fft_6_1}
    \end{subfigure}
    \begin{subfigure}{0.325\textwidth}
    {\includegraphics[width = 0.99\columnwidth,trim={28mm 0mm 4cm 0mm}, clip]{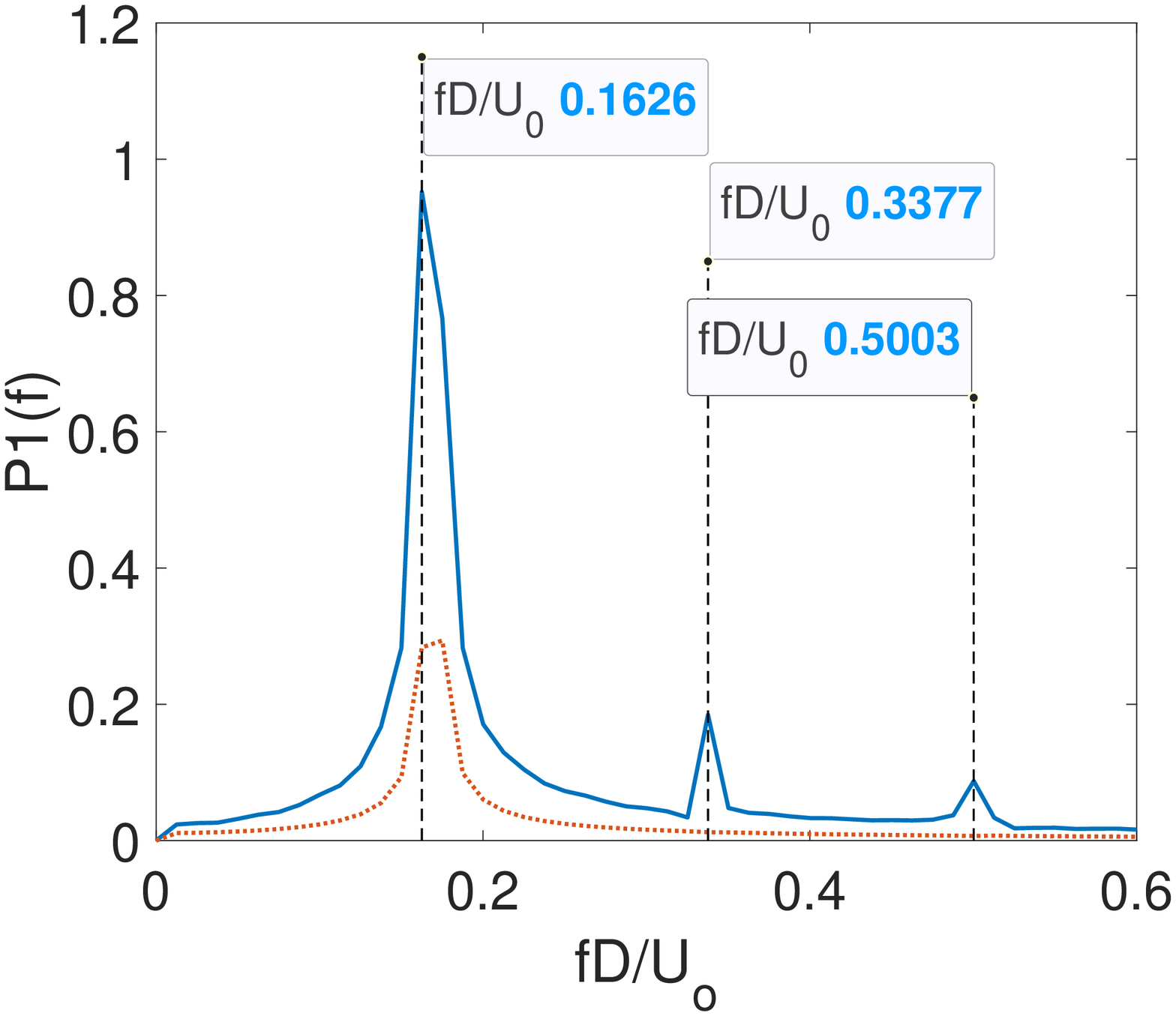}}
    \vspace{-12mm}
    \caption{}\label{fft_6_2}
    \end{subfigure}
    \begin{subfigure}{0.325\textwidth}
    {\includegraphics[width = 0.99\columnwidth,trim={28mm 0mm 4cm 0mm}, clip]
    {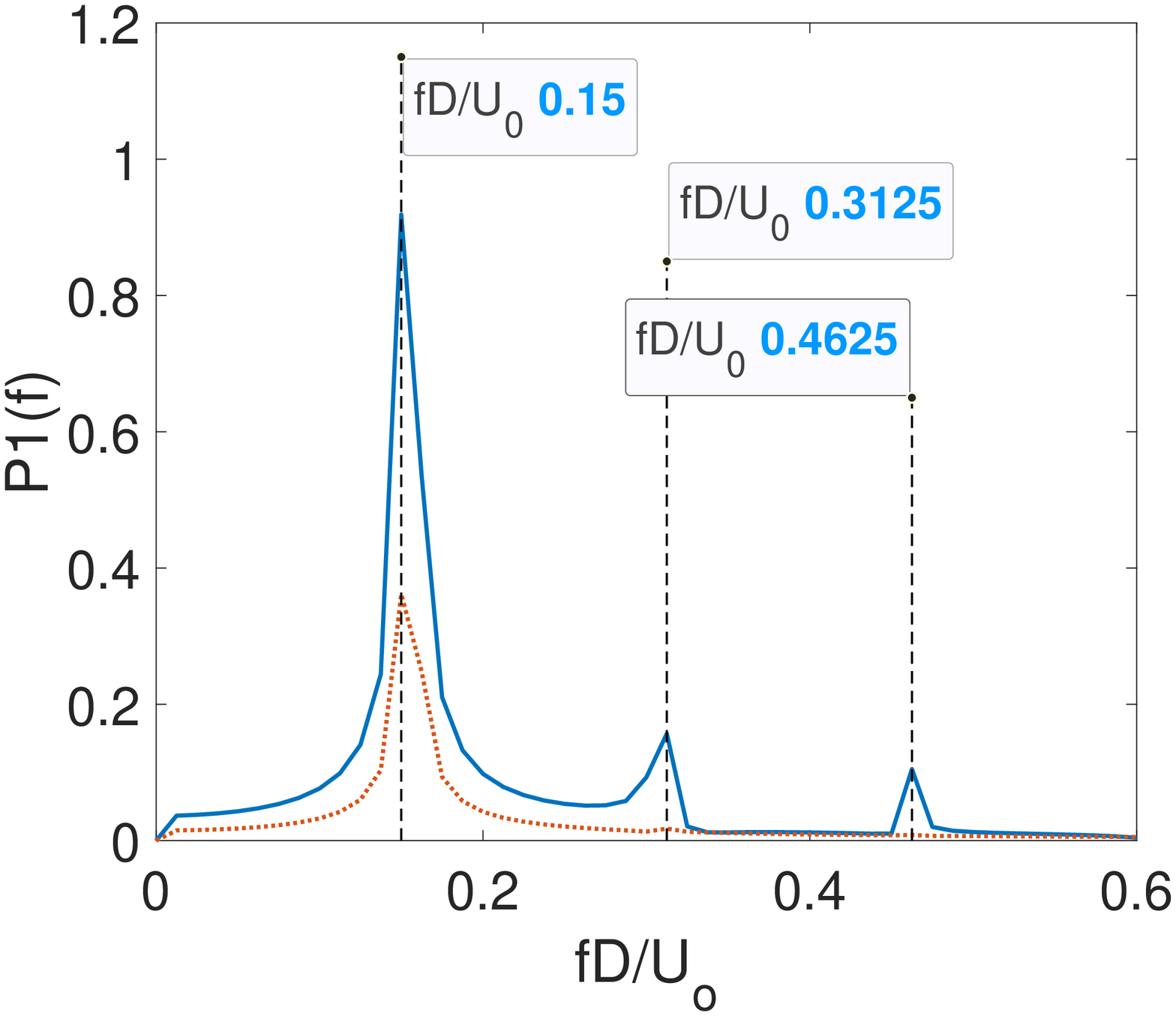}}
    \vspace{-12mm}
    \caption{}\label{fft_6_3}
    \end{subfigure}
    \begin{subfigure}{0.325\textwidth}
    {\includegraphics[width = 0.99\columnwidth,trim={28mm 0mm 4cm 0mm}, clip]{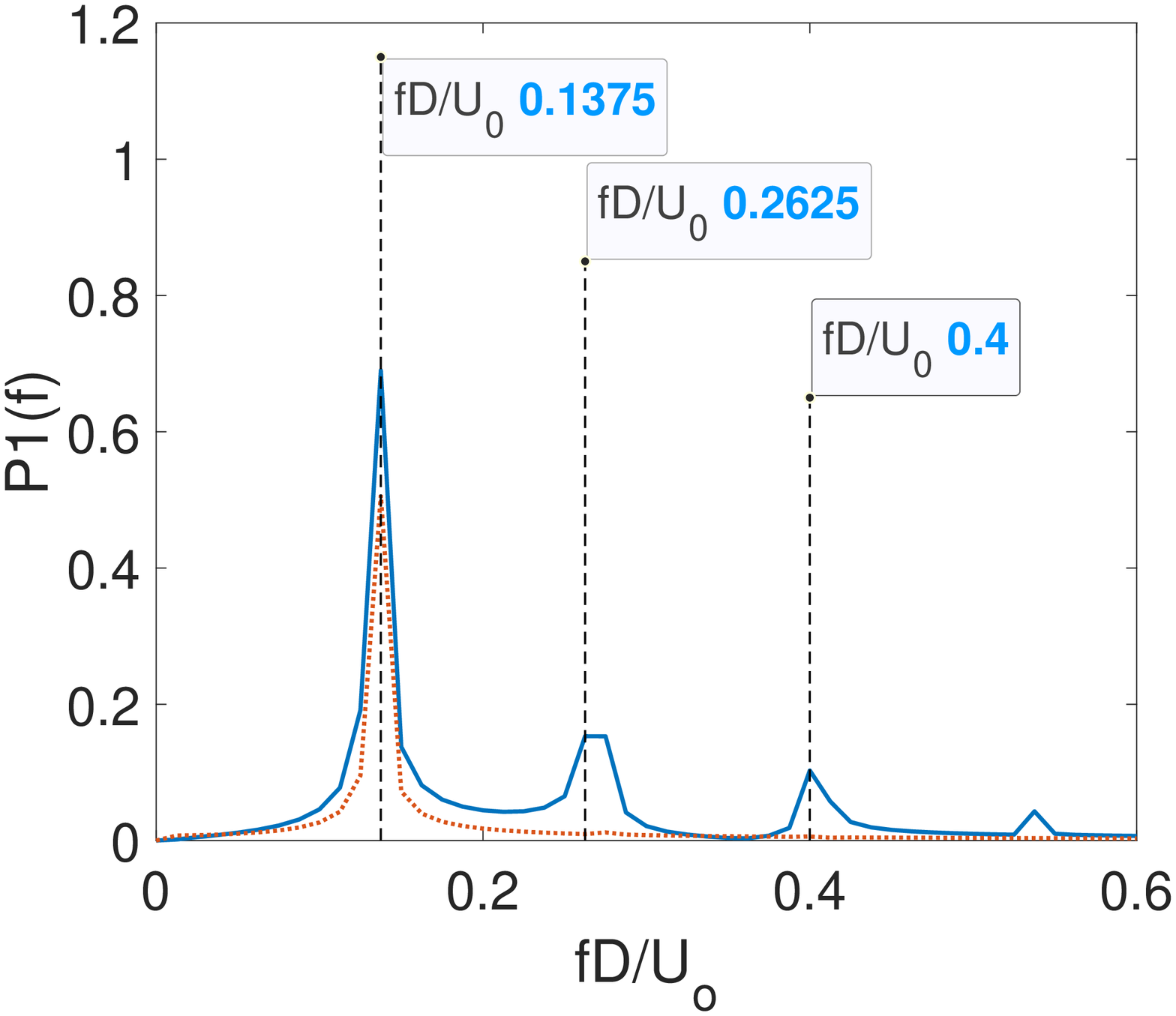}}
    \vspace{-12mm}
    \caption{}\label{fft_6_4}
    \end{subfigure}
    \begin{subfigure}{0.325\textwidth}
    {\includegraphics[width = 0.99\columnwidth,trim={28mm 0mm 4cm 0mm}, clip]{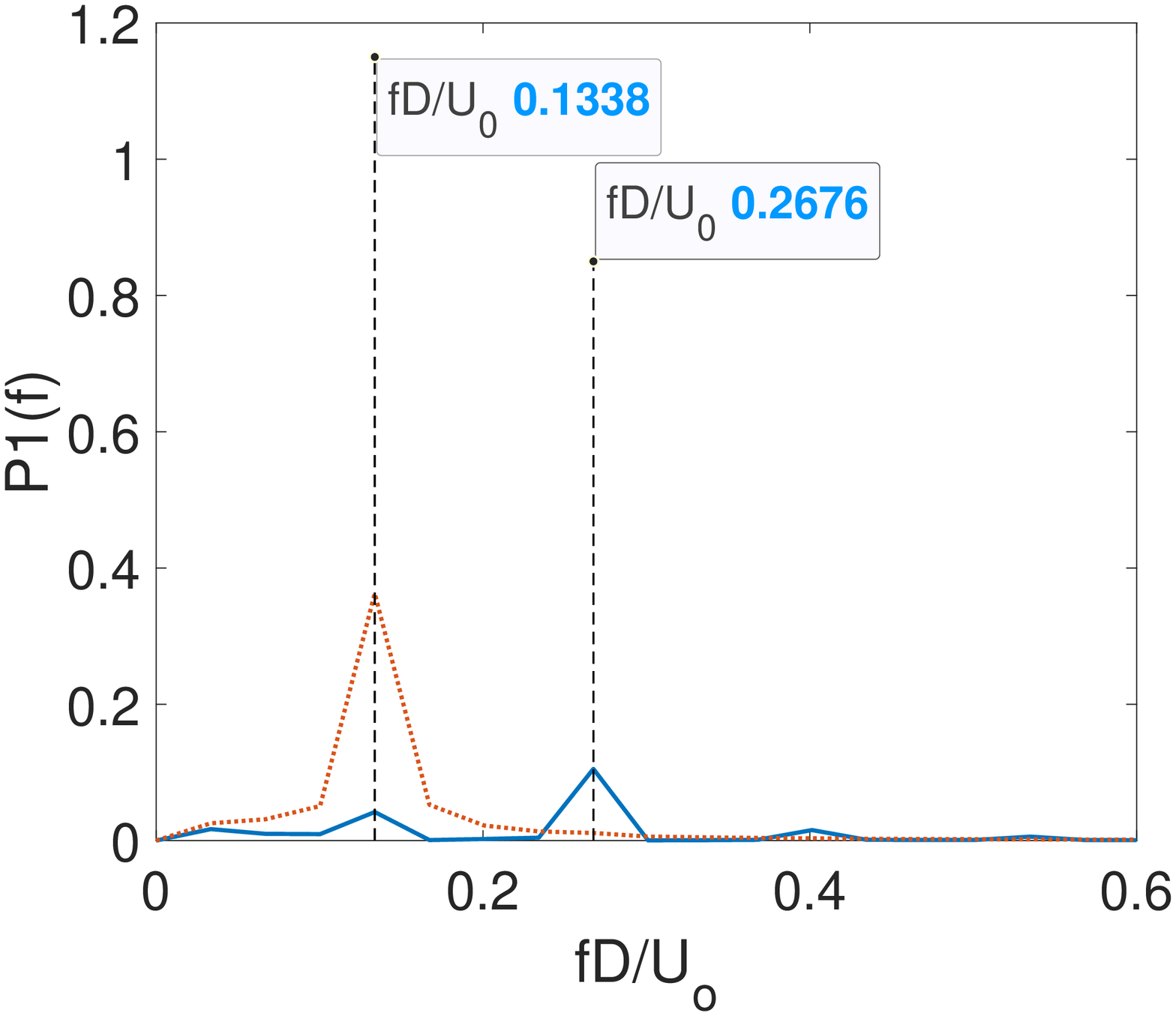}}
    \vspace{-12mm}
    \caption{}\label{fft_6_5}
    \end{subfigure}
    \begin{subfigure}{0.325\textwidth}
    {\includegraphics[width = 0.99\columnwidth,trim={28mm 0mm 4cm 0mm}, clip]{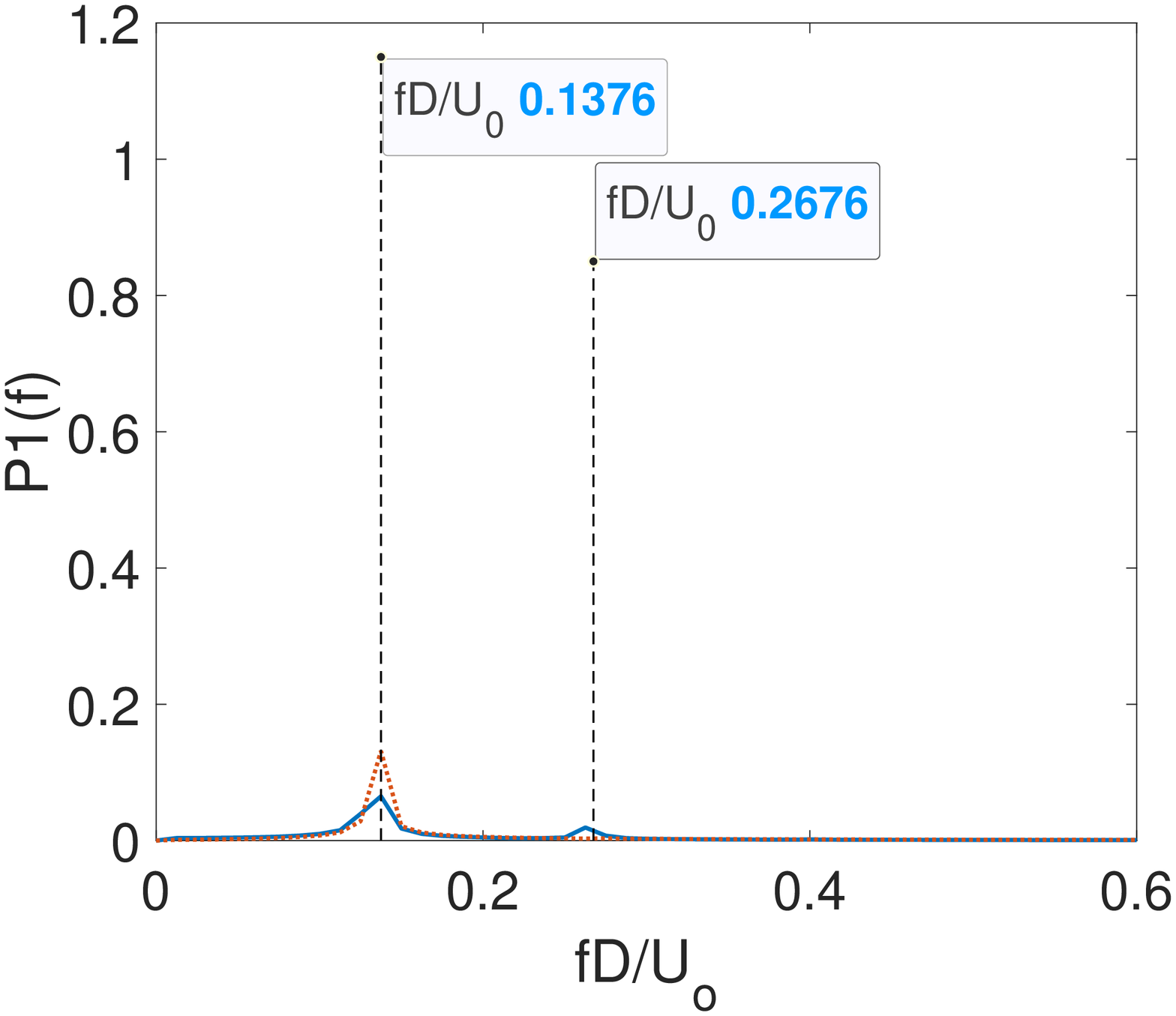}}
    \vspace{-12mm}
    \caption{}\label{fft_6_6}
    \end{subfigure}
\caption{Fast Fourier transform (FFT) of the lift force response for different $U^*$ values at a gap ratio of $g/D = 0.6$ for $m^* = 10$ and $Re = 100$. a) $U^*$ = 4.25, b) $U^*$ = 5.5, c) $U^*$ = 6, d) $U^*$ = 7, e) $U^*$ = 7.5, f) $U^*$ = 7.75. (bold line represents lift force and dotted line represents displacement) }
\label{fft_G6}
\end{figure}

Frequency response of the transverse displacement and $C_L$ are presented in figure~\ref{fft_G6} for $g/D = 0.6$ at $U^* = \{4.25,~5.5,~6,~7,~7.5,~7.75\}$. The transverse displacement exhibits a single frequency response while the $C_L$ exhibits multiple frequencies. The figure shows that the transverse displacement frequency locks-in with the first component of the multi frequency $C_L$ response. The successive frequency components of the $C_L$ are found to be higher harmonics of the first frequency. Figures \ref{fft_6_1}-\ref{fft_6_4} correspond to the frequency response plots from the $\mathrm{I^{st}}$ branch while the other two sub-figures correspond to the $\mathrm{II^{nd}}$ branch. In the $\mathrm{I^{st}}$ branch, the first frequency component of the lift is the strongest. However, as the phase $\Phi$ jumps from zero to $\pi$, i.e., the $\mathrm{I^{st}}$ branch transitions into the $\mathrm{II^{nd}}$ branch, the second frequency component of the lift force becomes stronger than the first component. The transverse displacement response however, is still locked-in with the first frequency component of the lift force. This behaviour is consistent with the observations for an isolated cylinder in [\onlinecite{prasanth_mittal_2008} where the phase jump was accompanied by overshadowing of the first frequency lift force component by the second frequency lift force component. As $U^*$ increases further in the $\mathrm{II^{nd}}$ branch, the first frequency component of lift force once again becomes the strongest component. Similar behavior was also observed in [\onlinecite{Chen2019}]. From figure \ref{fft_G6} it can also be observed that the lock-in frequency reduces with increasing $U^*$ values. This reduction is more prominent in the $\mathrm{I^{st}}$ branch as compared to the $\mathrm{II^{nd}}$ branch where the first frequency does not change considerably with increasing $U^*$. 

The response characteristics for gap $g/D = 0.5$ presented in figure~\ref{G5_branch} are quite similar to the ones presented above for $g/D=0.6$. In both these cases, the $\mathrm{I^{st}}$ response branch is wider with increasing transverse amplitudes while the $\mathrm{II^{nd}}$ response branch is very narrow and is marked by decreasing response amplitudes. The frequency response too is similar to the case of $g/D = 0.6$. This branching of the synchronization response is similar to the case of low-m$*\zeta$ that was explored thoroughly in [\onlinecite{govardhan2000}] where the upper branch was marked by large amplitude response and $\Phi=0$, and the lower branch was marked by low maximum amplitude values and a phase angle close to $\pi$.

\subsubsection{Gap ratios ($g/D = 0.4, 0.3$ and $0.2$)}

\begin{figure} 
\centering
{\includegraphics[width = 1\linewidth]{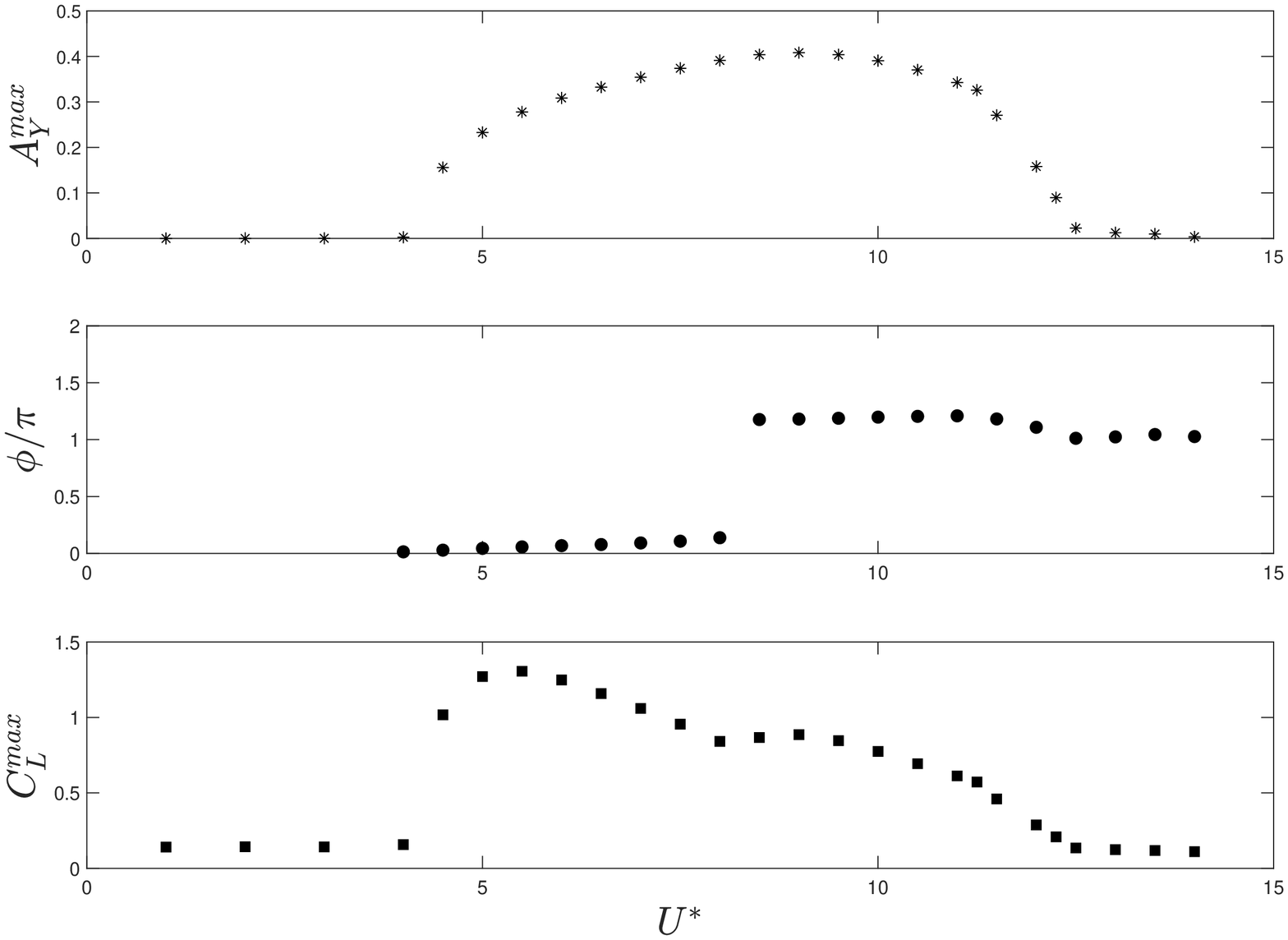}}
\caption{Variation of maximum transverse displacements$(A_{y}^{max}/D)$, $\Phi$ and $C_L^{max}$ at a gap ratio of $g/D = 0.4$ with $U^*$ for $m^* = 10$ and $Re = 100$}
\label{G4_branch}
\end{figure}

\begin{figure} 
\centering
{\includegraphics[width = 1\linewidth]{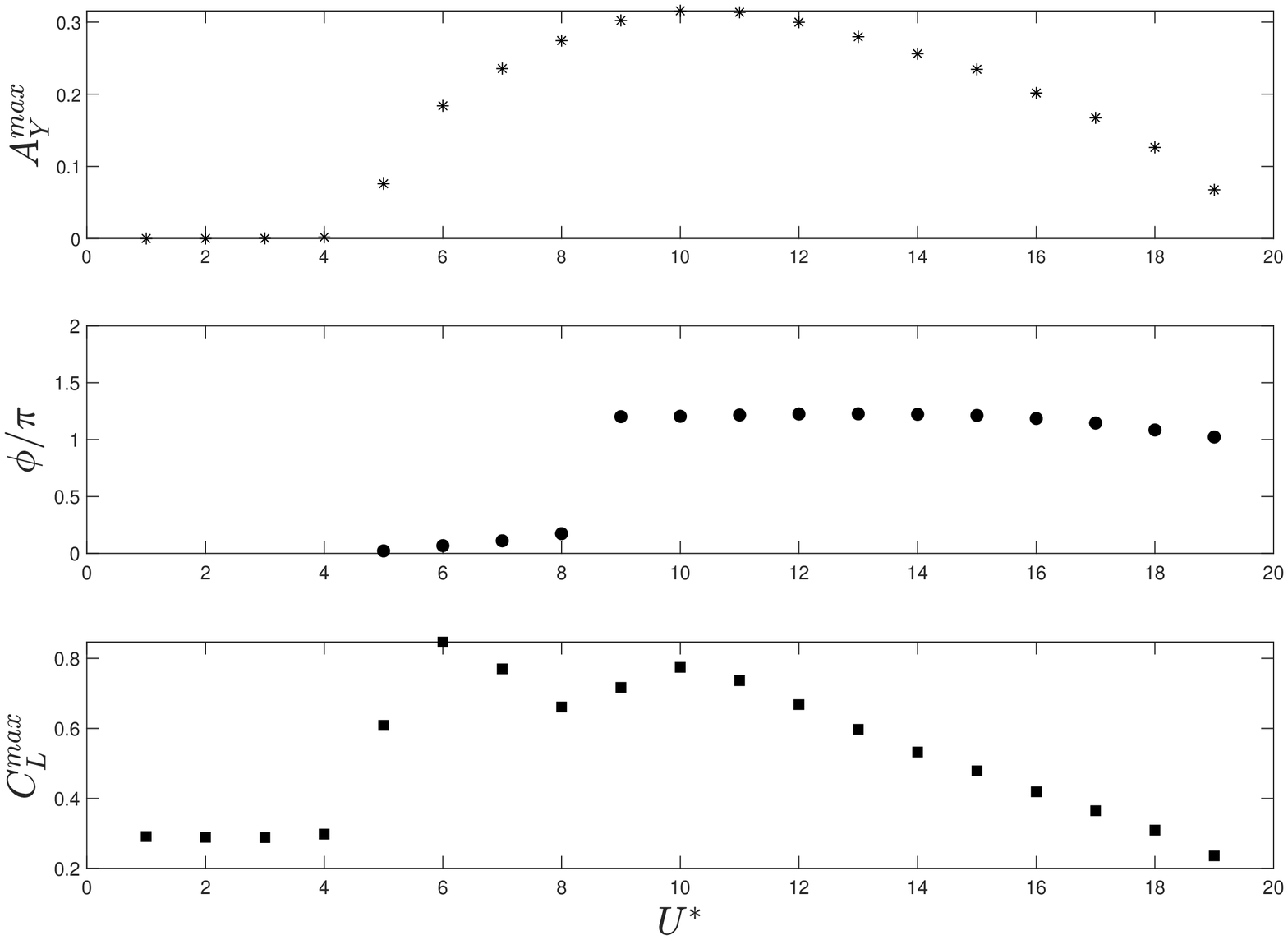}}
\caption{Variation of maximum transverse displacements$(A_{y}^{max}/D)$, $\Phi$ and $C_L^{max}$ at a gap ratio of $g/D = 0.3$ with $U^*$ for $m^* = 10$ and $Re = 100$}
\label{G3_branch}
\end{figure}

\begin{figure} 
\centering
{\includegraphics[width = 1\linewidth]{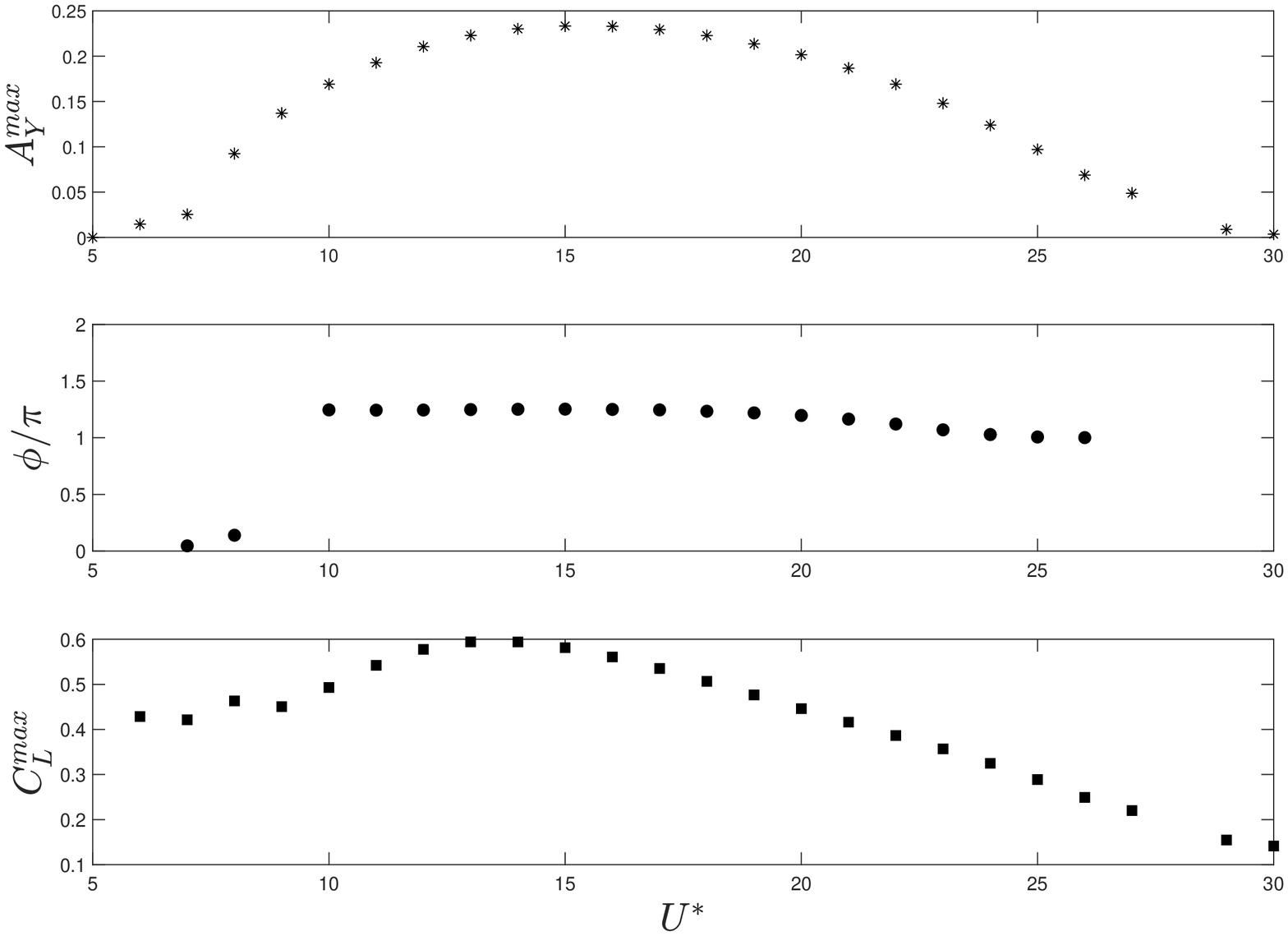}}
\caption{Variation of maximum transverse displacements$(A_{y}^{max}/D)$,  $\Phi$ and $C_L^{max}$ at a gap ratio of $g/D = 0.2$ with $U^*$ for $m^* = 10$ and $Re = 100$}
\label{G2_branch}
\end{figure}

 To understand the branching response characteristics for the gap ratio $g/D = 0.4$, $(A_{y}^{max}/D)$, $\Phi$ and $C_L^{max}$ are presented as a function of different $U^*$ values in figure~\ref{G4_branch}. In contrast to the $A_{y}^{max}$ values for $g/D=0.6$ and $0.5$ which involved a sharp jump, the $A_{y}^{max}$ in figure~\ref{G4_branch} exhibit a smooth variation over the lock-in range of $4 \le U^* \le 12.5$ where the $A_{y}^{max}$ value first increases slowly from zero before starting to reduce gradually back to zero. This however does not imply that there is no branching in the case of $g/D = 0.4$, there is a sudden jump in the $\Phi$ values from $0$ to a value that is slightly greater than $\pi$. This again characterises the VIV response for $g/D=0.4$ into two branches with the $\mathrm{I^{st}}$ branch marked by $\Phi = 0$ for $4 \le U^* \le 8$ and the $\mathrm{II^{nd}}$ marked by $\Phi$ values slightly greater than $\pi$ for $U^* \le 12.5$. Unlike the $\mathrm{I^{st}}$ and $\mathrm{II^{nd}}$ branches corresponding to the gaps $g/D=0.6$ and $0.5$ wherein the $(A_{y}^{max}/D)$ do not increase with $U^*$ for the $\mathrm{II^{nd}}$ branch, the first and second subplots of figure~\ref{G4_branch} clearly show that the $A_{y}^{max}$ values continue to increase even in the $\mathrm{II^{nd}}$ branch before gradually reducing to zero.  %The third subplot of figure~\ref{G4_branch} presents the $C_{L}^{max}$ values as a function of $U^*$ value further supports the two branched response discussed earlier.
 The variation of $C_L^{max}$ with $U^*$ subplot from figure~\ref{G4_branch} shows two distinct lobes and on further comparison between the second and third subplots, the transition between the lobes is accompanied by the jump in the $\Phi$ from $0$ to a value slightly greater than $\pi$ which characterizes the branching phenomenon.  
 
 The response characteristics for gap ratio $g/D = 0.3$ and $0.2$ presented in figure~\ref{G3_branch} and \ref{G2_branch} respectively are similar to the response characteristics observed for gap $g/D = 0.4$ in figure~\ref{G4_branch}. Comparing the branching response characteristics of $g/D = 0.2,~0.3$ and $0.4$ with that of $g/D = 0.5$ and $0.6$, it can be observed that with decreasing $g/D$, the $\mathrm{I^{st}}$ branch narrows and the $\mathrm{II^{nd}}$ branch widens. This behaviour can be observed both from the $\Phi$ as well as the $C_L^{max}$ subplots where the first lobe reduces in size and the second lobe becomes more prominent as the gap ratio reduces from $g/D = 0.4-0.2$.
 
\begin{figure} 

    \begin{subfigure}{0.325\textwidth}
    {\includegraphics[width = 0.99\columnwidth,trim={30mm 0mm 4.4cm 0mm}, clip]{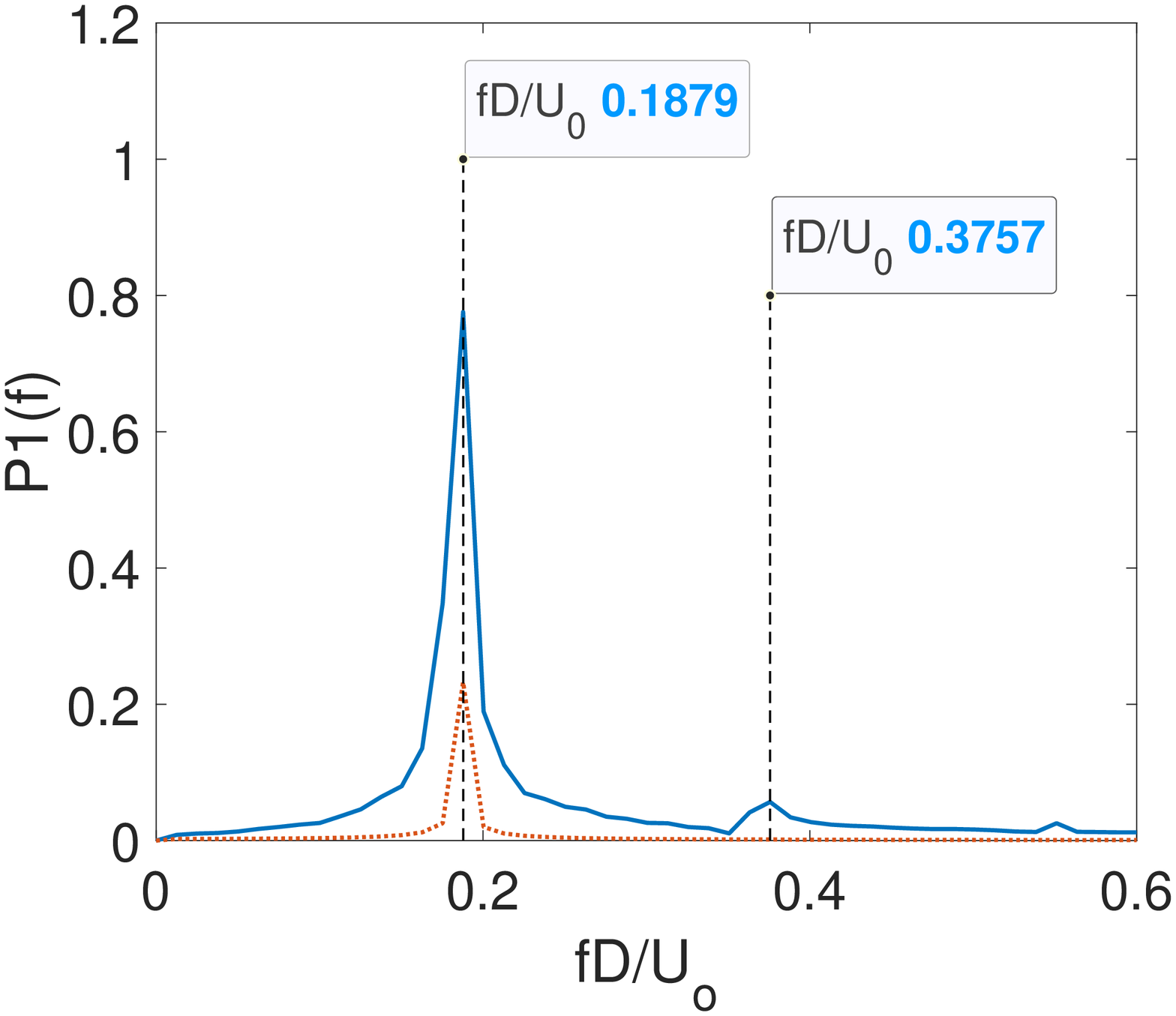}}
    %\centering
    \caption{}\label{fft_4_1}
    \end{subfigure}
    \begin{subfigure}{0.325\textwidth}
    {\includegraphics[width = 0.99\columnwidth,trim={28mm 0mm 4cm 0mm}, clip]{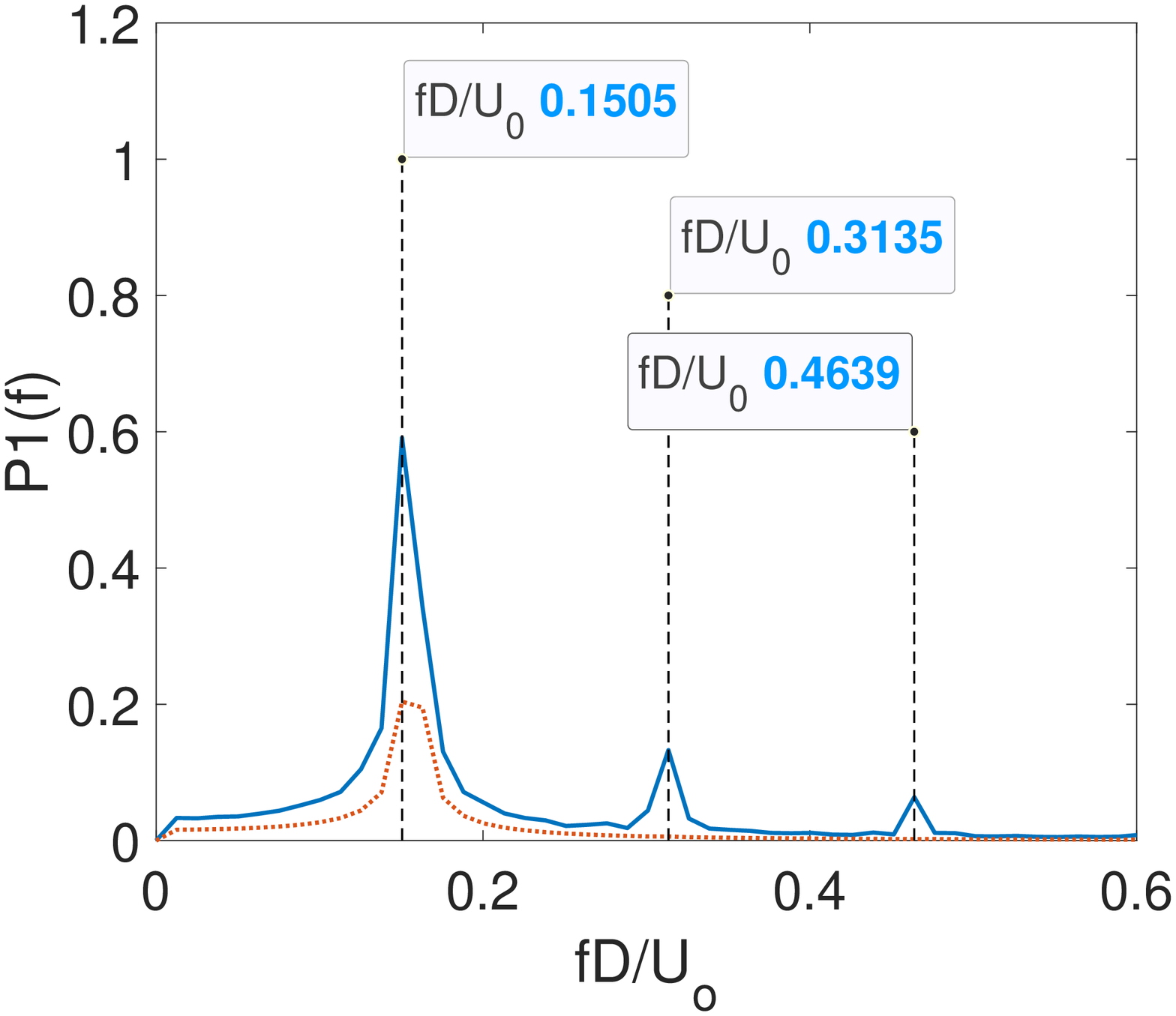}}
    %\centering
    \caption{}\label{fft_4_2}
    \end{subfigure}
    \begin{subfigure}{0.325\textwidth}
    {\includegraphics[width = 0.99\columnwidth,trim={28mm 0mm 4cm 0mm}, clip]{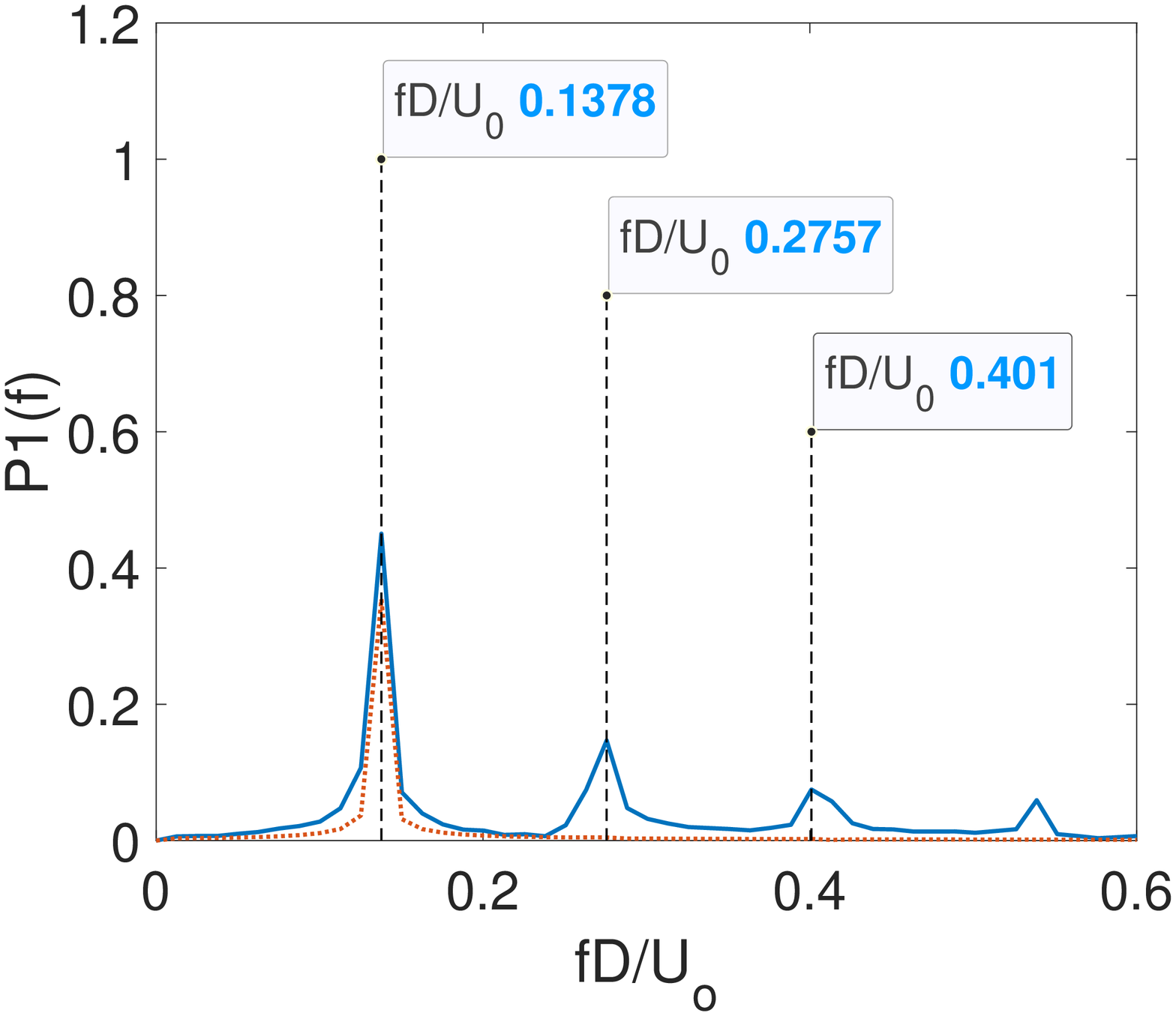}}
    \caption{}\label{fft_4_3}
    \end{subfigure}
    \begin{subfigure}{0.325\textwidth}
    {\includegraphics[width = 0.99\columnwidth,trim={28mm 0mm 4cm 0mm}, clip]{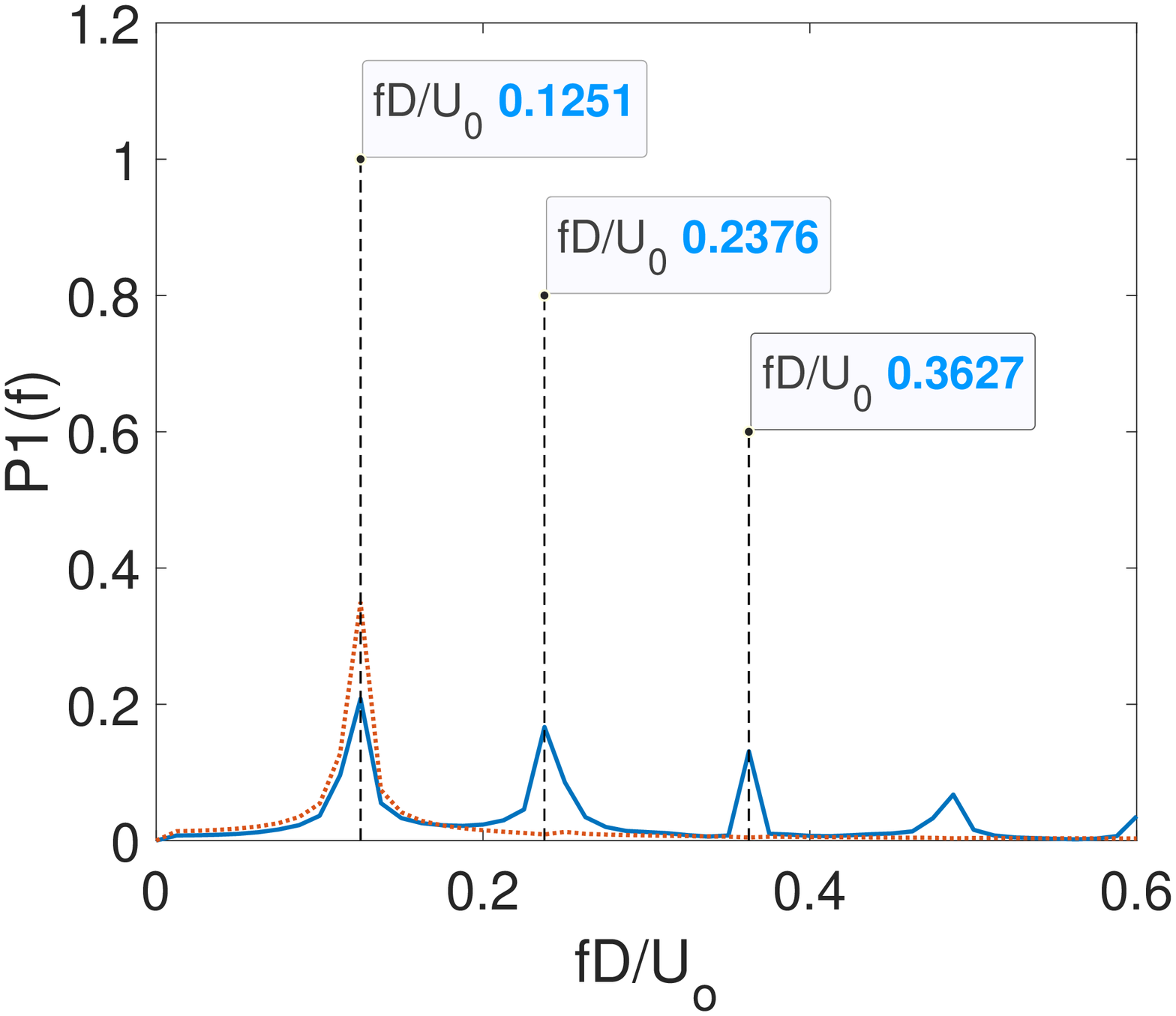}}
    \caption{}\label{fft_4_4}
    \end{subfigure}
    \begin{subfigure}{0.325\textwidth}
    {\includegraphics[width = 0.99\columnwidth,trim={28mm 0mm 4cm 0mm}, clip]{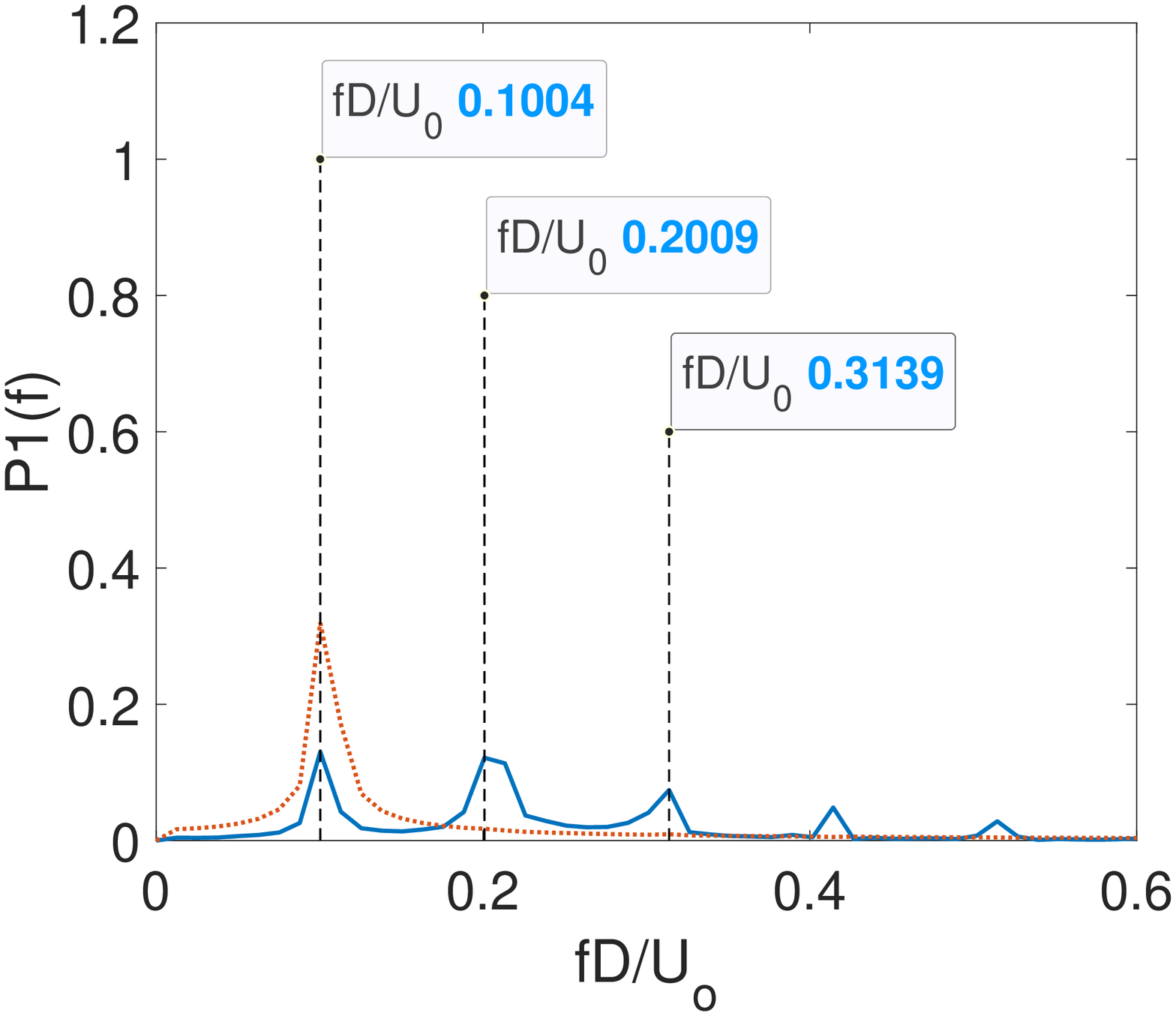}}
    \caption{}\label{fft_4_5}
    \end{subfigure}
    \begin{subfigure}{0.325\textwidth}
    {\includegraphics[width = 0.99\columnwidth,trim={28mm 0mm 4cm 0mm}, clip]{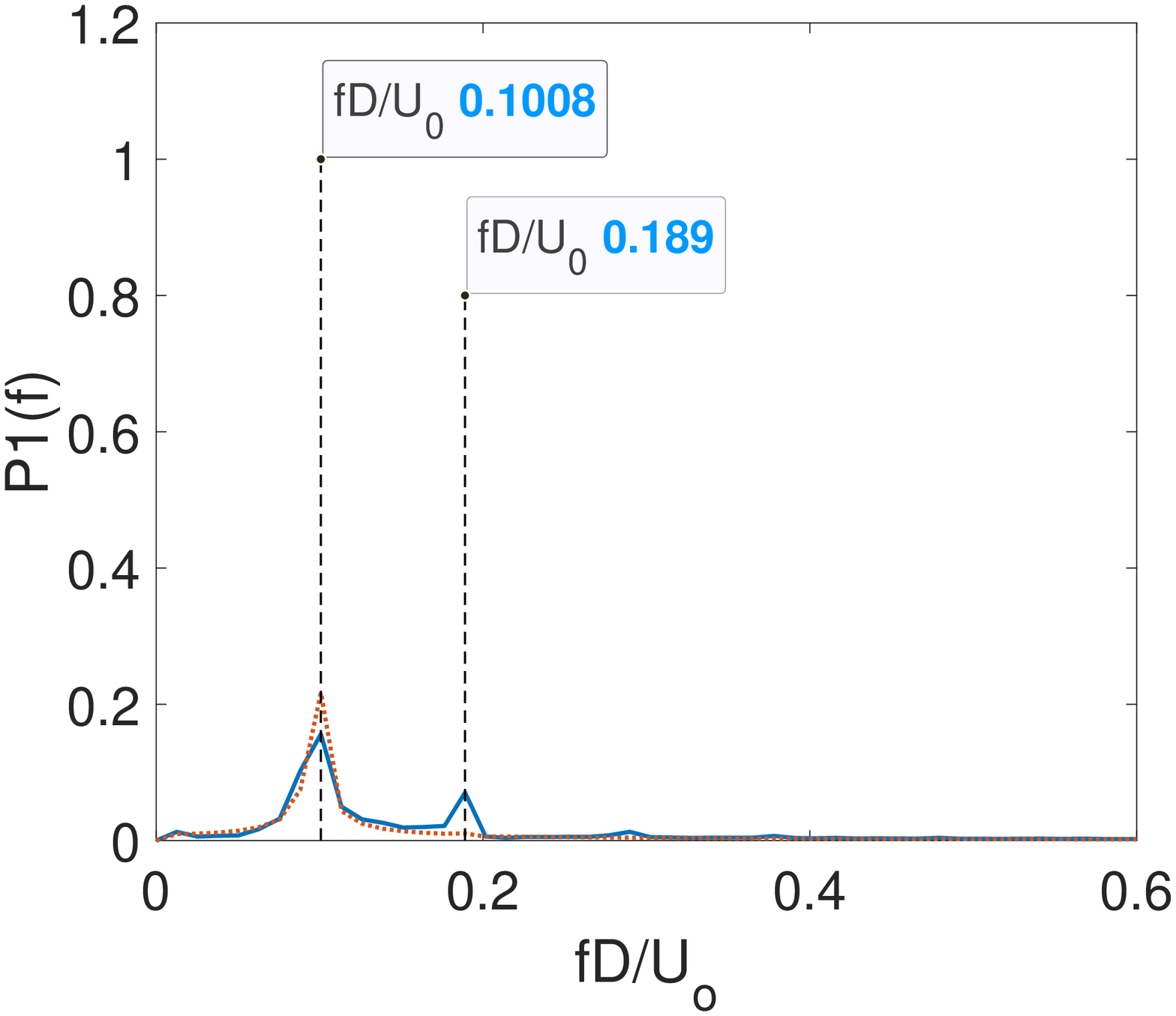}}
    \caption{}\label{fft_4_6}
    \end{subfigure}
\caption{Fast Fourier transform (FFT) of the lift force response for different $U^*$ values at a gap ratio of $g/D = 0.4$ for $m^* = 10$ and $Re = 100$. a) $U^*$ = 5, b) $U^*$ = 6, c) $U^*$ = 7, d) $U^*$ = 8, e) $U^*$ = 10, f) $U^*$ = 11.5. (bold line represents lift force and dotted line represents displacement) }
\label{fft_G4}
\end{figure}

 The frequency response for the gap ratio of $g/D = 0.4$ is presented in figure~\ref{fft_G4} for $U^* = 5,~6,~7,~8,~10$ and $11.5$. It is similar to the frequency response for the gap ratio of $g/D = 0.6$ with the single frequency displacement response locking in with the first frequency response of the multi-component lift force. Further, the transition from the $\mathrm{I^{st}}$ branch to the $\mathrm{II^{nd}}$ branch is accompanied by an increase in the strength of the second frequency component of the lift force as compared to the first frequency component similar to the case of $g/D = 0.6$. Sub-figures \ref{fft_4_1}-\ref{fft_4_4} correspond to the $\mathrm{I^{st}}$ branch while the other two sub-figures correspond to the $\mathrm{II^{nd}}$ branch. Interestingly, as the $U^*$ value increases further, it can be observed that the strength of the second frequenct component reduces and the first frequency component of the lift force becomes stronger again. Similar behaviour can be observed for both $g/D = 0.3$ and $0.2$.
 
 %the $\mathrm{II^{nd}}$ branch is wider than the $\mathrm{I^{st}}$ branch. As the gap ratio decreases from $g/D = 0.5$ to $0.4$ and then to $0.3$, the $\mathrm{I^{st}}$ branch narrows and the $\mathrm{II^{nd}}$ branch widens.
 
 %Figure~\ref{G2_branch} presents $(A_{y}^{max}/D)$, $\Phi$ and  $C_L^{max}$ as a function of different $U^*$ values for gap ratio $g/D = 0.2$. Similar to the case of $g/D = 0.3$ and $0.4$, the first subplot presents a contiguous $A_{y}^{max}$ response over the lock-in region and a sudden jump is observed in the second subplot at $U^* = 9$ where the $\Phi$ value shifts from $\Phi = 0$ to a value slightly greater than $\pi$. The effect of branching on the $C_{L}^{max}$ values is still distinct, however, the first lobe is barely present and the second lobe becomes predominant exhibiting higher $C_{L}^{max}$ before reducing to zero over a range of $U^*$ values. It can also be observed that the lock-in region has further widened from $g/D = 0.3$ to $0.2$ with the $\mathrm{I^{st}}$ branch narrowing and the $\mathrm{II^{nd}}$ branch widening.

 %Further, the widening of the lock-in region with decreasing gap ratios that was discussed in Section \ref{subsec: res_dyn} is a consequence of the considerable widening of the $\mathrm{II^{nd}}$ response branch.

\subsubsection{Gap ratios ($g/D = 0.1$)}

\begin{figure} 
\centering
{\includegraphics[width = 1\linewidth]{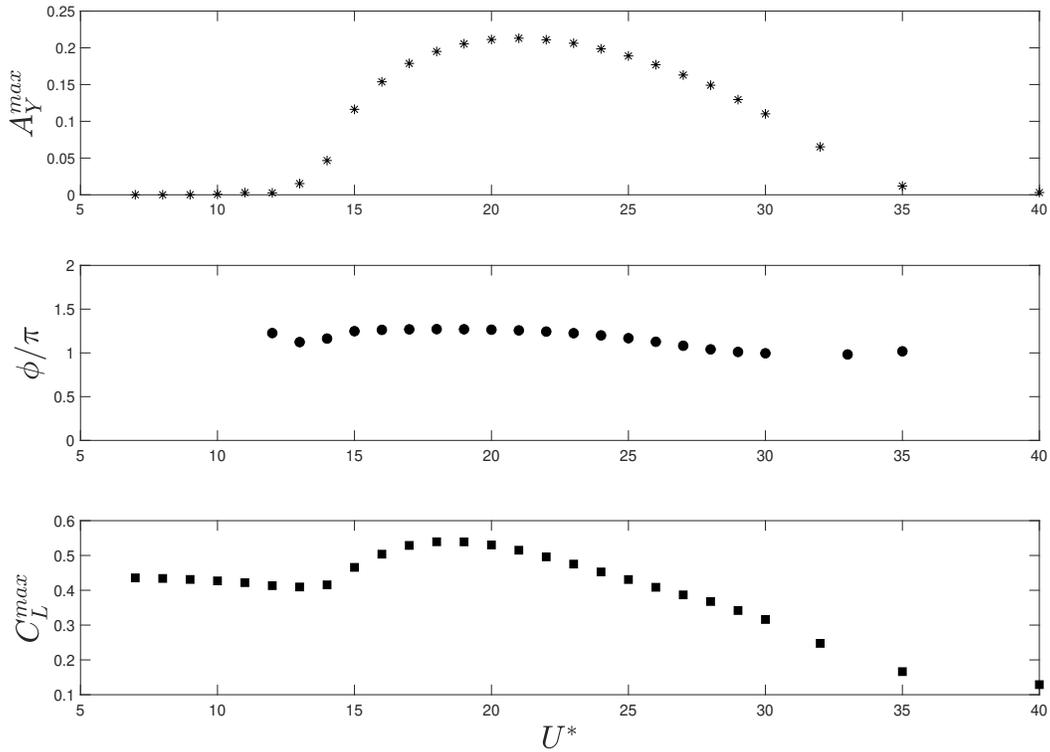}}
\caption{Variation of maximum transverse displacements $(A_{y}^{max}/D)$,  phase angle $\Phi$ and maximum lift coefficient $C_L^{max}$ at a gap ratio of $g/D = 0.1$ with $U^*$ for $m^* = 10$ and $Re = 100$}
\label{G1_branch}
\end{figure}

\begin{figure} 

    \begin{subfigure}{0.325\textwidth}
    {\includegraphics[width = 0.99\columnwidth,trim={30mm 0mm 4.4cm 0mm}, clip]{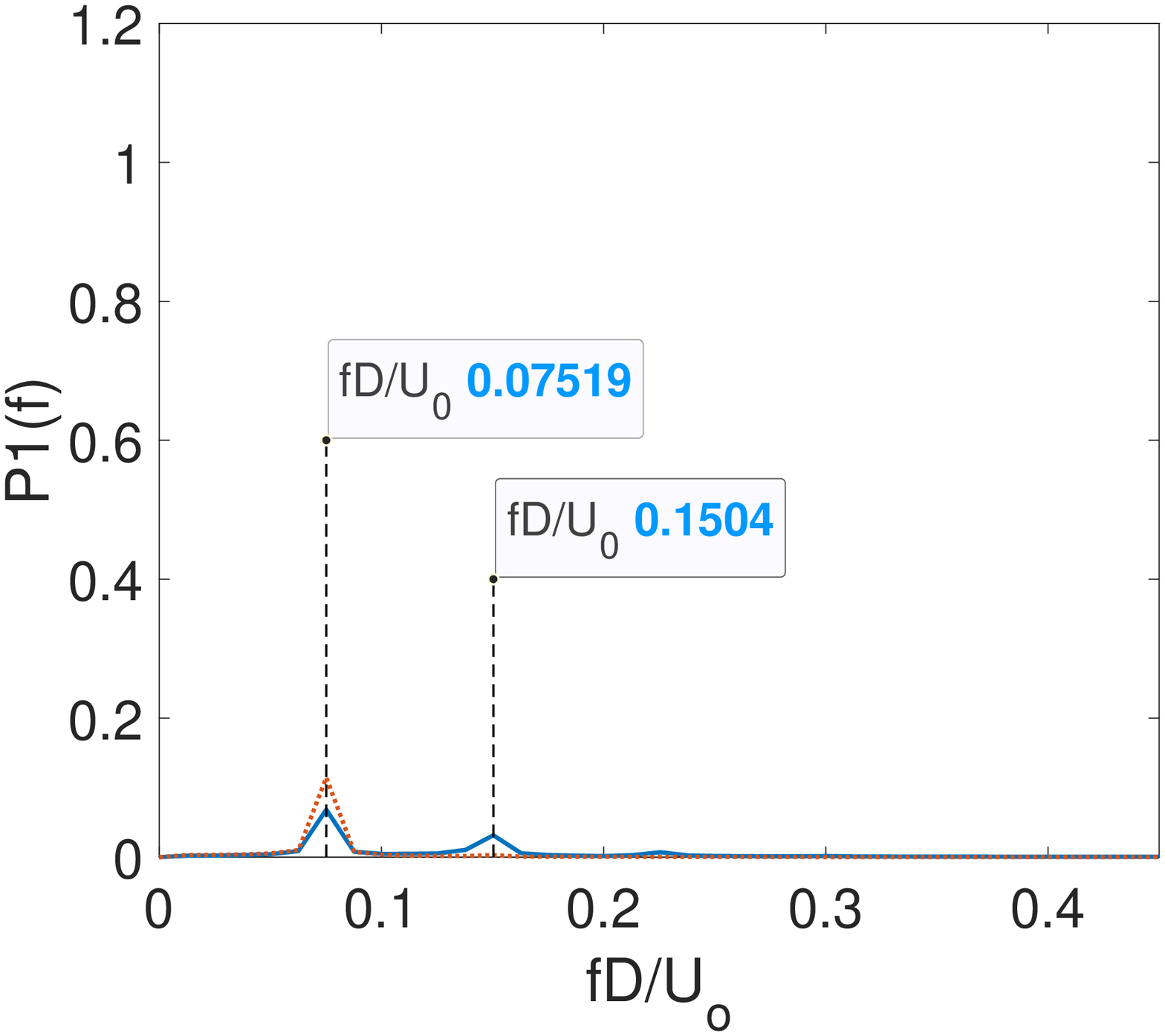}}
    %\centering
    \caption{}\label{fft_1_1}
    \end{subfigure}
    \begin{subfigure}{0.325\textwidth}
    {\includegraphics[width = 0.99\columnwidth,trim={30mm 0mm 4.4cm 0mm}, clip]{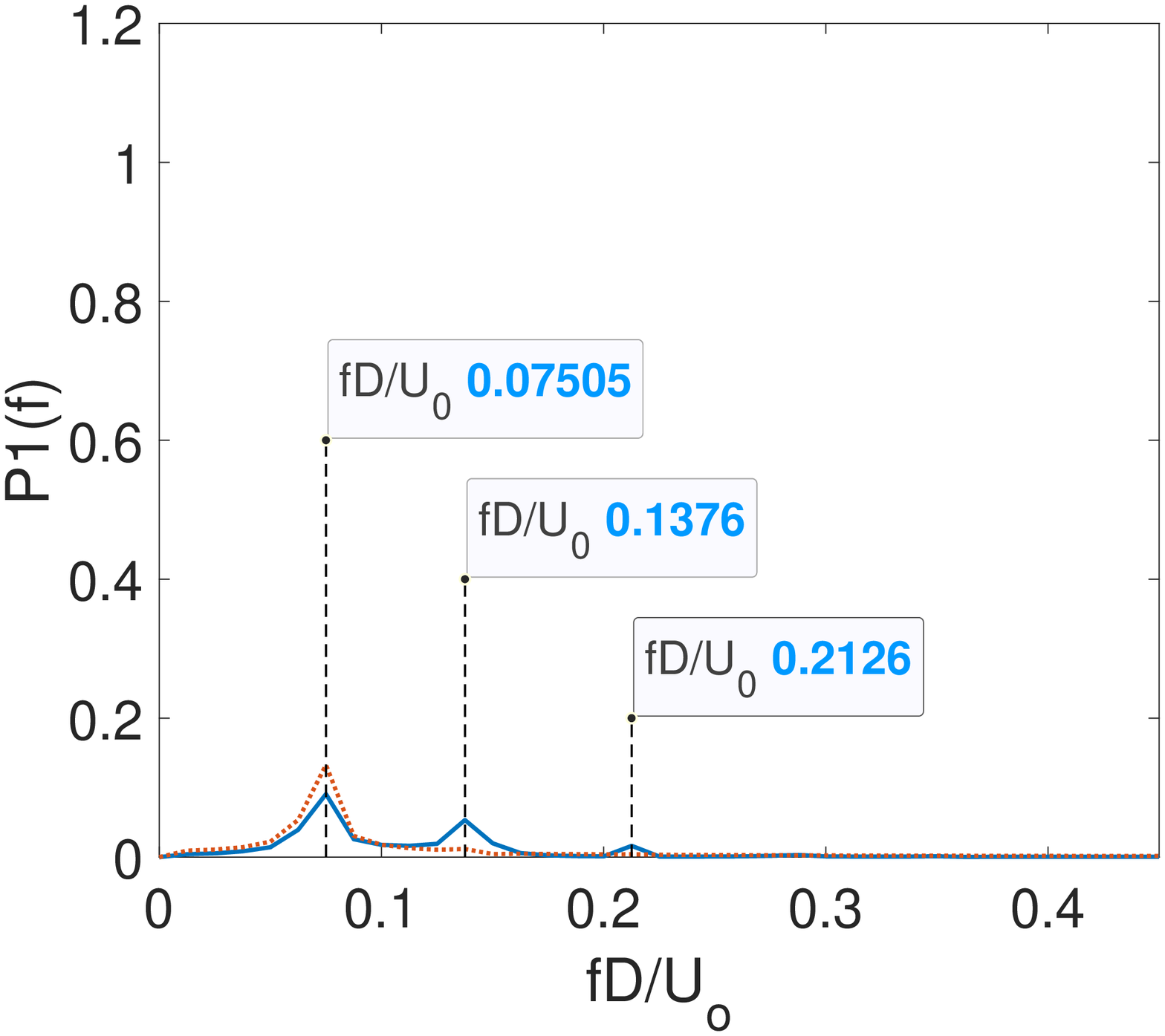}}
    %\centering
    \caption{}\label{fft_1_2}
    \end{subfigure}
    \begin{subfigure}{0.325\textwidth}
    {\includegraphics[width = 0.99\columnwidth,trim={30mm 0mm 4.4cm 0mm}, clip]{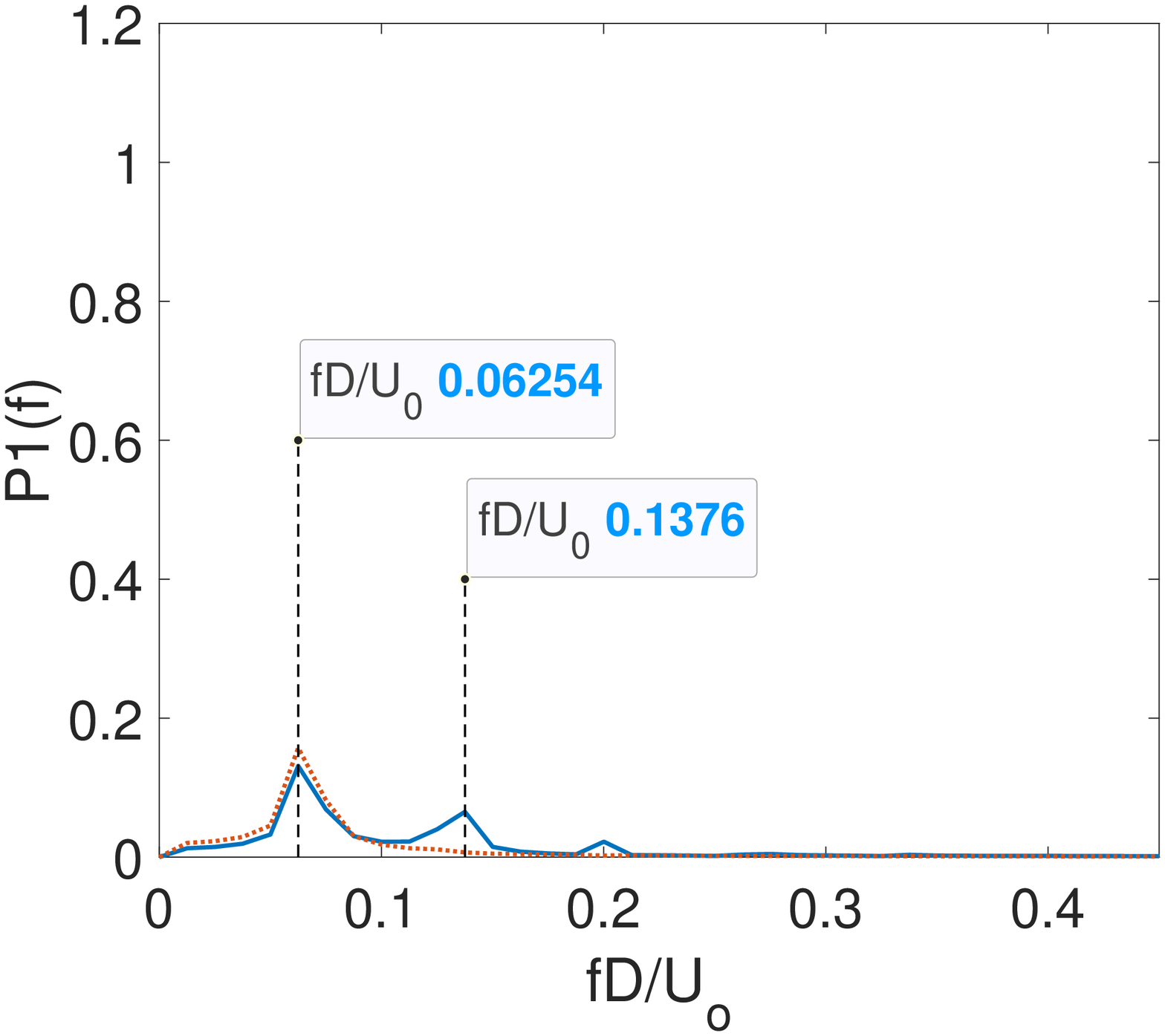}}
    \caption{}\label{fft_1_3}
    \end{subfigure}
    \begin{subfigure}{0.325\textwidth}
    {\includegraphics[width = 0.99\columnwidth,trim={30mm 0mm 4.4cm 0mm}, clip]{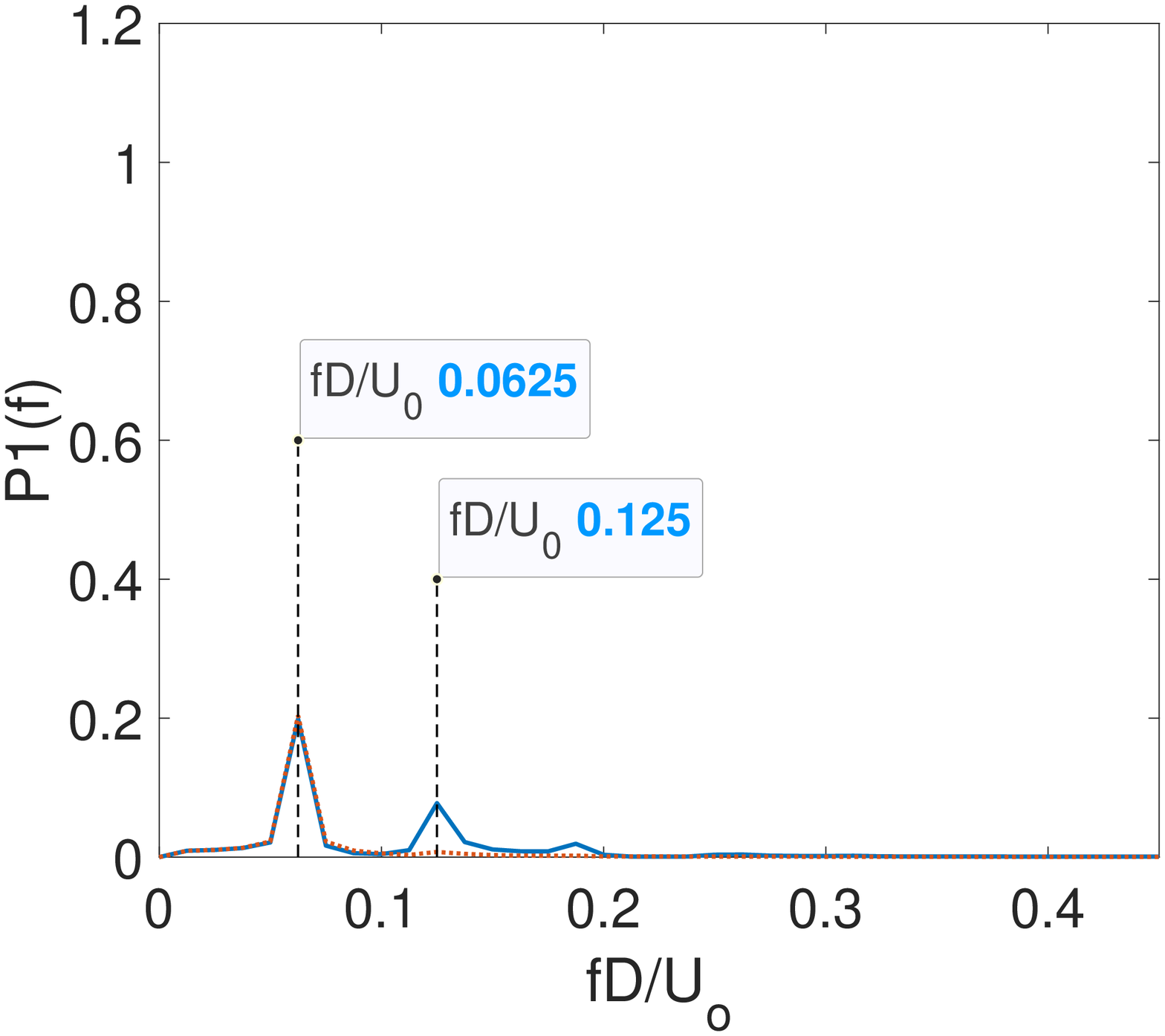}}
    \caption{}\label{fft_1_4}
    \end{subfigure}
    \begin{subfigure}{0.325\textwidth}
   {\includegraphics[width = 0.99\columnwidth,trim={30mm 0mm 4.4cm 0mm}, clip]{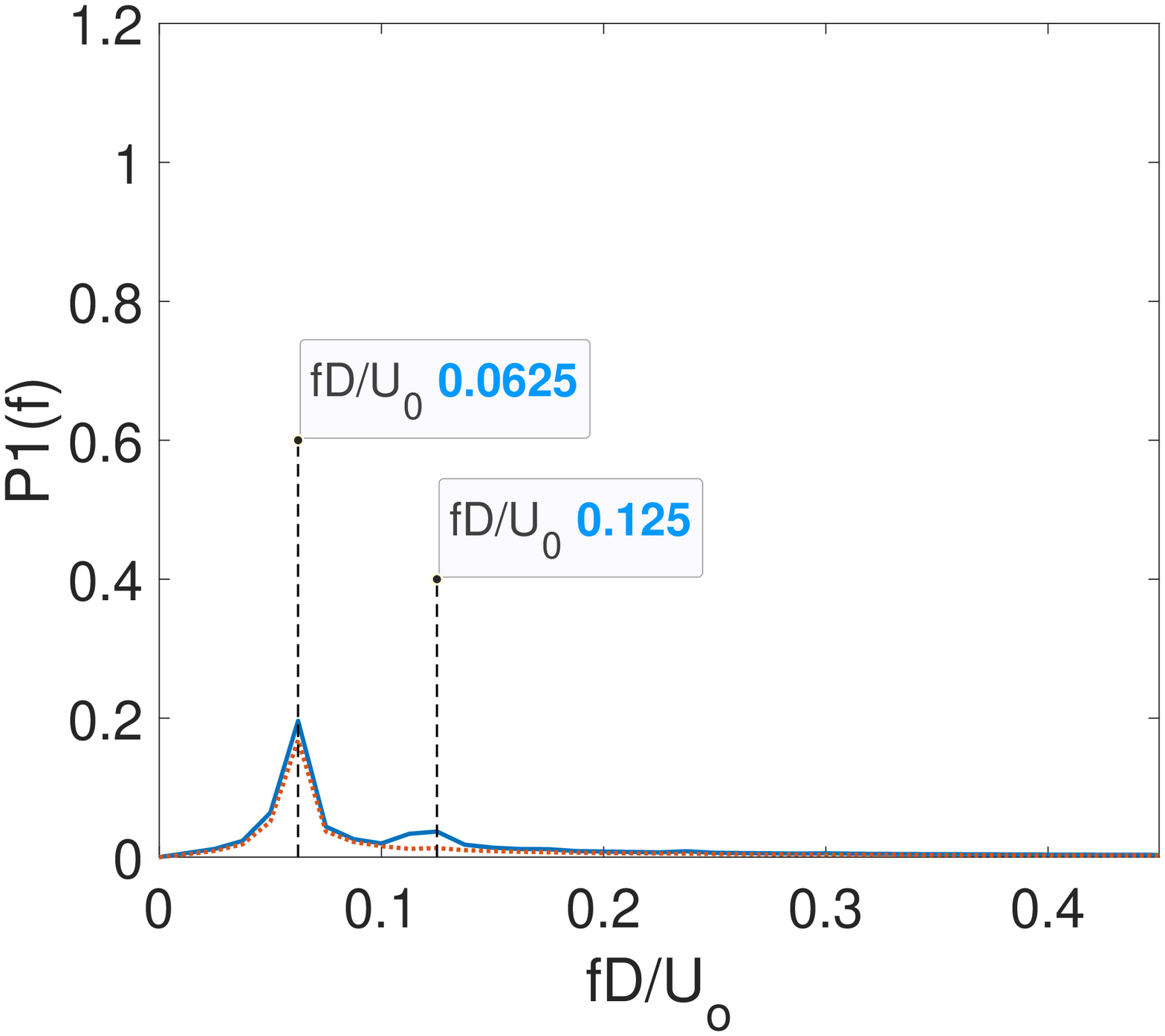}}
    \caption{}\label{fft_1_5}
    \end{subfigure}
    \begin{subfigure}{0.325\textwidth}
    {\includegraphics[width = 0.99\columnwidth,trim={30mm 0mm 4.4cm 0mm}, clip]{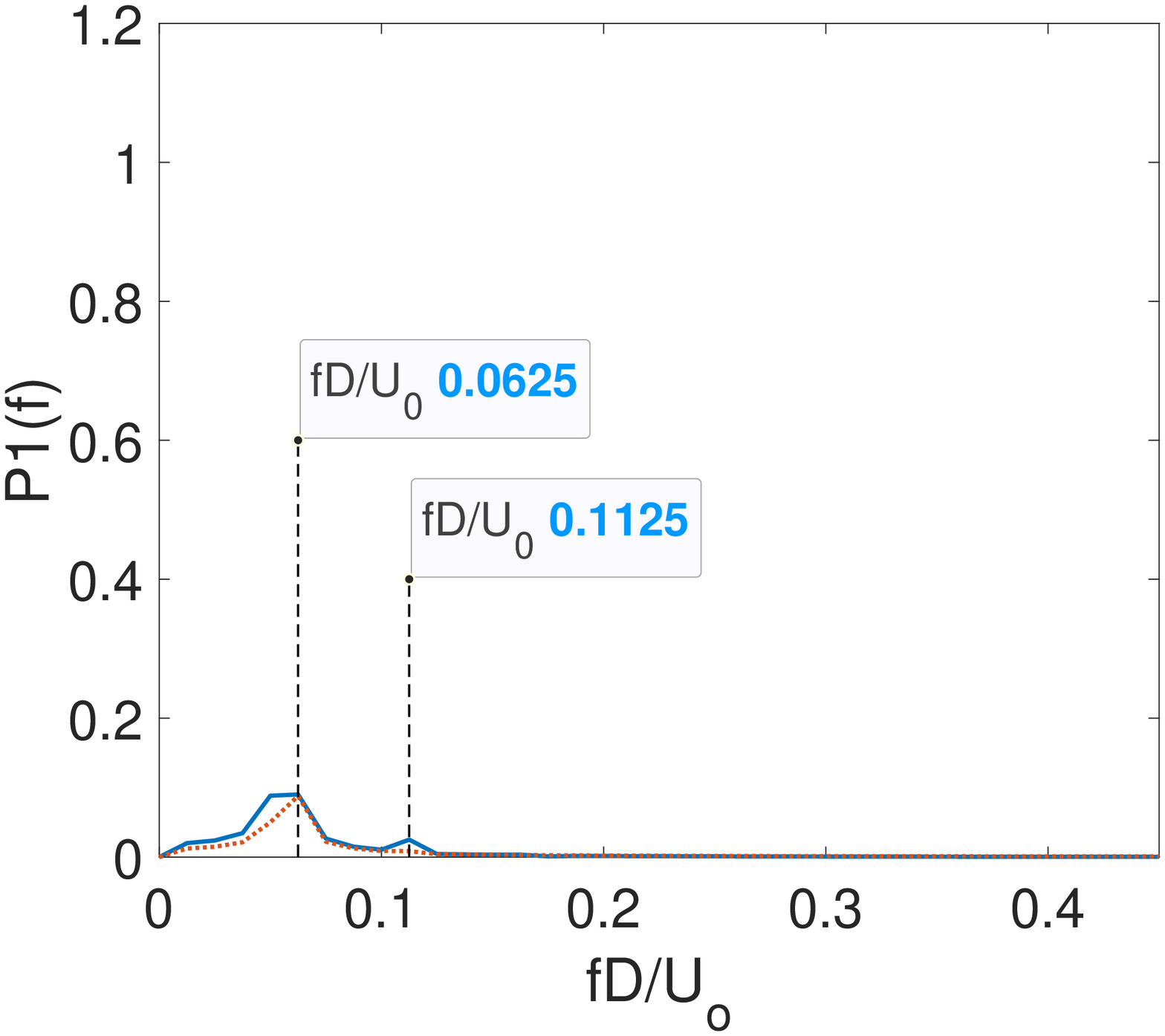}}
    \caption{}\label{fft_1_6}
    \end{subfigure}
\caption{Fast Fourier transform (FFT) of the lift force response for different $U^*$ values at a gap ratio of $g/D = 0.1$ for $m^* = 10$ and $Re = 100$. a) $U^*$ = 15, b) $U^*$ = 16, c) $U^*$ = 18, d) $U^*$ = 20, e) $U^*$ = 25, f) $U^*$ = 30. (bold line represents lift force and dotted line represents displacement) }
\label{fft_G1}
\end{figure}

Figure~\ref{G1_branch} presents $(A_{y}^{max}/D)$, $\Phi$ and  $C_L^{max}$ as a function of different $U^*$ values for gap ratio $g/D = 0.1$. Unlike the earlier cases for $g/D = 0.2, 0.3$ and $0.4$ no jump in the $\Phi$ values can be observed for $g/D = 0.1$. Over the entire lock-in region, the $\Phi$ value remains close to $\pi$. Further, the $C_L^{max}$ subplot presents a contiguous response over the lock-in range with $C_L^{max}$ values first increasing slowly before they gradually reduce back to zero. The entire response hence exhibits characteristics similar to the $\mathrm{II^{nd}}$ branch. The frequency response for $g/D = 0.1$ for $U^*$ values of $15,16,18,20,25$ and $30$ is presented in figure~\ref{fft_G1}. Similar to the frequency response obtained for $g/D = 0.2 - 0.6$, it can be observed that the for $g/D = 0.1$, the single frequency displacement response locks-in with the first frequency response of the lift force. Further, the frequency remains almost constant with increasing $U^*$ values in line with the earlier observation of a contiguous response with characteristics similar to the $\mathrm{II^{nd}}$ branch observed over the entire lock-in region. A comparison of the different gap ratios from $g/D = 0.6-0.1$ presents that the widening of the lock-in region is accompanied by the widening of the $\mathrm{II^{nd}}$ response branch and a narrowing down of the $\mathrm{I^{st}}$ response branch with decreasing gap ratios. As the gap ratio reduces to $g/D = 0.1$, the $\mathrm{I^{st}}$ response branch vanishes completely and a single contiguous lock-in region is observed.   

\section{Conclusions}\label{sec: conclusions}

Two dimensional numerical simulations have been performed to understand the effect of wall proximity on the 2-DOF VIV of an elastically mounted circular cylinder in close proximity to the wall for gap ratios $g/D = ~\{0.1,0.2,0.3,0.4,0.5,0.6\}$. The simulations were performed at $Re = 100$ and $m^*=10$ over a range of $U^*$ and the structural vibration response, vortex structures, lift and drag forces acting on the structure, distribution of pressure, phase difference between the hydrodynamic forces and displacements and finally different VIV branching response regimes have been presented. As the cylinder moves closer to the wall, it was observed that the maximum transverse vibration amplitude reduces, and the lock-in region widens. Further, as the gap ratio reduces, the peak maximum transverse amplitude is observed to occur at a higher $U^*$ value. Similar observations were made for the variation of the root-mean-squared stream-wise displacement response as a function of the different gap ratios. As the gap ratio reduces, it was also observed that the there is a substantial increase in the mean transverse amplitudes that are observed in the lock-in and post lock-in regions.

Periodic vortex shedding is observed for all gap ratios $g/D = 0.1$ to $0.6$ with a single ``S" vortex street shed from the top surface of the cylinder. As the gap ratios reduces, it was observed that the vortex is shed farther away from the cylinder resulting in a smaller low pressure region on the top surface of the cylinder. This results in the reduction of $C_L^{mean}$ values with decreasing gap ratios that is observed in the lock-in region. It was also observed that for gap ratios $g/D < 0.6$, a non-zero $C_L^{mean}$ can also be observed in the pre and post lock-in regions unlike for higher gap ratios where the effect of wall proximity on the $C_L^{mean}$ can only be observed in the lock-in region. $C_D^{mean}$ was observed to decrease with decreasing gap ratios over the pre lock-in, lock-in and post lock-in regions. However, at higher $U^*$, the $C_D^{mean}$ is observed to be constant for gap ratios of $g/D \le 0.3$. In the lock-in region, $C_L^{rms}$ was observed to decrease with the gap ratios, however for $g/D = 0.3$ and $0.4$ a double-lobed $C_L^{rms}$ response was observed. That is after the initial peak, the $C_L^{rms}$ values remained constant over a small range of values before reducing to zero. It is to be noted that for all gap ratios $0.1 \le g/D \le 0.6$, the $C_L^{rms} = $ in the pre and post lock-in regions. Similar behaviour was observed for the variation of $C_D^{rms}$ for the different gap ratios. Both the transverse and stream-wise frequency ratios were observed to be greater than one for smaller gap ratios. Further the ratio of the transverse and stream-wise frequency was observed to be close to one resulting in an oblique elliptical x-y trajectory of the cylinder.

The response dynamics of the vibrating cylinder can be characterized into two branches based on the maximum transverse amplitude, phase angle between transverse displacement and lift force $(\Phi) $and $C_L^{max}$ variation with $U^*$. While the branching response is not apparent from the transverse amplitude plots for lower gap ratios $g/D \le 0.4$, the variation of the phase angle and $C_L^{max}$ present the distinction between the branches. The $\mathrm{I^{st}}$ branch is marked by a $\Phi$ close to zero and the $\mathrm{II^{nd}}$ response branch is marked by a $\Phi$ slightly greater than $\pi$. The transition between the two branches was marked by the sudden shift in $\Phi$ from zero to a value slightly greater than $\pi$. Further, for gap ratios $g/D = 0.3$ and $0.4$, two distinct lobes can be observed in $C_L^{max}$ plots corresponding to the two different branches. It was also identified that during the transition between the branches, the second frequency component of the lift force becomes stronger than the first frequency component. Further, as the gap ratio reduces, the widening of the lock-in region is accompanied by the widening of the $\mathrm{II^{nd}}$ branch and a narrowing down of the $\mathrm{I^{st}}$ branch. As the gap ratio reduces to $g/D = 0.1$, the $\mathrm{I^{st}}$ response branch vanishes entirely, and a single contiguous lock-in region is observed.

\begin{acknowledgments}
The corresponding author would like to acknowledge the financial support from the Science and Engineering Research Board's Start-up Research Grant (SERB-SRG) with sanction order number SRG/2019/001249 and the BITS Pilani's OPERA award.
\end{acknowledgments}
\section*{Data Availability}
The data that support the findings of this study are available from the corresponding author upon request.
%\newpage
\bibliographystyle{unsrt}
\bibliography{References}

\begin{thebibliography}{10}

\bibitem{Sumer2006}
B~Mutlu Sumer.
\newblock {\em Hydrodynamics Around Cylindrical Structures}.
\newblock {World} {Scientific}, sep 2006.

\bibitem{Li2016}
Zhong Li, Weigang Yao, Kun Yang, Rajeev~K. Jaiman, and Boo~Cheong Khoo.
\newblock On the vortex-induced oscillations of a freely vibrating cylinder in
  the vicinity of a stationary plane wall.
\newblock {\em Journal of Fluids and Structures}, 65:495--526, aug 2016.

\bibitem{gurugubelliCAF2018}
PS~Gurugubelli, R~Ghoshal, V~Joshi, and RK~Jaiman.
\newblock A variational projection scheme for nonmatching surface-to-line
  coupling between 3d flexible multibody system and incompressible turbulent
  flow.
\newblock {\em Computers \& Fluids}, 165:160--172, 2018.

\bibitem{Williamson2004}
C.H.K. Williamson and R.~Govardhan.
\newblock Vortex-induced vibrations.
\newblock {\em Annual Review of Fluid Mechanics}, 36(1):413--455, jan 2004.

\bibitem{Bearman1978}
P.~W. Bearman and M.~M. Zdravkovich.
\newblock Flow around a circular cylinder near a plane boundary.
\newblock {\em Journal of Fluid Mechanics}, 89(1):33--47, nov 1978.

\bibitem{govardhan2000}
R.~Govardhan and C.~H.~K. Williamson.
\newblock Modes of vortex formation and frequency response of a freely
  vibrating cylinder.
\newblock {\em Journal of Fluid Mechanics}, 420:85--130, 2000.

\bibitem{Sarpkaya2004}
T.~Sarpkaya.
\newblock A critical review of the intrinsic nature of vortex-induced
  vibrations.
\newblock {\em Journal of Fluids and Structures}, 19(4):389--447, may 2004.

\bibitem{Anagnostopoulos1992}
P.~Anagnostopoulos and P.W. Bearman.
\newblock Response characteristics of a vortex-excited cylinder at low reynolds
  numbers.
\newblock {\em Journal of Fluids and Structures}, 6(1):39--50, 1992.

\bibitem{Anagnostopoulos1994}
Pierre Anagnostopoulos.
\newblock Numerical investigation of response and wake characteristics of a
  vortex-excited cylinder in a uniform stream.
\newblock {\em Journal of Fluids and Structures}, 8:367--390, 1994.

\bibitem{Leontini2006}
J.S. Leontini, M.C. Thompson, and K.~Hourigan.
\newblock The beginning of branching behaviour of vortex-induced vibration
  during two-dimensional flow.
\newblock {\em Journal of Fluids and Structures}, 22(6-7):857--864, aug 2006.

\bibitem{prasanth_mittal_2008}
T.~K. Prasanth and S.~Mittal.
\newblock Vortex-induced vibrations of a circular cylinder at low reynolds
  numbers.
\newblock {\em Journal of Fluid Mechanics}, 594:463–491, 2008.

\bibitem{Taneda1965}
Sadatoshi Taneda.
\newblock Experimental investigation of vortex streets.
\newblock {\em Journal of the Physical Society of Japan}, 20(9):1714--1721, sep
  1965.

\bibitem{Zdravkovich1985}
M.M. Zdravkovich.
\newblock Forces on a circular cylinder near a plane wall.
\newblock {\em Applied Ocean Research}, 7(4):197--201, oct 1985.

\bibitem{Buresti1992}
G.~Buresti and A.~Lanciotti.
\newblock Mean and fluctuating forces on a circular cylinder in cross-flow near
  a plane surface.
\newblock {\em Journal of Wind Engineering and Industrial Aerodynamics},
  41(1-3):639--650, oct 1992.

\bibitem{Lei1999}
C.~Lei, L.~Cheng, and K.~Kavanagh.
\newblock Re-examination of the effect of a plane boundary on force and vortex
  shedding of a circular cylinder.
\newblock {\em Journal of Wind Engineering and Industrial Aerodynamics},
  80(3):263--286, apr 1999.

\bibitem{Price2002}
S.J. Price, D.~Sumner, J.G. Smith, K.~Leong, and M.P. Paidoussis.
\newblock Flow visualization around a circular cylinder near to a plane wall.
\newblock {\em Journal of Fluids and Structures}, 16(2):175--191, feb 2002.

\bibitem{Wang2008}
X.K. Wang and S.K. Tan.
\newblock Comparison of flow patterns in the near wake of a circular cylinder
  and a square cylinder placed near a plane wall.
\newblock {\em Ocean Engineering}, 35(5-6):458--472, apr 2008.

\bibitem{Ong2010}
Muk~Chen Ong, Torbjørn Utnes, Lars~Erik Holmedal, Dag Myrhaug, and Bjørnar
  Pettersen.
\newblock Numerical simulation of flow around a circular cylinder close to a
  flat seabed at high reynolds numbers using a k–ε model.
\newblock {\em Coastal Engineering}, 57(10):931--947, 2010.

\bibitem{Ong2012}
Muk Ong, Torbj{\o}rn Utnes, Lars~Erik Holmedal, Dag Myrhaug, and Bj{\o}rnar
  Pettersen.
\newblock Near-bed flow mechanisms around a circular marine pipeline close to a
  flat seabed in the subcritical flow regime using a k–ε model.
\newblock {\em Journal of Offshore Mechanics and Arctic Engineering},
  134:021803, 05 2012.

\bibitem{Prisc2016}
Mia {Abrahamsen Prsic}, Muk~Chen Ong, Bj{\o}rnar Pettersen, and Dag Myrhaug.
\newblock Large eddy simulations of flow around a circular cylinder close to a
  flat seabed.
\newblock {\em Marine Structures}, 46:127--148, 2016.

\bibitem{Tsahalis1981}
Tsahali and Jones.
\newblock Vortex-induced vibrations of a flexible cylinder near a plane
  boundary in steady flow.
\newblock {\em 13th Offshore Technology Conference, OTCAt: Houston, Texas,
  USA}, 1981.

\bibitem{Fredsoe1987}
J.~Freds{\o}e, B.~M. Sumer, J.~Andersen, and E.~A. Hansen.
\newblock Transverse vibrations of a cylinder very close to a plane wall.
\newblock {\em Journal of Offshore Mechanics and Arctic Engineering},
  109:52--60, 1987.

\bibitem{Yang2009}
Bing Yang, Fuping Gao, Dong-Sheng Jeng, and Yingxiang Wu.
\newblock Experimental study of vortex-induced vibrations of a cylinder near a
  rigid plane boundary in steady flow.
\newblock {\em Acta Mechanica Sinica}, 25(1):51--63, jan 2009.

\bibitem{Hsieh2016}
Shih-Chun Hsieh, Ying~Min Low, and Yee-Meng Chiew.
\newblock Flow characteristics around a circular cylinder subjected to
  vortex-induced vibration near a plane boundary.
\newblock {\em Journal of Fluids and Structures}, 65:257--277, 2016.

\bibitem{Daneswar2020}
Sina Daneshvar and Chris Morton.
\newblock On the vortex-induced vibration of a low mass ratio circular cylinder
  near a planar boundary.
\newblock {\em Ocean Engineering}, 201:107109, 2020.

\bibitem{Barbosa2017}
João~Manuel {de Oliveira Barbosa}, Yang Qu, Andrei~V. Metrikine, and Eliz-Mari
  Lourens.
\newblock Vortex-induced vibrations of a freely vibrating cylinder near a plane
  boundary: Experimental investigation and theoretical modelling.
\newblock {\em Journal of Fluids and Structures}, 69:382--401, 2017.

\bibitem{Feher2021}
Rafael Fehér and Juan~Julca Avila.
\newblock Vortex-induced vibrations model with 2 degrees of freedom of rigid
  cylinders near a plane boundary based on wake oscillator.
\newblock {\em Ocean Engineering}, 234:108938, 2021.

\bibitem{Jin2016}
Yiming Jin and Ping Dong.
\newblock A novel wake oscillator model for simulation of cross-flow vortex
  induced vibrations of a circular cylinder close to a plane boundary.
\newblock {\em Ocean Engineering}, 117:57--62, 2016.

\bibitem{Zhao2011}
Ming Zhao and Liang Cheng.
\newblock Numerical simulation of two-degree-of-freedom vortex-induced
  vibration of a circular cylinder close to a plane boundary.
\newblock {\em Journal of Fluids and Structures}, 27(7):1097--1110, 2011.

\bibitem{Tham2015}
Daniel Mun~Yew Tham, Pardha~S. Gurugubelli, Zhong Li, and Rajeev~K. Jaiman.
\newblock Freely vibrating circular cylinder in the vicinity of a stationary
  wall.
\newblock {\em Journal of Fluids and Structures}, 59:103--128, nov 2015.

\bibitem{Chen2020}
L.F. Chen and G.X. Wu.
\newblock Flow-induced transverse vibration of a circular cylinder close to a
  plane wall at small gap ratios.
\newblock {\em Applied Ocean Research}, 103:102344, 2020.

\bibitem{Chen2021}
Linfeng Chen, Yuhong Dong, and Yitao Wang.
\newblock Flow-induced vibration of a near-wall circular cylinder with a small
  gap ratio at low reynolds numbers.
\newblock {\em Journal of Fluids and Structures}, 103:103247, may 2021.

\bibitem{Chen2019}
Weilin Chen, Chunning Ji, Dong Xu, and John Williams.
\newblock Two-degree-of-freedom vortex-induced vibrations of a circular
  cylinder in the vicinity of a stationary wall.
\newblock {\em Journal of Fluids and Structures}, 91:102728, 2019.

\bibitem{Chen2022}
Weilin Chen, Chunning Ji, Dong Xu, and Zhimeng Zhang.
\newblock Three-dimensional direct numerical simulations of vortex-induced
  vibrations of a circular cylinder in proximity to a stationary wall.
\newblock {\em Phys. Rev. Fluids}, 7:044607, Apr 2022.

\bibitem{jieJCP}
Jie Liu, Rajeev~K Jaiman, and Pardha~S Gurugubelli.
\newblock A stable second-order scheme for fluid-structure interaction with
  strong added-mass effects.
\newblock {\em Journal of Computational Physics}, 270:687--710, 2014.

\bibitem{Hughes1981}
Thomas~J.R. Hughes, Wing~Kam Liu, and Thomas~K. Zimmermann.
\newblock Lagrangian-eulerian finite element formulation for incompressible
  viscous flows.
\newblock {\em Computer Methods in Applied Mechanics and Engineering},
  29(3):329--349, dec 1981.

\bibitem{donea2004}
Jean Donea, Antonio Huerta, J.-Ph. Ponthot, and A.~Rodríguez-Ferran.
\newblock {\em Arbitrary Lagrangian–Eulerian Methods}, chapter~14.
\newblock John Wiley \& Sons, Ltd, 2004.

\bibitem{jaiman2015fully}
Rajeev~K Jaiman, Subhankar Sen, and Pardha~S Gurugubelli.
\newblock A fully implicit combined field scheme for freely vibrating square
  cylinders with sharp and rounded corners.
\newblock {\em Computers \& fluids}, 112:1--18, 2015.

\bibitem{Jiankang2021}
Jiankang Zhou, Xiang Qiu, Jiahua Li, and Yulu Liu.
\newblock The gap ratio effects on vortex evolution behind a circular cylinder
  placed near a wall.
\newblock {\em Physics of Fluids}, 33(3):037112, 2021.

\end{thebibliography}
\end{document}